\documentclass{aa}
\usepackage[varg]{txfonts}
\usepackage{natbib}
\usepackage{lscape}
\usepackage{graphicx}
\usepackage{soul}
\usepackage{csquotes}
\usepackage[utf8]{inputenc}
\begin{document}

\title{Chemical Features in the Circumnuclear Disk of the Galactic Center \thanks{Figures 5-9 are only available in electronic form via
http://www.edpsciences.org} \thanks{Based on observations carried out with the IRAM 30m Telescope. IRAM is supported by INSU/CNRS (France), MPG (Germany) and IGN (Spain).}} 

\subtitle{} 

\author{N. Harada \inst{1, 2}
\and D. Riquelme \inst{1}
\and S. Viti \inst{3}
\and I. Jim{\'e}nez-Serra \inst{3,4} 
\and M. A. Requena-Torres \inst{1} 
\and K. M. Menten \inst{1}
\and S. Mart{\'i}n \inst{5,6}
\and R. Aladro \inst{6,7}
\and J. Martin-Pintado \inst{8}
\and S. Hochg\"{u}rtel \inst{1}
} 

\offprints{Nanase Harada, \email{harada@asiaa.sinica.edu.tw}}

\institute{Max Planck Institute for Radio Astronomy, Auf dem Huegel 69, D-53121 Bonn, Germany 
\and Academia Sinica Institute of Astronomy and Astrophysics, No.1, Sec. 4, Roosevelt Rd, Taipei 10617, Taiwan, R.O.C.
\and University College London, Gower Street, London, WC1E 6BT, UK
\and European Southern Observatory, Karl-Schwarzschild-Str. 2, D-85748 Garching, Germany
\and Institut de Radio Astronomie Millim{\'e}trique, 300 rue de la Piscine, Dom. Univ., F-38406, St. Martin d'H{\`e}res, France
\and European Southern Observatory, Avda. Alonso de C{\'o}rdova 3107, Vitacura, Santiago, Chile
\and Department of Earth and Space Sciences, Chalmers University of Technology, Onsala Observatory, 439 92 Onsala, Sweden
\and Centro de Astrobiolog{\'i}a (CSIC-INTA), Ctra. de Torrej{\'o}n Ajalvir km 4, E-28850 Torrej{\'o}n de Ardoz, Madrid, Spain}
\date{Received .../ Accepted ...}

\abstract {}
{ The circumnuclear disk (CND) of the Galactic Center is exposed to many energetic phenomena coming from the supermassive black hole Sgr A* and stellar activities. These energetic activities can affect the chemical composition in the CND by the interaction with UV-photons, cosmic-rays, X-rays, and shock waves. We aim to constrain the physical conditions present in the CND by chemical modeling of observed molecular species detected towards it.}
 {We analyzed a selected set of molecular line data taken toward a position in the southwest lobe of the CND with the IRAM 30m and APEX 12-meter telescopes and derived the column density of each molecule using a large velocity gradient (LVG) analysis. The determined chemical composition is compared with a time-dependent gas-grain chemical model based on the UCL\_CHEM code that includes the effects of shock waves with varying physical parameters.} 
{Molecules such as CO, HCN, HCO$^+$, HNC, CS, SO, SiO, NO, CN, H$_2$CO, HC$_3$N, N$_2$H$^+$ and H$_3$O$^+$ are detected and their column densities are obtained. Total hydrogen densities obtained from LVG analysis range between $2 \times 10^4$ and $1 \times 10^6\,$cm$^{-3}$ and most species indicate values around several $\times 10^5\,$cm$^{-3}$, which are lower than values corresponding to the Roche limit, which shows that the CND is tidally unstable. The chemical models show good agreement with the observations in cases where the density is $\sim10^4\,$cm$^{-3}$, the cosmic-ray ionization rate is high, $>10^{-15} \,$s$^{-1}$, or shocks with velocities $> 40\,$km s$^{-1}$ have occurred.} 
{Comparison of models and observations favors a scenario where the cosmic-ray ionization rate in the CND is high, but precise effects of other factors such as shocks, density structures, UV-photons and X-rays from the Sgr A* must be examined with higher spatial resolution data.}

\maketitle

\section{Introduction}
Located at $d \sim 8.4 \,$kpc from the Sun \citep{2009ApJ...700..137R}, the center of the Milky Way is the nearest galactic nucleus.
The supermassive black hole at its center is not currently active with an Eddington ratio of only $10^{-9}$, but X-ray observations suggest possible higher activity in the past  \citep[e.g., ][]{2013ASSP...34..331P}.
Even within the central parsec, at a close vicinity of Sgr A*, a massive star cluster is found with a total stellar mass of  $\sim 10^6 M_\odot$ \citep{2009A&A...502...91S}. Star formation near the black hole is difficult to explain due to tidal shear. It has been proposed that these massive stars may have formed in situ or farther away from the black hole and migrated inwards \citep[][ and references therein]{2006ApJ...643.1011P}.
 The black hole is surrounded by the circumnuclear disk (CND) composed of molecular and atomic gas \citep{1987ApJ...318..124G}.  The CND is actually a ring-like structure surrounding the very central region with an inner radius of 1.5 pc ($\sim$30 arcsec) that extends up to 7 pc at negative Galactic longitudes \citep{2011ApJ...732..120O}.

Studying the molecular gas in the CND might yield information as to what extent stars can be formed in situ or not. The molecular inventory of the CND has been mapped in different tracers such as CO, CS and HCN
 \citep{1986A&A...169...85S,1986Sci...233.1394L,1987ApJ...318..124G}. 
 \citet{2005ApJ...623..866B} found a high ($ T_{\rm kin} \sim 200-300$\,K) kinetic temperature of the molecular gas, and proposed shocks as the dominant heating source. 
 Higher resolution images of the HCN and HCO$^{+}$ $(1-0)$ lines were obtained by \citet{2005ApJ...622..346C} and they concluded that the molecular gas has high enough densities to be tidally stable. 
 On the other hand, \citet{2012A&A...542L..21R} conducted a multi-$J$ transition analysis of CO, and derived a much lower density of $n_{\rm H} \sim 6\times 10^4\,$cm$^{-3}$ for most of the material. Besides the study of physical conditions by above authors, \citet{2013ApJ...779...47M} and \citet{2014MNRAS.437.3159S} used multi-$J$ lines of HCN to estimate the excitation conditions of the gas. Their derived densities range somewhere between those derived by \citet{2005ApJ...622..346C} and \citet{2012A&A...542L..21R}. \citet{2012A&A...539A..29M} explored the chemical complexity by mapping CN, SiO, H$_{2}$CO, HC$_3$N in the CND. Recently,  a line survey in the CND of the 3mm wavelength range was conducted by \citet{2014ApJS..214....2T}.  
 
 Probing the physical conditions of molecular material is a complex problem. The abundance of a molecule may be influenced by a variety of phenomena. Moreover, even for a \textit{given} molecule, different transitions with different frequencies and intrinsic strengths may require a wide range of critical densities for their excitation. In turn, analysis of multi-transition datasets with, e.g., a large velocity gradient (LVG) analysis using a radiative transfer program may yield estimates of the kinetic temperature and density. In general, it is necessary to observe a variety of molecules as wide as possible over the broadest feasible frequency range. Riquelme et al. (in preparation) conducted a line survey using the IRAM 30 meter telescope towards a position in the southwest lobe of the CND, covering 80 to 115 GHz and 129 to 175 GHz. 
Hochg\"{u}rtel (2013, PhD thesis) surveyed a similar region using the FLASH+ receiver \citep{klein_flash} on the APEX 12-meter telescope \citep[270 to 379 GHz and 385 to 509 GHz;][]{2006A&A...454L..13G}.
Based on their results, this paper studies possible interpretations of the chemical features observed in the CND.

 In section 2, we discuss the processes that can affect the CND's physical conditions and chemistry. Column densities from the LVG analysis are presented in Section 3. The details of the chemical model are presented in Section 4, while we discuss the results of the chemical modeling in Section 5. The interpretation of the results is discussed in Section 6, and main results are summarized in Section 7.
 
\section{Physical conditions and the heating of the CND}
\label{sec:phys_cond}
In this section we summarize the environment within and surrounding the CND that can affect the chemical composition of the molecular gas. Ionizing sources such as UV-photons, cosmic rays, and X-rays can all dissociate molecules as well as heat the gas. 
Shock waves may also provide very efficient heating. Fast shocks also ionize the gas. These factors are also dependent on the density distribution of the molecular clump.
\subsection{Kinetic Temperature and Heating Mechanisms}
Observed temperatures can be elevated by any of above heating sources. Similar to the results by \citet{2005ApJ...623..866B}, \citet{2012A&A...542L..21R} constrained the temperatures of the molecular clouds in the southwest lobe of the CND to $T_{\rm kin} = 200^{+300}_{-70}\,$K for a low excitation component and $T_{\rm kin} = 500^{+100}_{-210}\,$K for a high excitation one. Possible heating mechanisms are as follows:

{\bf Shocks: } \citet{2005ApJ...623..866B} stated that this high temperature cannot be explained by UV-photons because of the high extinction in the clump, and proposed shock heating. In addition to the gas heating, fast shock waves can sputter molecules off the dust surface \citep{1997A&A...321..293S}, which may dramatically change the chemical composition in the gas phase and in particular sputtering is known to increase the abundance of SiO \citep{1997ApJ...482L..45M}. 

{\bf X-rays: }A strong X-ray flux can affect the chemistry by ionizing and heating its surroundings creating an X-ray dominated region (XDR). The current value of unabsorbed X-ray flux from Sgr A* in its quiescent state, i.e, not during a flare, is estimated to be $L\sim 2 \times 10^{33} \,$erg s$^{-1}$ in $2-10\,$keV range (Baganoff et al 2003). There are other X-ray sources nearby (Degenaar et al. 2012), but those are either too weak or too far to affect the molecular 
cloud. An estimation of the X-ray ionization rate as a function of the intervening total hydrogen column density is presented in Section \ref{sec:appen_xray} and its effect is discussed in Section \ref{sec:disc}. Past high X-ray luminosities caused by increased activity of the central black hole have been suggested in the literature \citep[e.g., ][]{2004A&A...425L..49R, 2007PASJ...59S.245K,2010PASJ...62..423N,2010ApJ...719..143T,2010ApJ...714..732P,2012A&A...545A..35C}. Although higher X-ray fluxes could affect the chemistry, the flares causing them are likely of short duration, and may have a minimal effect on the chemistry given that the average flux is much lower.

{\bf UV-photons: }The star clusters in the center of the CND provide a strong UV field that could affect the chemistry.  Taking $L_*=1.1\times10^{7}\,L_{\odot}$ and $d=1-3\,$pc, the UV radiation field would be $G_{0} \sim10^4-10^5$ Habing field units if we use the rough estimation described in Section \ref{sec:app_uv}.
This high radiation field can affect the chemistry via photo-dissociation or photo-ionization of molecules. 
On the other hand, the effect of UV-photon enhancement on the abundances of certain species can be suppressed for total hydrogen column densities  $> N_{\rm H} \sim 10^{22}\,$cm$^{-3}$, i.e., $A_v \sim 5$ (see Section \ref{sec:disc_uv}). 
Since the density of molecular clouds is expected to be high ($n_{\rm H}>6\times 10^4 \,$cm$^{-3}$), the depth of the photon-dominated regions (PDRs), i.e., the interface between the ionized and the molecular regions, must be less than $0.06\,$pc wide. 
Interferometric data show that the source size of our target molecular cloud has a size of $\sim$ 0.3 pc \citep{2005ApJ...622..346C}. Considering it to be a 0.3-pc diameter sphere illuminated from one direction, 27\% of the material could be considered to be a PDR ($A_{\rm V} < 5$) when $n_{\rm H}=6\times10^4\,$cm$^{-3}$, and 9 \% for $n_{\rm H}=2\times 10^5\,$cm$^{-3}$. Therefore, PDRs may have some effect on the chemistry if the density is low, but the fraction of PDR is insignificant within the single-dish beam of our observations (from 0.5 to 1.1 pc) when the density is as high as what we derive from the LVG analysis for most of the species as described in a later section of this paper.
The detailed discussion of the effects of a PDR is discussed separately in Section \ref{sec:disc_uv}.

{\bf Cosmic-rays:}  An enhancement of the cosmic-ray ionization rate has been proposed as an alternative source of heating for the gas \cite[e.g., ][]{2013ApJ...762...33Y}. Although cosmic rays are likely to be an additional heating source, its ionization rate must be extremely high $\zeta >10^{-13}\,$s$^{-1}$ in order to achieve a high temperature of $T_{\rm kin}>280\,$K \citep[see also ][ for a comparison of different heating mechanisms]{2013A&A...550A.135A}. Cosmic-rays also provide direct ionization and dissociation. The electrons produced by the ionization can also produce UV-photons by exciting H$_2$ molecules to the Lyman and Werner bands \citep{1983ApJ...267..603P}. Both X-rays and cosmic-rays can generate these secondary UV-photons; the effects of X-rays on the chemistry are similar to those of cosmic rays although X-rays heat the gas more efficiently than cosmic rays. Although high-energy $\gamma$-ray emission from the very center of the Galaxy provides strong evidence for cosmic rays \citep{2011ApJ...726...60C}, it is hard to constrain the cosmic-ray ionization rates of molecules from the $\gamma$-ray observations. This is due to the strong dependence of cosmic-ray ionization rates on the lower energy term of the cosmic-ray spectrum \citep{2009A&A...501..619P,2009ApJ...694..257I}, which is not well-known. It has been claimed that in the Central Molecular Zone (CMZ), a region of $\sim$ a few $\times$ 100 pc around Sgr A$^*$, $\zeta$ is higher than in regular spiral-arm Galactic giant molecular clouds (GMCs), with $\zeta \gtrsim 1 \times 10^{-15}\,$s$^{-1}$ \citep{2008ApJ...688..306G, 2013JPCA..117.9919G,2014ApJ...786...96G,2013ApJ...762...33Y}. Observations of high-energy gamma rays show higher flux around the Sgr A* (see above), and a model of gamma-ray emission by \citet{2011ApJ...726...60C} shows that an injection of lower energy cosmic-rays $10^4\,$yrs ago best reproduces the observed spectrum.

\subsection{Volume Densities }
A large range of volume densities\footnote{In this paper, number densities are presented in terms of total hydrogen densities, which are common in theoretical papers instead of $H_2$ densities, which are conventionally used in observational papers.}, $n_{\rm H}$, has been reported for the molecular gas in the CND. A large velocity gradient (LVG) analysis of multi-$J$ CO lines yields to $n_{\rm H}= (0.4-8)\times 10^5\,$cm$^{-3}$ \citep{2005ApJ...623..866B,2012A&A...542L..21R}, lower values than those estimated from the HCN emission \citep[$n_{\rm H} \gtrsim 10^7\,$cm$^{-3}$,][]{2009ApJ...695.1477M,2005ApJ...622..346C}. However, recent analysis of multi-$J$ HCN lines reduced the discrepancy with the densities derived from CO observations \citep[$n_{\rm H} = (0.2 - 20) \times 10^6\,$cm$^{-3}$, ][]{2014MNRAS.437.3159S,2013ApJ...779...47M} for most of the CND clumps because the above-mentioned previous studies needed to assume the kinetic temperature for their single-line analysis. An exception is a clump with a very high density of $n_{\rm H}=(0.6-8)\times 10^7$ cm$^{-3}$ found by \citet{2013ApJ...779...47M} in the southwest lobe in which a vibrationally excited HCN line was observed.

\section{Observations and LVG analysis}\label{sec:obs}
\begin{figure}[htbp]
\begin{center}
\includegraphics[width=0.4\textwidth]{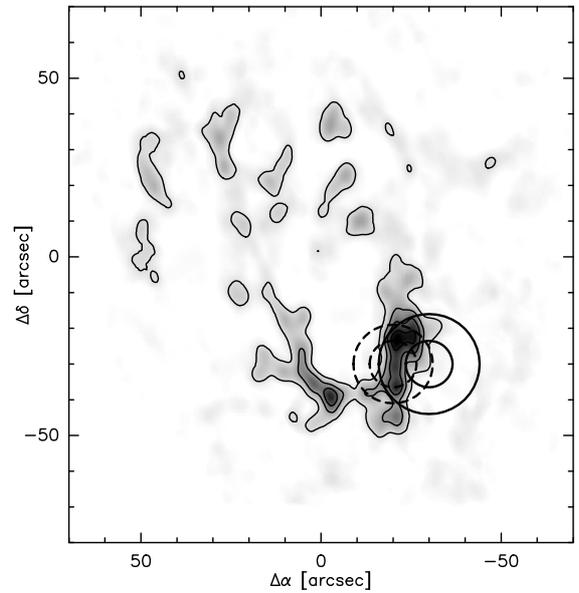}
\caption{A contour map of HCN (4-3) emission in the CND from \citet{2009ApJ...695.1477M}. Circles in solid lines show minimum and maximum beam sizes for the IRAM 30m telescope, and circles in dotted lines are for APEX FLASH instrument. The central position is the location of Sgr A* at EQ J2000 $\alpha=$17:45:39.99, $\delta=$-29:00:26.6.
One arcsecond corresponds to 0.04 pc using the distance to the Galactic Center as $d=8.4\,$kpc}
\label{fig:cnd}
\end{center}
\end{figure}

The observations were conducted with 3- and 2-mm receivers of the IRAM 30m telescope and the FLASH+ receiver of the APEX telescope. Detailed information on the data will be presented in Riquelme et al. (in preparation). The observed positions of our survey are shown in Figure \ref{fig:cnd}.
The survey position of the IRAM 30m telescope is at an offset $\Delta \alpha=-30"$, $\Delta \delta  =-30"$ and this telescope has an FWHM of 28" at 86 GHz and 16" at 145 GHz. The APEX survey was taken with the FLASH+ receiver observing a position with an offset of $\Delta \alpha=-20.1"$, $\Delta \delta = -30.1"$ with a FWHM of 22" at 262 GHz and 12" at 516 GHz. All offsets are relative to the position of Sgr A* (the absolute coordinates: $\alpha_{J2000}=$17:45:39.99, $\delta_{J2000}=$-29:00:26.6). The maximum and minimum beam sizes are shown in solid lines (IRAM 30m) and in dotted lines (APEX - FLASH+) in Figure \ref{fig:cnd}. The molecular transitions used in this analysis are listed in Table \ref{tab:lines}.

\begin{figure}[htbp]
\begin{center}
\includegraphics[width=0.48\textwidth]{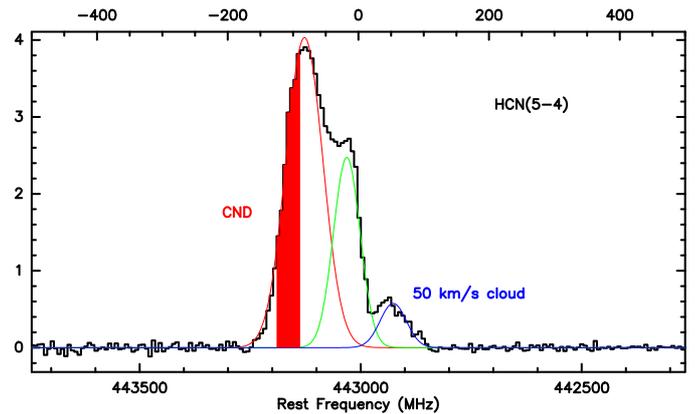}
\caption{A spectrum of HCN(5-4) is shown as an example of typical velocity components in the line of sight. A component centered at $\sim 50$ km s$^{-1}$ is from \enquote{50 km s$^{-1}$ cloud,} whereas other two components are from the CND. The velocity range used for the analysis, which should come from a cloud in the southwest lobe, is highlighted.}
\label{fig:hcn54}
\end{center}
\end{figure}

There are multiple velocity components along the same line of sight. Figure \ref{fig:hcn54} shows a spectrum of HCN(5-4) as an example of those velocity components. The emission from positive velocities arises from M-0.02-0.07 (\enquote{20 km s$^{-1}$ cloud}) or M-0.13-0.08 (\enquote{50 km s$^{-1}$ cloud}), which may be interacting with the CND. The emission at negative velocities is thought to come from the CND \citep{2011A&A...526A..54A}. Interferometric data \citep[e.g.,][]{2005ApJ...622..346C} show a few peaks in our beam, namely, clumps O, P, and Q in their notation, each centered at $-108$, $-73$, and $-38$ km s$^{-1}$. We select the velocity range for our study based on the spectra shown in \citet{2009ApJ...695.1477M}. Clump Q is centered at around $\sim -40\,$km s$^{-1}$, and may be affected by self-absorption. Clump P is centered around $-70$ km s$^{-1}$ while Clump O is centered at around $-110$ km s$^{-1}$. Since emission with the velocity $v > -90\,$km s$^{-1}$ is likely to arise from all three clouds, we use the velocity cut from $v = -90$ to $-120\,$km s$^{-1}$ for our analysis to target only Clump O, located at the Galactic offset around $\Delta \alpha  = -20"$, $\Delta \delta = -20"$. Smaller beams in higher frequency observations of each telescope will only partially cover our target clump. Thus, we treated the line intensities differently for lines above 100 GHz for IRAM observations and lines above 400 GHz for APEX observations. When there are more than 3 transitions observed in other frequency ranges for the species (CS, SO, and SiO), these lines are omitted from the analysis, otherwise we assumed large ( 50\%) uncertainties.

Since our lower $J$ $^{12}$CO data may be affected by high optical depth, we also used the $^{13}$CO(3-2), CO(6-5) and CO(7-6) transition for the LVG analysis of CO. When using $^{13}$CO, an isotopic ratio of $^{12}C/^{13}C =$ 25 is assumed, following  \citet{2012A&A...542L..21R}. Data of CO (6-5) and (7-6) are taken from \citet{2012A&A...542L..21R} where they mapped the whole CND with the CHAMP+ receiver \citep{2006SPIE.6275E..0NK} at the APEX telescope. It is also expected that HCN is optically thick, and H$^{13}$CN lines are used in the LVG analysis assuming the same isotopic ratio.

With the obtained velocity-integrated intensities, we conduct an LVG analysis using the non-LTE radiative transfer code RADEX \citep{2007A&A...468..627V} covering the parameter ranges: $n_{\rm H} = 2 \times 10^2 - 2\times10^7 \,$cm$^{-3}$, column densities of species  X $N(X) = 10^{11} - 10^{19}\,$cm$^{-3}$, and fixed kinetic temperatures of $T_{\rm kin}=100$, 300, and 500\,K. The interferometric HCN (1-0) data show that the cloud of our interest has a size of 0.33 pc \citep{2005ApJ...622..346C}, while a smaller source size of 0.16 pc is reported from mapping of higher-$J$ lines \citep{2009ApJ...695.1477M}. Thus, we run our analysis for different source sizes of 0.1, 0.2, and 0.3 pc. Since our LVG analysis did not yield valid solutions for a source size $< 0.3$\,pc, we assum a source size of 0.3 pc for the results shown in the following sections. The brightness temperature was derived as $T_{b}=T_{MB} {\theta_s^2}/({\theta_s^2 + \theta_b^2})$, where $T_{MB}$ is the measured main beam temperature, $\theta_{s}$ is the source size, and $\theta_b$ is the beam size of the telescope.
 The observed lines in the CND are listed in Table\ref{tab:lines} while results from the LVG analysis are shown in Table \ref{tab:obs_spc}. The derived values of column densities have little dependence on temperatures. Other than the listed species, CH$_{3}$OH and C$_{2}$H were also detected with strong intensities, but their lines were blended, making the LVG analysis very uncertain. These blended lines are omitted from the analysis. Since only one line of NO was detected, a temperature of $T_{\rm kin}=300\,$K and a density range of $n_{\rm H}=2\times 10^5 - 2\times 10^6 \,$cm$^{-3}$ are assumed for this molecule to obtain the range of column density. The column densities obtained from our LVG analysis at $T_{\rm kin}=300\,$K as well as the best-fit H$_2$ densities are listed in Table \ref{tab:obs_spc}. Due to the uncertainty in the high frequency observations mentioned above, volume densities for these species (CN, H$_{2}$CO, N$_{2}$H$^+$, HCN, HC$_3$N, and H$^{13}$CO$^+$) are very uncertain and therefore not shown.

\section{Chemical Models}\label{sec:chem_model}
 To model the chemistry of the observed molecular cloud in the CND we use the time-dependent gas-grain code, UCL\_CHEM \citep{2004MNRAS.354.1141V}, complemented with the parametric shock model of \citet{2008A&A...482..549J}. Gas-phase reactions in this model are taken from the Rate06 version of the UMIST database \citep{2007A&A...466.1197W}. Details of the coupled code can be found in \citet{2011ApJ...740L...3V}. For our study we run the code in two phases. In Phase I, a diffuse cloud with an initial density of 100 cm$^{-3}$ collapses to a core of chosen pre-shock densities. Models are run assuming that initially half of hydrogen is in atomic form while the other half in molecular form, C$^+$, S$^+$, Mg$^+$, Si$^+$ are in ionic form, and O, N, and He are in the atomic form.
      We assume free fall collapse and a cosmic ray ionization rate of $\zeta = 10^{-17}\,$s$^{-1}$ and a standard interstellar radiation field, $G_0=1$.
 The models are run for 3 different pre-shock densities, $n_{\rm H}=2\times 10^4$, $2\times 10^5$, and $2\times 10^6\,$cm$^{-3}$. 
In Phase II, the code is run for 4 shock velocities of $v_s=10, 20, 30$ and 40 km s$^{-1}$ and the molecular abundances are followed along the propagation of the shock.  We also run models with an increased cosmic-ray ionization rate but without shocks. Maximum temperatures and post-shock densities for corresponding velocities and densities are summarized in Table \ref{tab:param}. 
Cosmic-ray ionization rates are varied from $\zeta = 10^{-17}$ to $10^{-13}\,$s$^{-1}$. In this paper, values of $\zeta$ are expressed with respect to H$_{2}$, not H (i.e., $\zeta_{H2}$). In Phase II, $\zeta$ is enhanced at the same time as the shock waves pass. 

\begin{figure*}[ht]
\centerline{\includegraphics[angle=0,width=.4\textwidth]{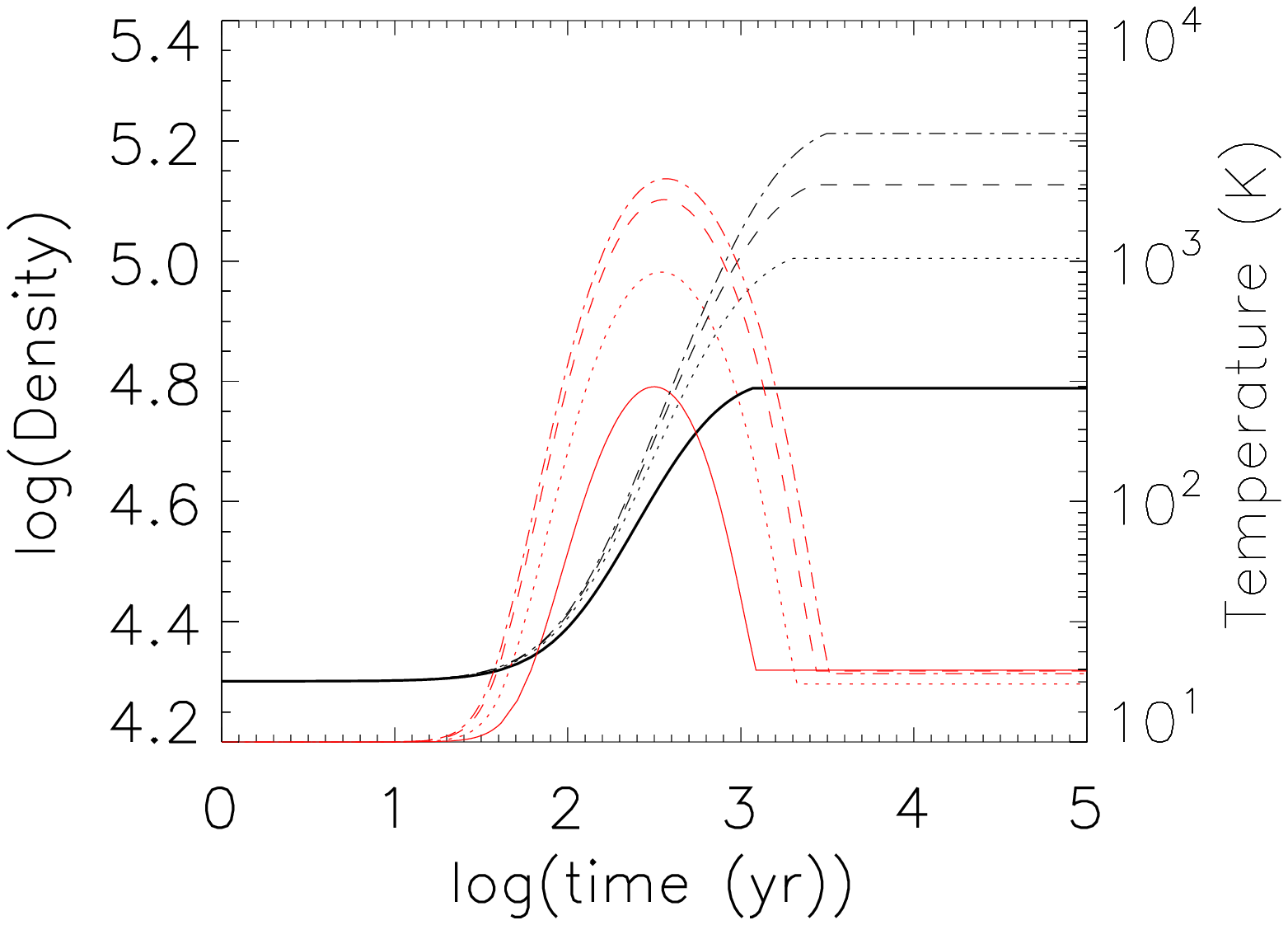}~~~~~
\includegraphics[angle=0,width=.4\textwidth]{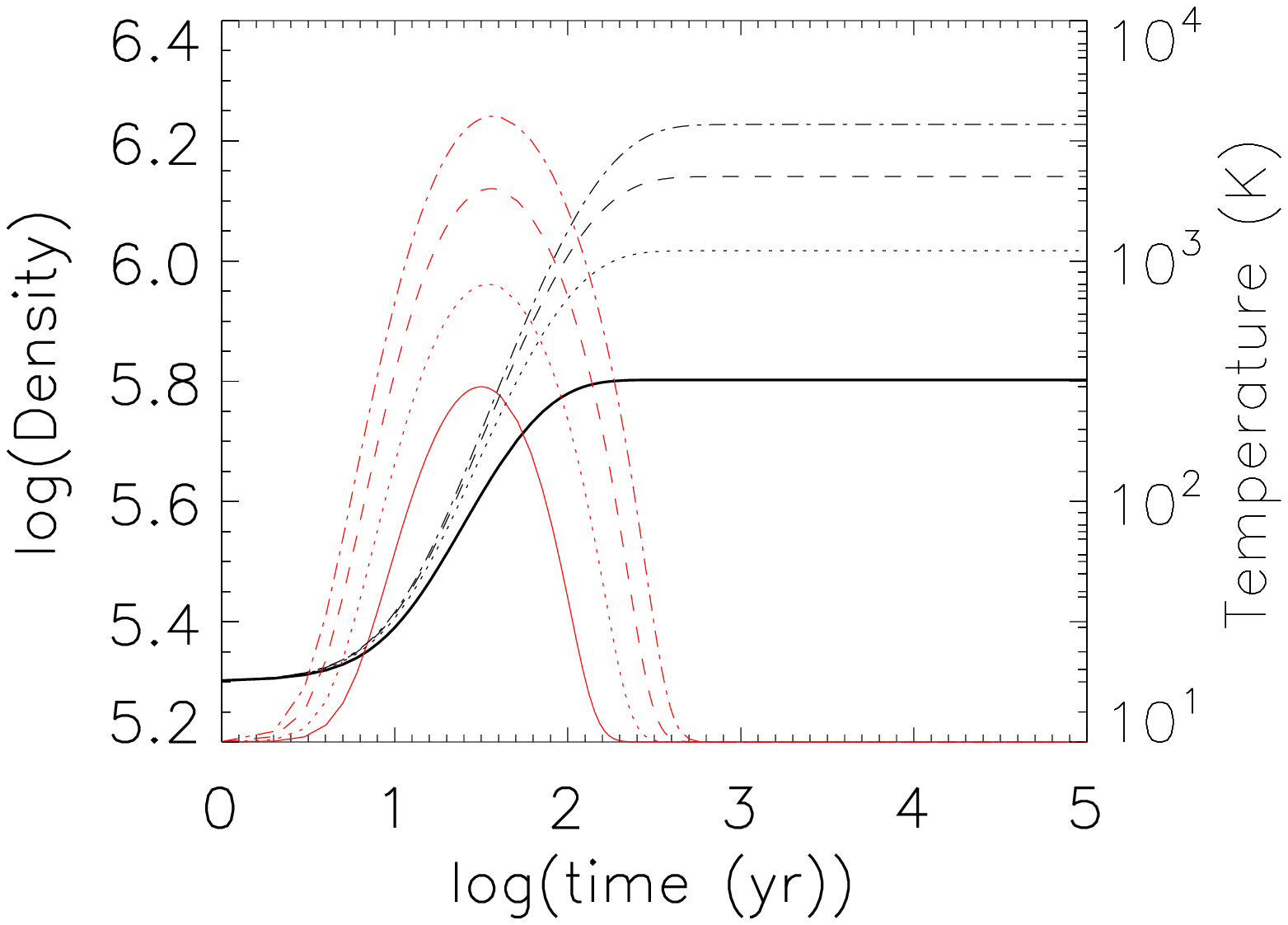}}
\caption{The temperatures (red lines with the scale on the right ordinate) and densities (black lines with the scale on the left ordinate) after the shock are shown for $n_H=2\times10^4\,$cm$^{-3}$ ({\it left panel}) and $n_H=2\times10^5\,$cm$^{-3}$ ({\it right panel}) with varying shock velocities of $v_s =10$, 20, 30, and 40 km s$^{-1}$. Different velocities are indicated with solid lines (10 km s$^{-1}$), dotted lines (20 km s$^{-1}$), dashed lines (30 km s$^{-1}$), and dash-dotted lines (40 km s$^{-1}$). 
}
\label{fig:physcon}
\end{figure*}

 The effect of sputtering is included by considering saturation time-scales for silicon, times for which the silicon abundance in the gas-phase changes by less than 10\% between two time steps by sputtering \cite[see ][]{2008A&A...482..549J}. Saturation time scales are also shown in Table \ref{tab:param} and temperatures and densities for $n_{\rm H, pre}=2\times10^4\,$cm$^{-3}$ and $n_{\rm H, pre}=2\times10^5\,$cm$^{-3}$ with shock velocities of $v_{s} =10$, 20, 30, and 40 km s$^{-1}$ are shown in Figure \ref{fig:physcon}.

 Sputtering by shock waves can vary the elemental abundances of the molecular clouds. For different shock velocities, different elemental abundances E1 and E2 are used (Table \ref{tab:elem}). When the shock velocities are smaller than 20 km s$^{-1}$, the elemental abundances of E1, "low-metal abundances", were applied. For higher shock velocity where the core sputtering of grains needs to be considered, E2 was used. This set of elemental abundances considers the depletion of metals onto the dust grains and is widely used for cold core chemical modeling as in \citet{2008ApJ...680..371W}. Since ice mantles are sputtered by shocks only partially for $v_{s} = 10\,$km s$^{-1}$, only 10 \% of the mantles are considered to be sputtered. For shock velocities $v_{s} \geq 30\,$km s$^{-1}$, sputtering from the core of dust grains starts to occur. Thus, the higher elemental abundances of E2 were used for shock velocities of $v_{s}=30$ and 40 km s$^{-1}$ considering this core sputtering. The fraction of core sputtering of Si in E2 is taken from the results by \citet{2008A&A...482..549J} for the degree of core sputtering caused by $v_{s}=30$ and 40 km s$^{-1}$. The value of Mg in E2 is estimated by assuming that 3 \% of elemental Mg is sputtered from the grain core, which is a similar percentage as that of elemental Si. The sum of the elemental abundances for Mg and Fe was used for the Mg elemental abundance since the ionization of both Mg and Fe inject large amounts of free electrons which affects the chemistry, and this model currently does not contain Fe chemistry. The total elemental abundances of Si, Mg, and Fe abundances are taken from \citet{1989GeCoA..53..197A}.
 
 The treatment of Si core sputtering for $v_s \geq 30\,$km s$^{-1}$ is only approximated in our model. Precise fractional abundances should be obtained with models such as those by \citet{2008A&A...482..809G,2008A&A...482..549J, 2012MNRAS.421.2786F}, especially for SiO. In addition, physical parameters such as the temperature and the density are parameterized in our model. For precise values of physical parameters, see \citet{2008A&A...482..809G,2012MNRAS.421.2786F}.

 In order to study the effects of UV-photons, we use the publicly available Meudon PDR code {\it version 1.4.4} \citep{2006ApJS..164..506L}, which we run separately from the shock model code. This code calculates steady-state abundances without the time dependence. Since the chemistry can reach steady state at a very low $A_{\rm V}$ within about 100 years, results in low $A_{\rm V}$ should be reasonable.

\section{Results of Chemical Modeling}
\subsection{Dependence on the Physical Conditions}
As previously mentioned, chemical abundances are dependent on their pre-shock values, the physical structures of shock waves, cosmic-ray ionization rates, elemental abundances, densities, UV-photons, and their time evolution. In this subsection, we summarize the behavior of chemical abundances in the shock model with the variation of these parameters. First, the fractional abundances in pre-shock conditions are listed in Table \ref{tab:preshock} for each pre-shock density of $n_{\rm H}=2\times 10^4,~2\times10^5$, and $2\times10^6\,$cm$^{-3}$, as a result of Phase I modeling. Since observed regions include different time steps of the shock passage, fractional abundances averaged over time are calculated. Plots of these time-averaged fractional abundances of observed species and H$_{2}$O in the post-shock phase (Phase II) are shown as examples in Figures \ref{fig:n2e5_v20_em2_1} \& \ref{fig:n2e5_v20_em2_2} for $n_{\rm H}=2\times10^5\,$cm$^{-1}$, $v=20\,$km s$^{-1}$, and varying cosmic-ray ionization rates of $\zeta = 1 \times 10^{-17}\,$s$^{-1} -  1 \times 10^{-13}\,$s$^{-1}$ to see the effects of cosmic-ray ionization.  In order to demonstrate the effects of varying shock velocities, Figures \ref{fig:n2e5_z-16_1} and \ref{fig:n2e5_z-16_2} show post-shock time-averaged fractional abundances of observed species and H$_{2}$O for pre-shock densities of $n_{\rm H}=2\times 10^5\,$cm$^{-3}$, a cosmic-ray ionization rate of $\zeta = 1 \times 10^{-16}\,$s$^{-1}$, with varying shock velocities of 10, 20, 30, and 40 km s$^{-1}$. As shown in Table \ref{tab:param}, higher shock velocities mean higher maximum temperatures.  Abundance ratios of H$_{3}$O$^{+}$/H$_{2}$O and HNC/HCN are also shown in Figure \ref{fig:n2e5_v20_em2_2} and \ref{fig:n2e5_z-16_2}, respectively.

\subsubsection{Pre-shock abundances}

Molecules on the grain mantle are ejected after the shock waves pass, producing an increase of their gas-phase abundances in the post-shock phase. Significant amounts of some species such as HCN, HNC, H$_{2}$CO and NO are depleted on the grain surface before the shock passes (Table \ref{tab:preshock}). For other species, even if the fractional abundances in the ice phase are not very high, their post-shock abundances in the gas phase can be increased from the ejection of other species. For example, ionic species (e.g., HCO$^+$ and H$_3$O$^+$) are assumed to be neutralized when they are depleted on grains, but their abundances can be enhanced after shock when their neutral counterparts (e.g., CO and H$_2$O) that have high abundances on ice mantles are released in the gas phase. Water, CO, and N$_{2}$ have high fractional abundances on the ice ($\sim 3\times 10^{-5} - 1\times 10^{-4}$, $\sim 3\times 10^{-7} - 4\times 10^{-5}$, $\sim 2\times 10^{-7}  - 6\times 10^{-6}$, respectively, with higher ice abundances for higher pre-shock densities), which enhance the abundances of their protonated species H$_{3}$O$^{+}$, HCO$^{+}$, and N$_{2}$H$^{+}$ after the shock passes. In addition, the fractional abundance of H$_{2}$S on the ice is very high since our model favors hydrogenation reactions on grains. Thus, the sputtering of H$_{2}$S ice also causes the increase of the fractional abundances of sulfur species such as CS and SO. Similarly, major Si-containing species on ice (SiH$_{4}$ in our model) can react to form SiO after its ejection into the gas-phase \citep{1997A&A...321..293S}. It has been shown from the study of low-mass pre-stellar cores that the evaporation of CH$_{4}$ can lead to the formation of carbon-chain molecules including HC$_{3}$N \citep{2008ApJ...672..371S}. The pre-shock fractional abundance of CH$_{4}$ on the grain surface is also high in this model, around $10^{-5}$. In this model, radicals such as CN are assumed to be hydrogenated immediately after they are depleted on grains, and their pre-shock abundances are not shown in the table. 

\subsubsection{Dependence of shock velocities} 

After the shock, the HCN fractional abundance can be further increased by the high temperature chemistry induced by the shock waves due to endothermic reactions that convert HNC and CN to HCN (Figure \ref{fig:n2e5_z-16_2}). The following reaction involving CN
\begin{equation}\label{hydro_hcn}
{\rm H_2 + CN \longrightarrow HCN + H}
\end{equation}
 has a reaction barrier of 820 K. When the temperature becomes high enough for the collisional dissociation of molecular hydrogen to occur ($\sim 4000\,$K) via reaction of the form ${\rm H_{2} + M \longrightarrow H + H + M}$ where M is any given species (most often H$_{2}$ or H), 90 \% of hydrogen is rapidly converted into the atomic form. Such collisional dissociation occurs in our model when $n_{\rm H} \geq 2 \times 10^5\,$cm$^{-3}$ and $v_{s} = 40$ km s$^{-1}$. Then, the reverse reaction of the reaction (\ref{hydro_hcn}) becomes more important,  increasing the CN abundance. The fractional abundance of HNC decreases by the reaction:
 \begin{equation}
{\rm H + HNC \longrightarrow H + HCN.}
\end{equation}
The rate coefficient of this reaction has a strong temperature dependence\footnote{The temperature dependence of the rate coefficient of this reaction is derived by fitting experimental values. A temperature dependence that has a physical meaning should be in the form of exp(-$\gamma$/T) where $\gamma$ is a reaction barrier \citep[see ][ for detailed analysis of HNC/HCN ratio]{2014ApJ...787...74G}.} ($k \propto T^{4.5}$), and the ratio HCN/HNC increases when the maximum temperature caused by the shock waves is higher, and the abundance ratio of HCN/HNC $>10$ is reached for the model $v_s  > 30\,$km s$^{-1}$ for all densities in our models (see Figure \ref{fig:n2e5_z-16_2}).  Abundances of ionic species decrease after the shock because metal species such as Mg are sputtered into the gas-phase and become atomic ions, taking over the positive charge from molecular ions. Therefore, the case of $v_{s}=10\,$km s$^{-1}$ with less sputtering has a higher HCO$^{+}$ and N$_{2}$H$^{+}$ abundance. As for H$_{3}$O$^{+}$, which has higher proton affinity than HCO$^{+}$ and N$_{2}$H$^{+}$, the fractional abundance after the shock decreases less than that of these species. When $v_{s} = 40\,$km s$^{-1}$, the ionization fraction increases and the fractional abundances of HCO$^{+}$, N$_{2}$H$^{+}$, and H$_{3}$O$^{+}$ increase. Since a high abundance of sulfur is sputtered from grains, the CS and SO fractional abundances increase after the shock especially for $v_s > 30\,$km s$^{-1}$. The increase of a fractional abundance due to the sputtering is even more noticeable in the case of SiO. As mentioned in Section \ref{sec:chem_model}, the amount of the core elements sputtered from shock waves with $v_s > 30\,$km s$^{-1}$ is a rough estimate in this model, and a more careful treatment should be used for a better prediction of SiO's fractional abundances. The fractional abundances of NO increase slowly after the shock passes because the evaporation of NH$_{3}$ can initiate the reaction:
 \begin{equation}
{\rm NH + O \longrightarrow NO + H,}
 \end{equation}
where NH is produced through the dissociation of NH$_{3}$. There is no strong dependence on the shock velocity except for the case of $v_s = 40\,$km s$^{-1}$ for NO. The fractional abundance of H$_{2}$CO does not strongly depend on shock velocities. The only difference is that for increasing $v_s$, H$_{2}$CO is injected into the gas phase at earlier times due to the shorter saturation times. The fractional abundance of HC$_{3}$N for $v_s \leq 30\,$km s$^{-1}$ only increases after $10^4$ yrs while the increase happens earlier for $v_s=40\,$km s$^{-1}$, around 1000 yrs or so.   Higher maximum temperatures, $T_{\rm max}$, caused for higher shock velocities increase the HC$_{3}$N fractional abundances since they accelerate some endothermic reactions such as 
 \begin{equation}
{\rm C_2H + HCN \longrightarrow HC_3N + H.}
\end{equation}  

\subsubsection{Dependence on cosmic-ray ionization rate} 

In the models we run, the increase in $\zeta$ in Phase II causes the peak value of the HCN abundance ($\sim 3-10 \times 10^{-7}$ ) to occur earlier in time. However, a high $\zeta$ also decreases the fractional abundance of HCN faster in the post-shock region than low values of $\zeta$. Regarding the dependence on the cosmic-ray ionization rate, HNC has a similar behavior to HCN; a higher cosmic-ray ionization rate helps the formation of HNC on a shorter time scale. This trend by the higher cosmic-ray ionization rate to shorten the chemistry time scale also applies to other species such as CS, SO, SiO, NO, and HC$_{3}$N. Although varying $\zeta$ does not change the peak value itself, it shortens the time span in which the higher fractional abundances are attained.
While the HCN/HNC ratio has some temperature dependence, it decreases with higher cosmic-ray ionization rate because a high ionization fraction makes the HCNH$^{+}$ abundance high and a following reaction brings the HCN/HNC ratio closer to unity:

\begin{eqnarray}
{\rm HCNH^+ + e^- \longrightarrow HCN + H} \\
{\rm \longrightarrow HNC + H}\\
{\rm \longrightarrow CN + H + H}
\end{eqnarray}
 whose branching ratios are 34\%, 34\%, and 33\%, respectively. Although CN is an unstable radical, a high ionization fraction caused by a high cosmic-ray ionization rate can keep the CN fractional abundance high via reaction (7) above. A higher cosmic-ray ionization rate also increases the abundances of ionic species such as N$_{2}$H$^{+}$, HCO$^{+}$, and H$_{3}$O$^{+}$. The proton affinity of H$_{2}$O is higher than those of CO and N$_{2}$, and CO has a higher value than N$_{2}$. This means that CO is more likely to be protonated than N$_{2}$, and a reaction such as N$_{2}$H$^{+}$ + CO $\longrightarrow$ N$_{2}$ + HCO$^{+}$ is much faster than its reverse reaction. The fractional abundance of a protonated species of a lower proton affinity such as N$_{2}$H$^{+}$ has a stronger dependence on the cosmic-ray ionization rate. The increase of a peak fractional abundance of H$_{2}$CO at a higher cosmic-ray ionization rate is caused by the dissociation of CH$_{3}$OH via the gas-phase reactions
 \begin{equation}
{\rm CH_{3} + O \longrightarrow H_{2}CO + H,}
\end{equation}
and the cosmic-ray-photodissociation reaction
\begin{equation}
{\rm CH_{3}OH \longrightarrow H_{2}CO + H_{2}.}
\end{equation}

\subsubsection{ Dependence on UV-field strength }\label{sec:disc_uv}
Results obtained with the Meudon code \citep{2006ApJS..164..506L} show that the fractional abundance of HCO$^{+}$ and H$_{3}$O$^{+}$ can be enhanced up to $10^{-8}$ for $A_{\rm V}<1$ when $G_{0} =10^4$ Habing field units and $n_H=2\times 10^5\,$cm$^{-3}$ (Figure \ref{fig:pdr}). The peak value of CN is achieved at a larger value than those of HCO$^{+}$ and H$_{3}$O$^{+}$, at $A_V \sim 3$. Unlike HCO$^{+}$ and H$_{3}$O$^{+}$, the fractional abundance of N$_2$H$^{+}$ is not enhanced in the low $A_V$ regime (Figure \ref{fig:pdr}). For other species, there is no enhancement in their fractional abundances in $A_{\rm V} < 1$ regions.

\subsection{Comparison with Observations}\label{sec:cmp_obs}
We compare the observed column densities with the predictions of our chemical models in this section. Since chemical models give us fractional abundances, we opt for comparing ratios of column densities with ratios of fractional abundances (and hence make the implicit assumption that all molecules arise from the same gas component). All our ratios are with respect to CO.
The results of chemical modeling with different physical conditions mentioned in Section \ref{sec:chem_model} are compared following the method used in \citet{2007A&A...467.1103G}. A confidence level for each species $i$ is defined by 
\begin{equation}
\kappa_{i}=erfc \left(\frac{|log(X_{i, calc})-log(X_{\rm i, obs})|}{\sqrt{2} \sigma} \right)
\end{equation}
where $erfc$ is a complementary error function in a form $erfc(x)=\frac{2}{\sqrt{\pi}}\int^{\infty}_{x} e^{-t^2}dt$, $X_{i, calc}$, $X_{i,obs}$ are fractional abundances with respect to CO for molecules $i$ predicted from models and observations respectively, and in the above formula, a value of $\sigma=1$ is used. The degree of agreement between model predicted and observed abundances is measured by an average value of $\kappa _{i}$, $\bar{\kappa}$, for all the observed species. A value of $\bar{\kappa}=1$ means a perfect match, and $\bar{\kappa}=0.317$ means an average discrepancy of one order of magnitude. Since our calculation for SiO is only an approximation, values of $\kappa$ are calculated for cases with and without SiO included. We found that there is no significant differences between the cases with and without SiO.

The maximum values of confidence level in each model over the post shock time of $t=0-10^5$ yrs are plotted in Figure \ref{fig:fit}. Models in a reasonable agreement ($\kappa > 0.7$) with the observed abundances can be summarized in the following three categories: 1) low density models ($n_H=2\times 10^4\,$cm$^{-3}$, $v_{s} > 30\,$km s$^{-1}$, $\zeta < 10^{-13}\,$s$^{-1}$), 2) high velocity shock models ($n_{\rm H}=2\times 10^5\,$cm$^{-3}$, $v_{s}=40\,$km s$^{-1}$), and 3) high cosmic-ray ionization models ($n_{\rm H}=2\times 10^5\,$cm$^{-3}$, $\zeta > 10^{-15}\,$s$^{-1}$). In case 1), the chemistry is still in an early phase because a significant amount of atomic carbon is present, and because the radical and ionic species are abundant, resulting in a good agreement with the observed chemical features. In case 2), the maximum temperature is so high that the shock becomes dissociative, which means that the majority of molecular hydrogen is destroyed by collisions with molecular or atomic hydrogen. Over the short time period during which atomic hydrogen is abundant, the abundances of radical and ionic species increase. In case 3), the high cosmic-ray ionization rate keeps the abundances of radical and ionic species high. Example plots of fractional abundances and confidence levels are shown in Figures \ref{fig:n2e4_model2} for case 1), Figure \ref{fig:n2e5_model7} for case 2), and Figure \ref{fig:n2e5_model12} for case 3). For the cases without shocks, the agreement with the observations is better for the high cosmic-ray ionization rates, achieving a similar degree of agreement to models with shocks. The plausibility of each scenario is discussed in the next section.

\section{Discussion}
\label{sec:disc}
Although there are several different models that show a modest degree of agreement with observations, these cases can be constrained using other information. 
\subsection{Gas density}
 Our derived total hydrogen densities for the species we observed range from $n_{\rm H} =2 \times 10^4 -1.2 \times 10^6\,$cm$^{-3}$. \citet{2013ApJ...779...47M} derived $n({\rm H}) = 3 \times 10^7\,$cm$^{-3}$ from multi-$J$ observations of HCN for similar velocity range when including the high-lying $J=8-7$ transition. \citet{2014MNRAS.437.3159S} derived a total hydrogen density of $2.6 \times 10^{6}\,$cm$^{-3}$ also from HCN lines although they used the entire velocity component around $-80$ km s$^{-1}$ instead of the velocity slice that we have used. 
  The density derived from our observations of HCN are $n_{\rm H}=(6.94 - 10.5) \times 10^{5}\,$cm$^{-3}$ assuming $T_{\rm kin} = 300\,$K, and for the case of $T_{\rm kin}=100\,$K, $n_{\rm H}= (1.40 - 2.20) \times 10^{6}\,$cm$^{-3}$.
 Since the temperatures used by  \citet{2013ApJ...779...47M} and \citet{2014MNRAS.437.3159S} are 61 and 150 K respectively, our derived densities are roughly consistent considering the degeneracy between the derived density and the temperature. For the CO lines, the best-fit density in our analysis is $1 \times 10^4\,$cm$^{-3}$, which is similar to the value derived from multi-$J$ CO lines by \citet{2012A&A...542L..21R} and \citet{2005ApJ...623..866B}.  Since we used the high-$J$ CO line data from \citet{2012A&A...542L..21R}, this similarity is expected. The derived densities of other molecules are $n_{\rm H}>2\times 10^5\,$cm$^{-3}$ except for H$_{3}$O$^{+}$. For the case of H$_{3}$O$^{+}$, the density is derived from two lines whose energy levels are close together, and the solution has a large uncertainty. If we use the density of other species ($n_{\rm H}>2\times 10^5\,$cm$^{-3}$), the low-density scenario of $n_{\rm H}\sim 2\times 10^4\,$cm$^{-3}$ in Section \ref{sec:cmp_obs} is excluded.

 The tidal stability of the core in the CND is determined by the following equation \citep{2005ApJ...622..346C}
\begin{equation}
n({\rm H_2})=2.87 \times 10^7 \left [\left (\frac{R}{\rm pc}\right)^{-3} + 0.4\left(\frac{R}{\rm pc}\right)^{-1.75}\right] {\rm cm^{-3}}
\end{equation}
where $R$ is a Galactocentric radius. The core becomes tidally unstable when $n_{\rm H} < 2.8\times 10^7\,$cm$^{-3}$ for $R=1.5\,$pc and when $n_{\rm H} < 1.4\times 10^7\,$cm$^{-3}$ for $R=2\,$pc. Densities derived by our analysis are still lower than this limit even for HCN with the large dipole moment whose lines have higher critical densities. This result suggests that the core is tidally unstable. On the other hand, \citet{2013ApJ...779...47M} found a clump that may be gravitationally stable in their observation of the HCN emissions in the CND. Their use of higher transition lines ($J=9-8$ and a vibrationally excited line) may have resulted in their tracing a higher excitation or higher density portion, while we are tracing the bulk material of the clump. Since \citet{2013ApJ...779...47M} concluded that the bulk of the gas except for one clump is tenuous and gravitationally unstable, our results are overall consistent with theirs.  

\subsection{High velocity shocks}
It has been shown in Figure \ref{fig:fit} that a high shock velocity case ($v_s =40\,$km s$^{-1}$) apparently gives a good agreement with the observations. However, the time span required for attaining an agreement is very short ($t\sim 3\times 10^2$ yrs) as shown in Figure \ref{fig:n2e5_model7}. In this time span, the temperature should still be over 1000 K, which is not supported by previous studies found in the literature. However, if such high velocity shocks are occurring in short periods of time, that may also give a good agreement with the observations. \citet{2006A&A...455..971R} suggested shocks in the Galactic Center as often as $5\times 10^4$ yrs although this time scale is much longer than the time suggested by our high velocity shock case. 

\subsection{Cosmic-ray dominated gas}
If we assume a density $\gtrsim $ a few $\times 10^5\,$cm$^{-3}$ as our results of the LVG analysis for most species show, models with a higher cosmic-ray ionization rate of $\zeta > 10^{-15}\,$s$^{-1}$ show the best agreement with the observed column densities. This is in agreement with previous studies in other parts of the Galactic Center region \citep[e.g., ][]{2013ApJ...764L..19Y}. The level of agreement stays high until $t \sim 10^3 - 10^4\,$yrs after the passage of the shock wave. The derived cosmic-ray ionization rate is about 2-4 orders of magnitude higher than the regular value in dense molecular clouds in the spiral arms of $(1-5)\times 10^{-17}$ s$^{-1}$. A high cosmic-ray ionization rate in the CND is not very surprising since analysis of high-energy $\gamma$-ray emission in the Galactic Center claimed that most of this $\gamma$-ray emission comes from within 3 pc from Sgr A* \citep{2012ApJ...753...41L}. Some previous studies have also claimed high values for the cosmic-ray ionization rate in the Galactic Center. For example, \citet{2013JPCA..117.9919G, 2014ApJ...786...96G} studied the cosmic-ray ionization rate in the CND. In their earlier work, they derived $\zeta \sim 1.2 \times 10^{-15}\,$s$^{-1}$ from their H$_{3}^{+}$ observations while $\gamma$-ray observations indicate 4 orders of magnitude higher values. In their more recent work, they claimed that $\zeta$ is well above $10^{-15}\,$s$^{-1}$ \citep{2014ApJ...786...96G}. For a cosmic-ray ionization rate of $\zeta > 10^{-12}\,$s$^{-1}$, as suggested by $\gamma$-ray observations, the molecules start to dissociate in a short time scale ($\sim 10^3\,$ yrs). Since cosmic rays have high penetration depth, an enhancement of the cosmic-ray ionization rate as in the CND should also appear in other locations in the Galactic Center. This is consistent with other studies such as that by \citet{2006A&A...454L..99V} who claimed higher cosmic-ray ionization rate of $\zeta = 4\times 10^{-16}\,$s$^{-1}$ in Sgr B2 as well as the work of \citet{2007ApJ...656..847Y}, who claimed $\zeta = 5\times 10^{-13}$s$^{-1}$ ($\zeta_H = 0.5\zeta_{H2}$) on average in Sgr B1, Sgr B2, Sgr C, Radio Arc and Arches Cluster.

 We note our LVG analysis gives different volume densities
for CO than for any other species. In fact it is possible that CO is tracing a larger gradient in density than any other species analysed in this study because its emission may come from extended regions. Although the high-$J$ transitions of CO used in our analysis is likely to trace relatively compact and dense regions as other species, the uncertainty of the derived CO column density needs to be investigated. 
To test how sensitive
our results would be to variations in CO column densities, we have
calculated the confidence level for CO column densities a factor of 3 higher and
lower than what we derived. We find that when a higher
CO column density is used, the best fit cosmic-ray ionization rate
is lower by approximately one order of
magnitude, while the opposite is true when a lower CO column density is used. Since it is likely that the CO emission
is  extended, the real CO column density in
the region traced by the other molecules will be lower. Hence the upper value of our best fit cosmic ray ionization rate is in fact conservative and our conclusion that the gas chemistry is dominated by cosmic rays is still valid. Of course, our results should be confirmed with higher angular resolution observations of the CMZ.

\subsection{Effects of UV-photons}
Depending on the geometry of the molecular cloud, the surface area exposed to the central star cluster might be large, which means that the molecular mass affected by UV photons might be significant. In that case, species such as CN and HCO$^+$~become more abundant. These species are enhanced for higher cosmic-ray ionization rates as well. In addition, other factors to be considered are the beam size and the uncertainty of the source size that we used. The largest beam size is 28", which might include emission from more tenuous and extended parts of the CND. Those parts are more likely to be influenced by UV-photons, which may increase the abundance of HCO$^+$~and CN. Increase of these radicals and ions by UV-photons may misleadingly give a good agreement of observations with a higher cosmic-ray ionization rate than the actual value. Since cosmic-rays can penetrate into larger column densities than UV-photons, higher resolution map of these species as well as that of N$_{2}$H$^{+}$, which is not abundant in PDRs, can help differentiating between the two scenarios. Furthermore, the enhancement of the cosmic-ray ionization rate should affect larger regions in the CMZ than the UV-photons, and comparison with other positions in our survey can also help constraining the value of cosmic-ray ionization rate.

\subsection{Cosmic-rays or X-rays?}
\begin{figure}[ht]
\includegraphics[angle=0,width=.40\textwidth]{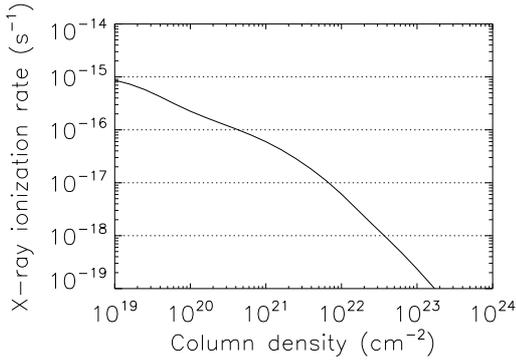}
\caption{The solid line shows the X-ray ionization rate derived by using a current X-ray luminosity $L_X = 2 \times 10^{33}\,$erg s$^{-1}$ in 2-10 keV range \citep{2003ApJ...591..891B} as a function of intervening column density at a distance of $d = 2$ pc from the X-ray source. X-ray flux in lower energy range is estimated by AtomDB program \citep{2012ApJ...756..128F} following a similar procedure in \citet{2013JPCA..117.9919G} using a plasma temperature of $kT=1.9$ keV.}
\label{fig:zeta_x}
\end{figure}

\begin{figure*}
\centerline{\includegraphics[angle=0,width=.4\textwidth]{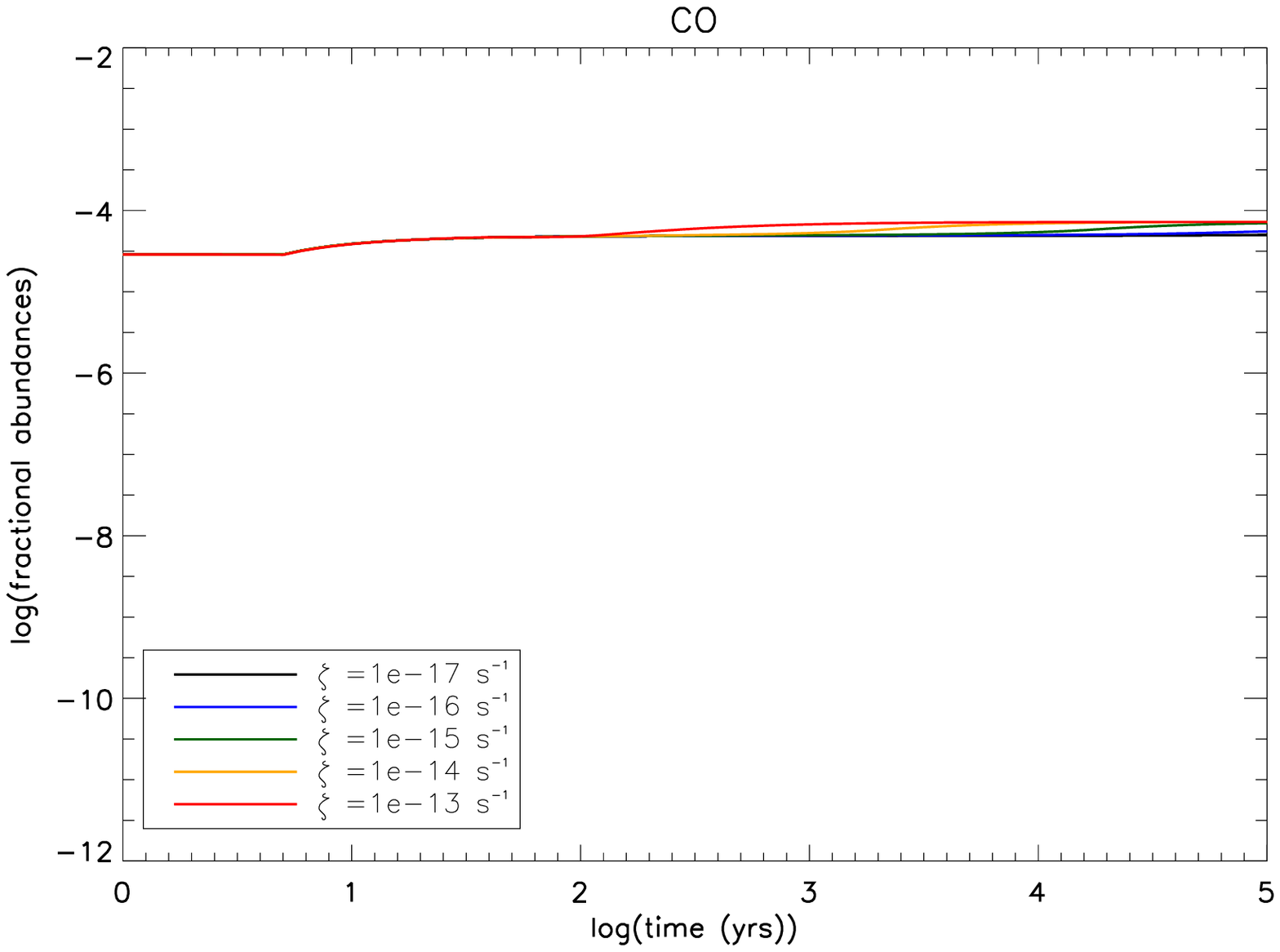}~~
\includegraphics[angle=0,width=.4\textwidth]{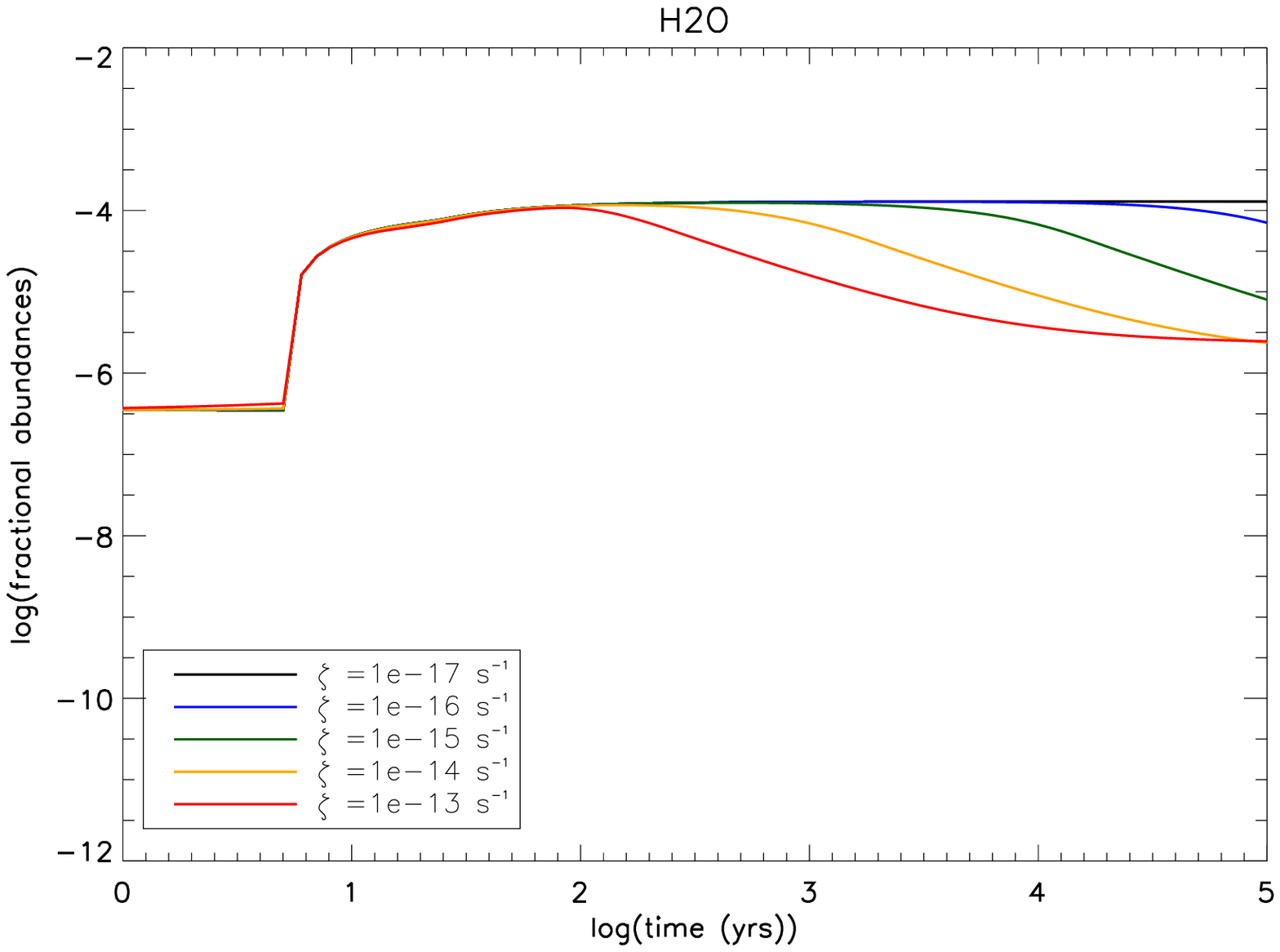}}
\centerline{\includegraphics[angle=0,width=.4\textwidth]{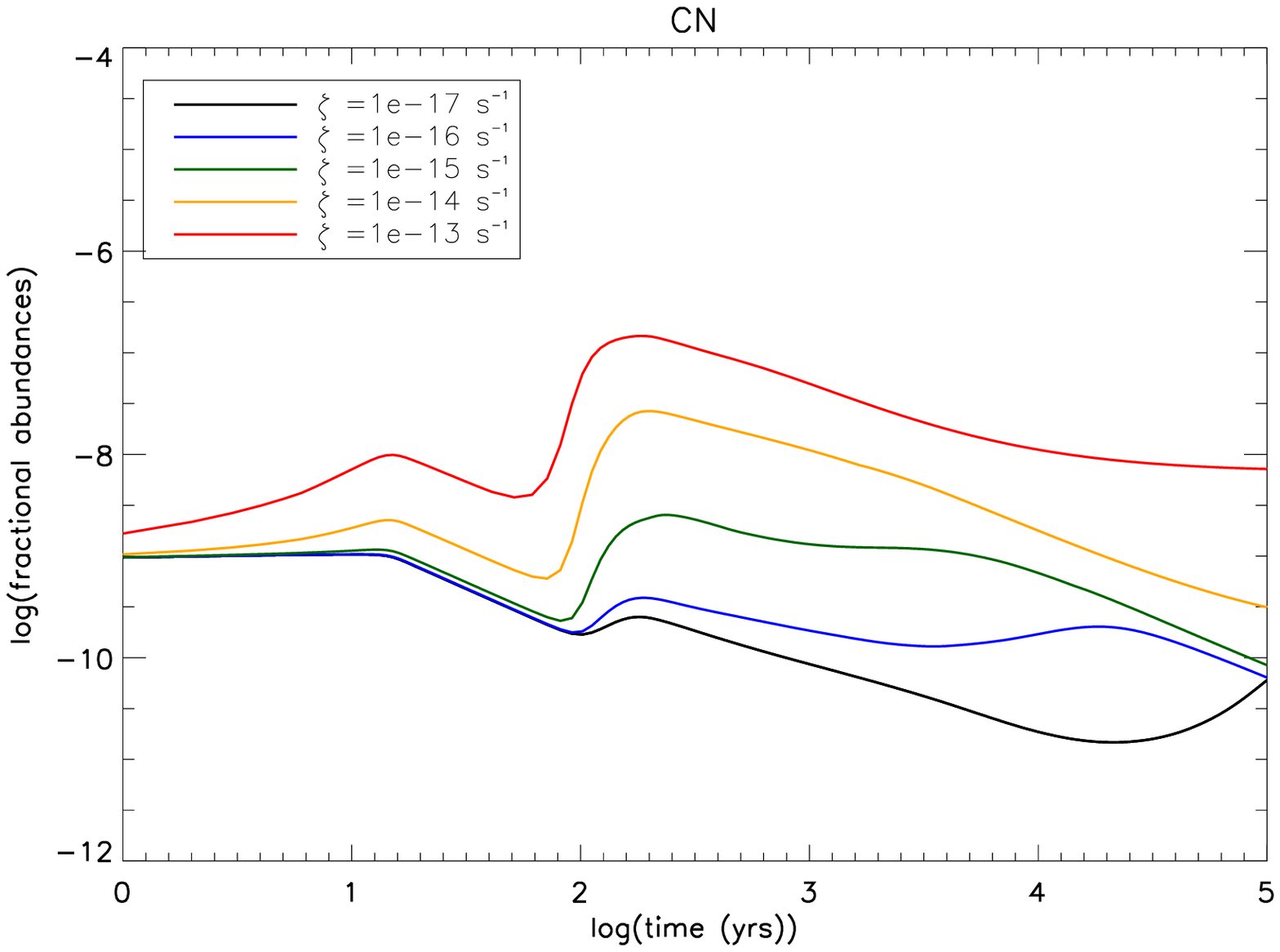}~~
\includegraphics[angle=0,width=.4\textwidth]{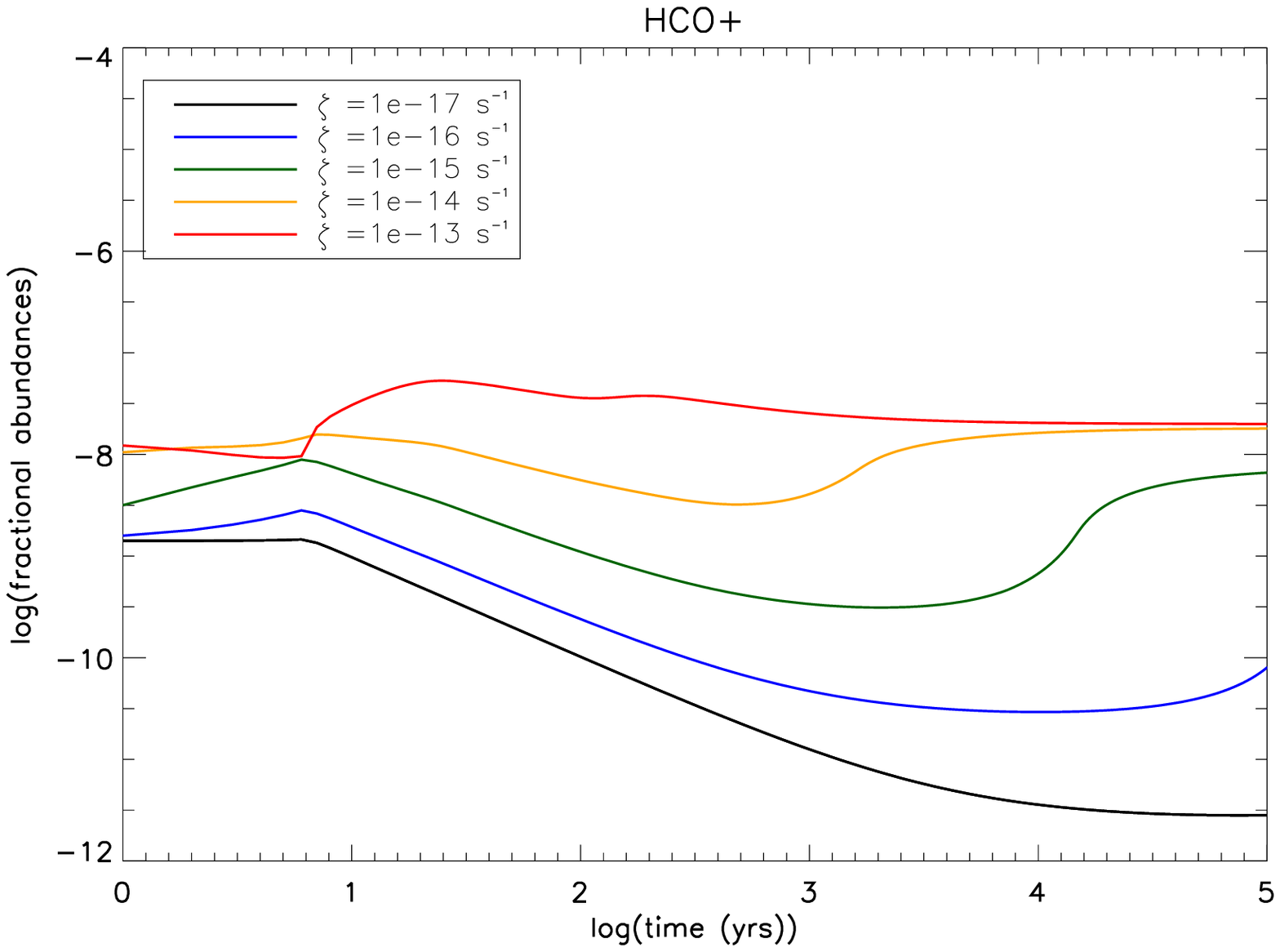}}
\centerline{\includegraphics[angle=0,width=.4\textwidth]{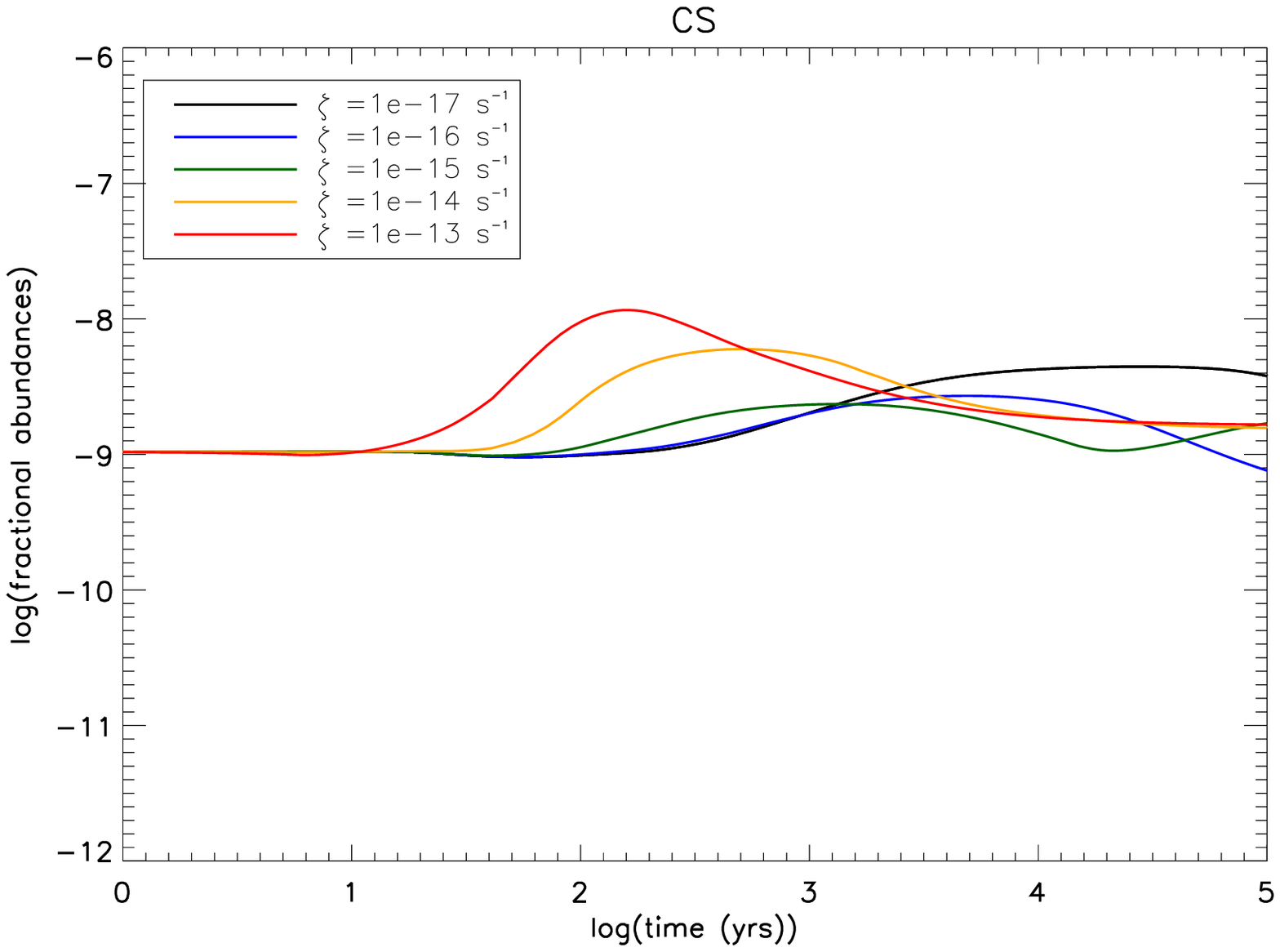}~~
\includegraphics[angle=0,width=.4\textwidth]{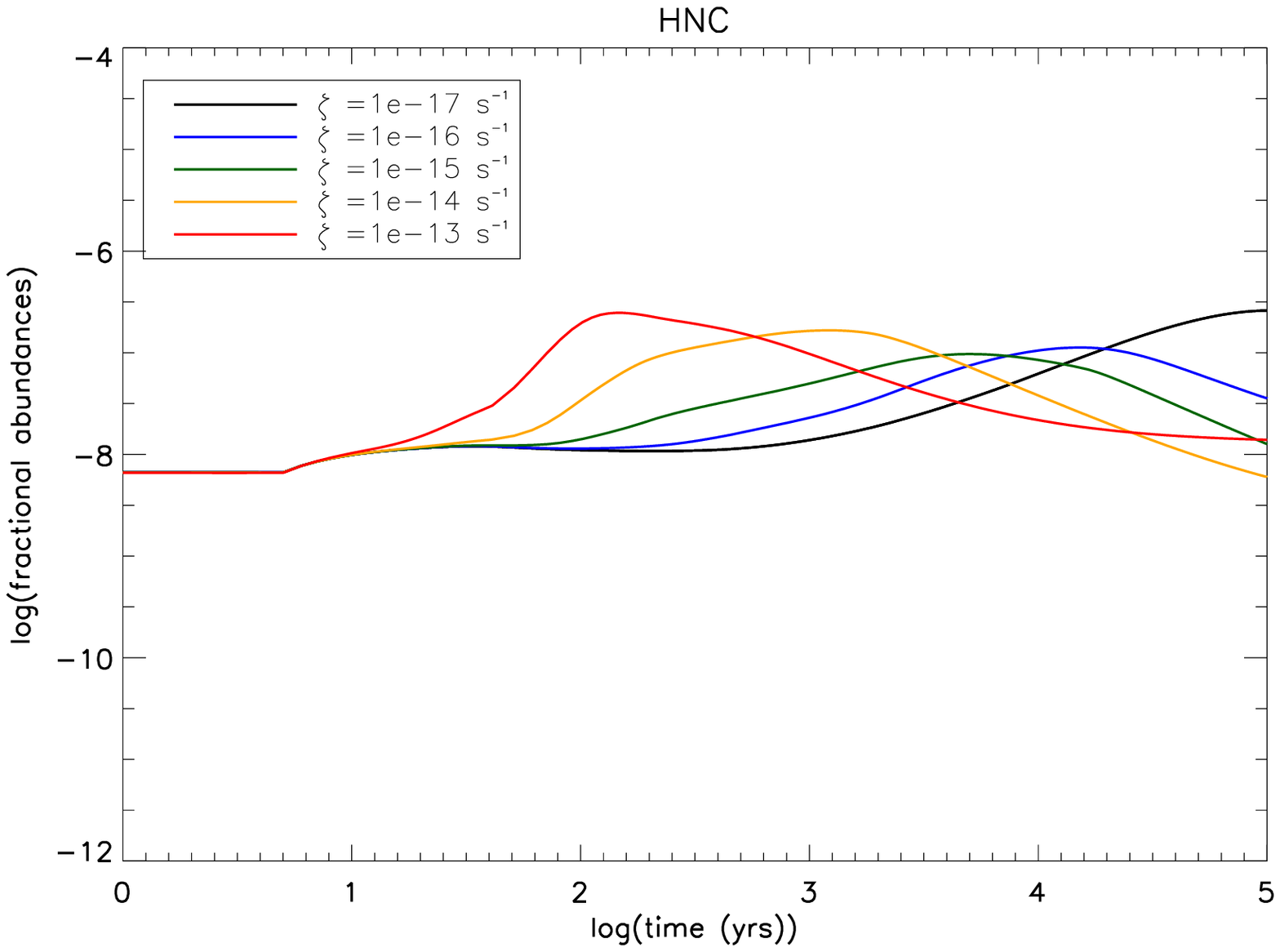}}
\centerline{\includegraphics[angle=0,width=.4\textwidth]{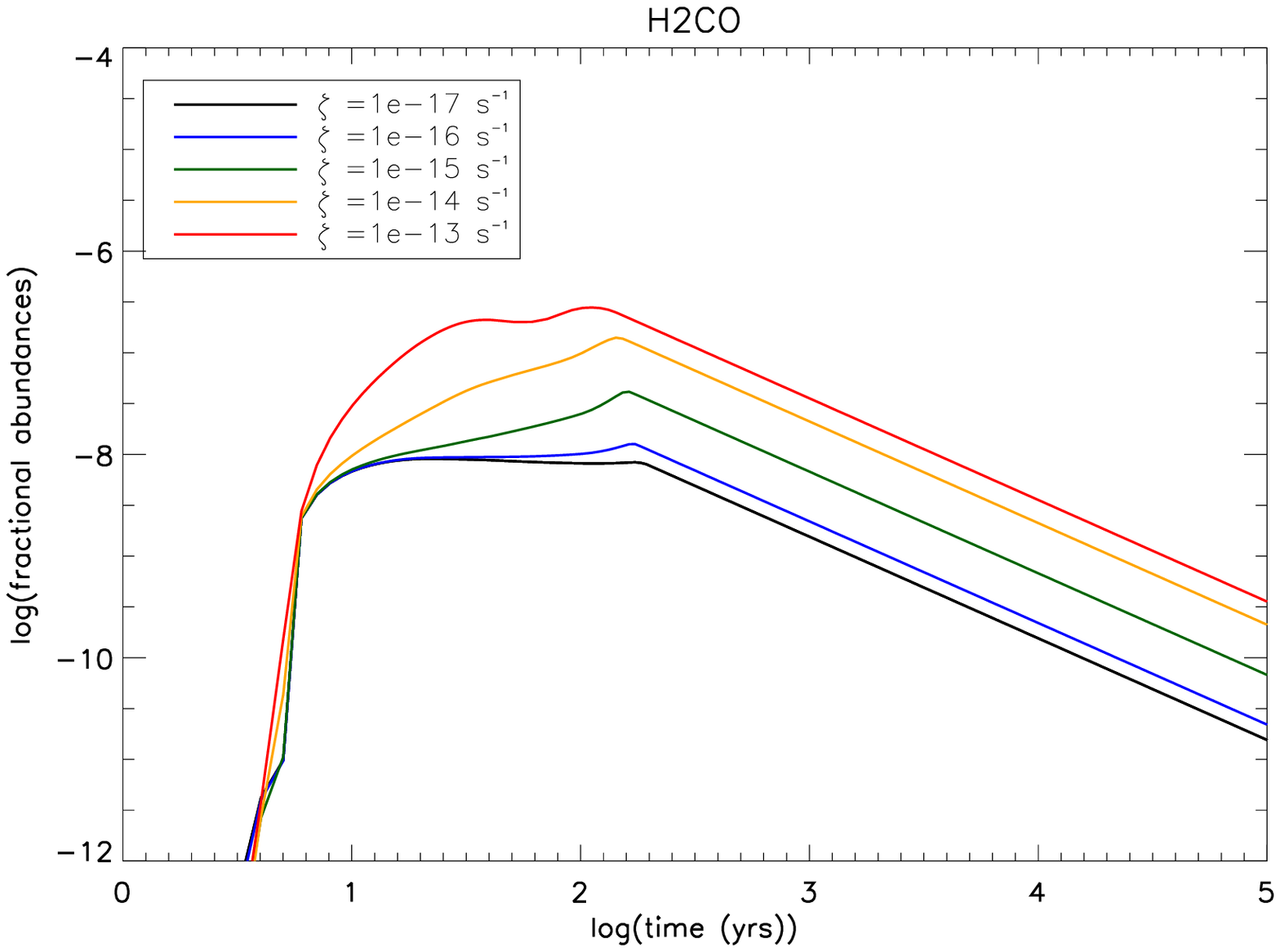}~~
\includegraphics[angle=0,width=.4\textwidth]{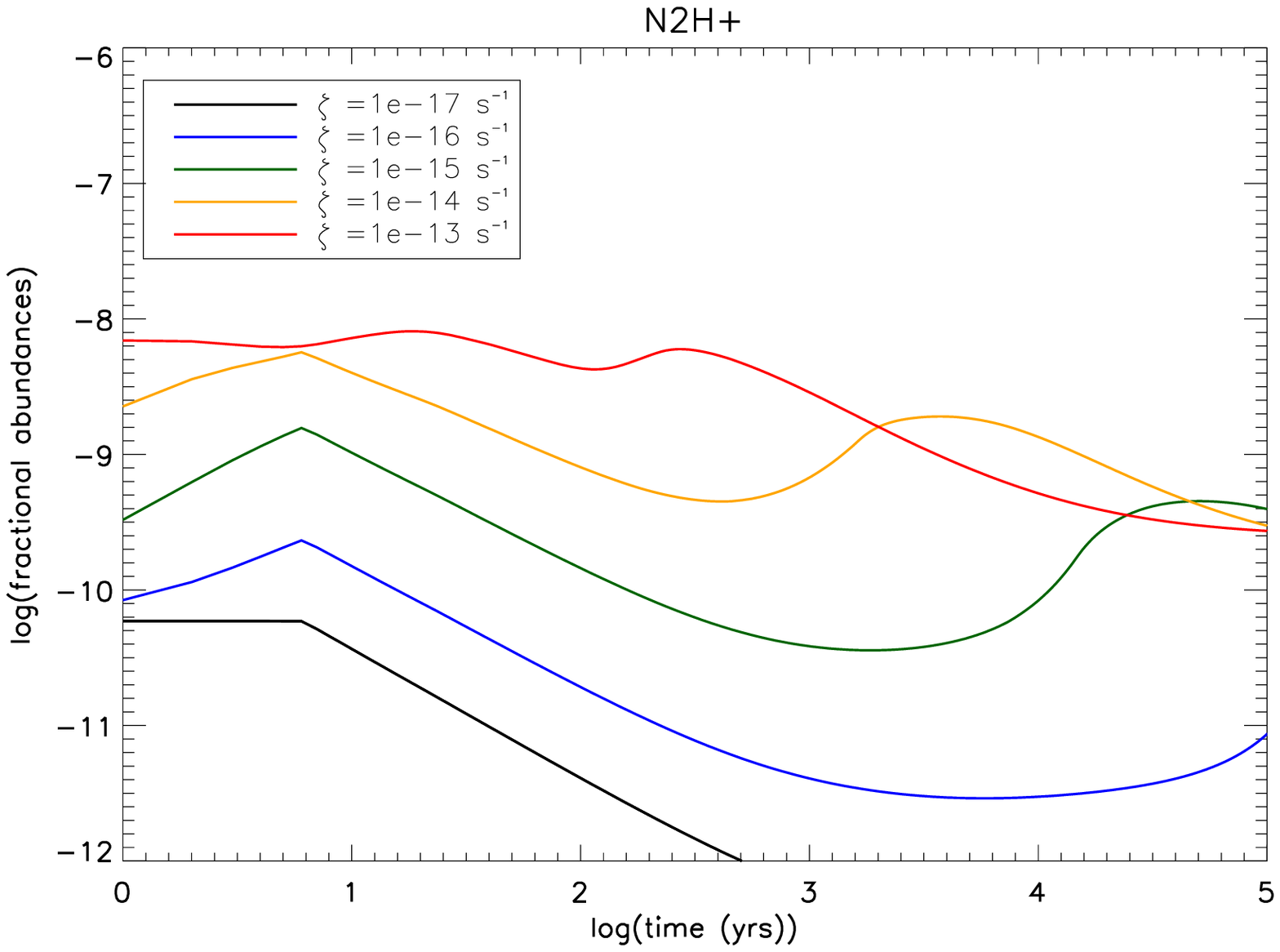}}
\caption{ Time-averaged fractional abundances of observed species are shown for $n_H=2\times10^5\,$cm$^{-3}$, $v=20\,$km s$^{-1}$, and varying cosmic-ray ionization rates of $\zeta = 1 \times 10^{-17}\,$s$^{-1} -  1 \times 10^{-13}\,$s$^{-1}$.}
\label{fig:n2e5_v20_em2_1}
\end{figure*}

\begin{figure*}
\centerline{\includegraphics[angle=0,width=.4\textwidth]{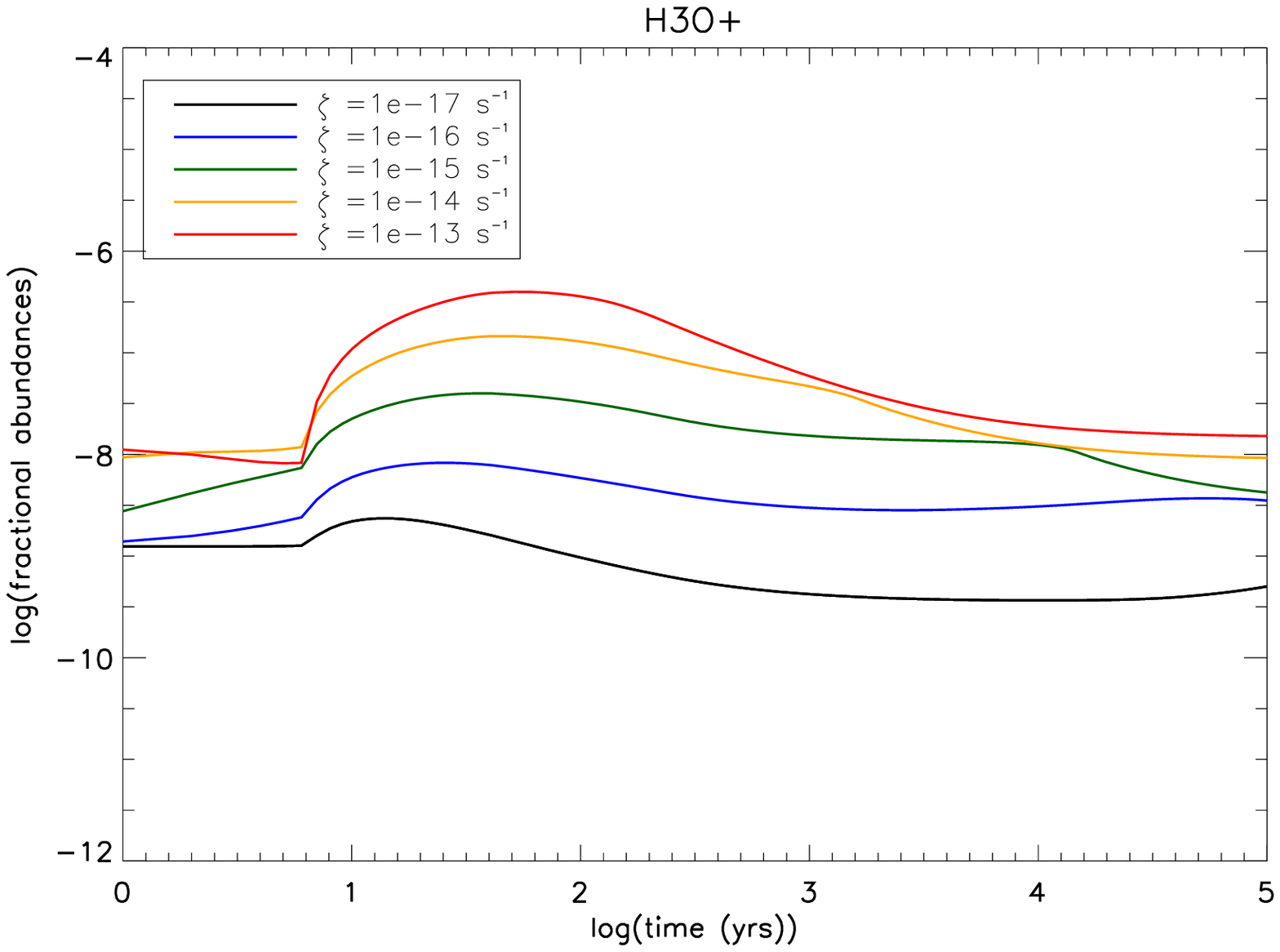}~~
\includegraphics[angle=0,width=.4\textwidth]{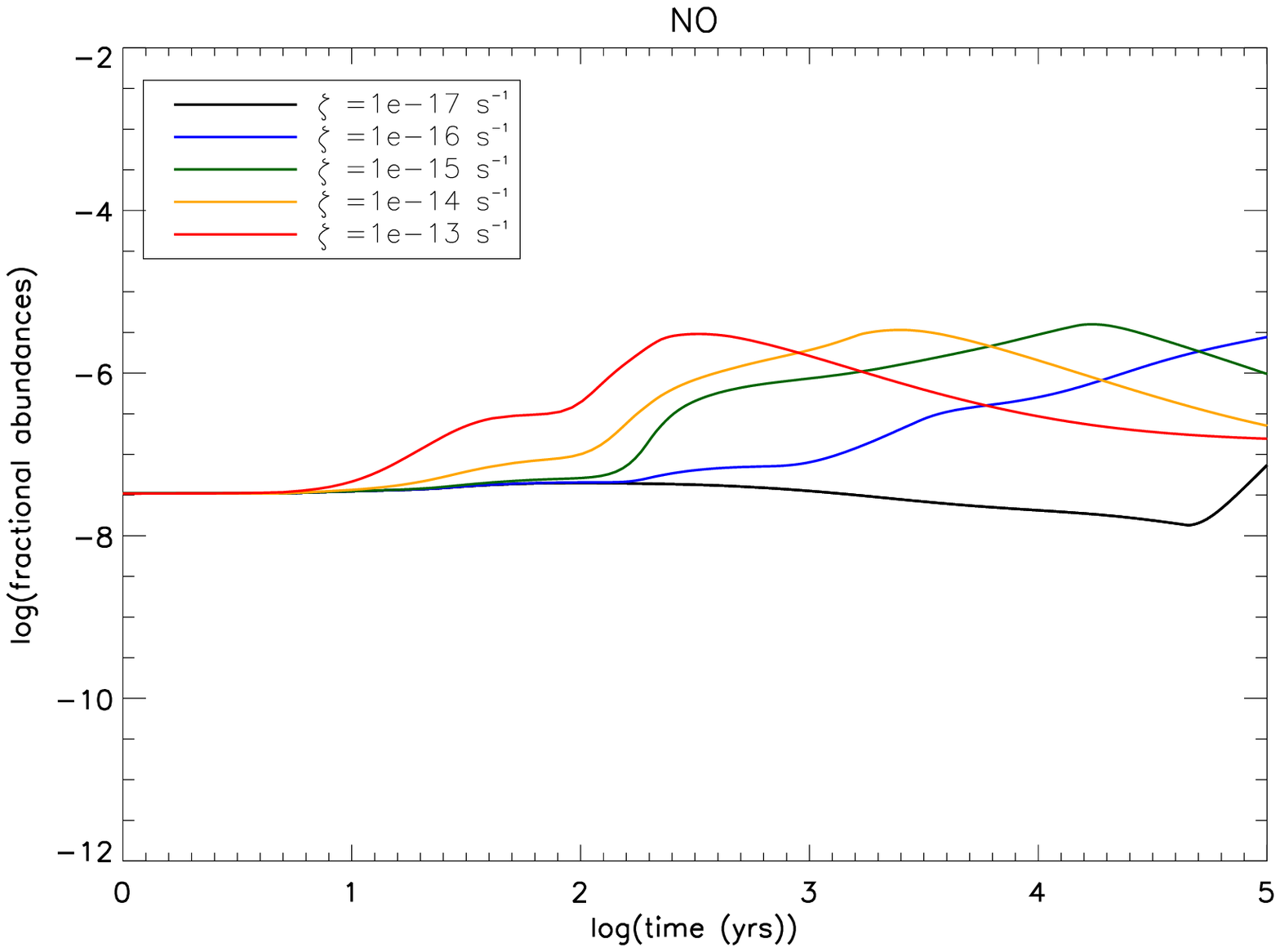}}
\centerline{\includegraphics[angle=0,width=.4\textwidth]{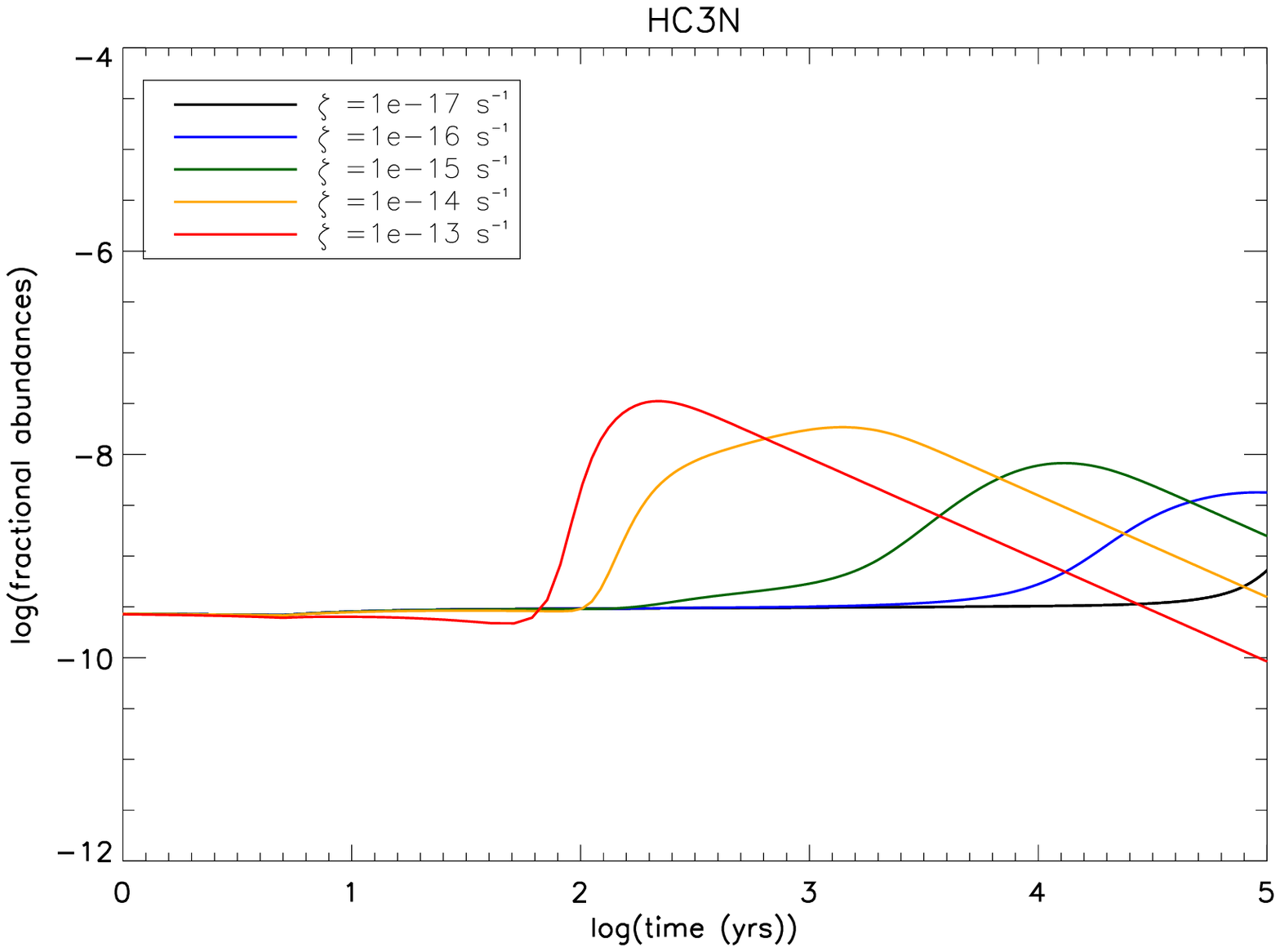}~~
\includegraphics[angle=0,width=.4\textwidth]{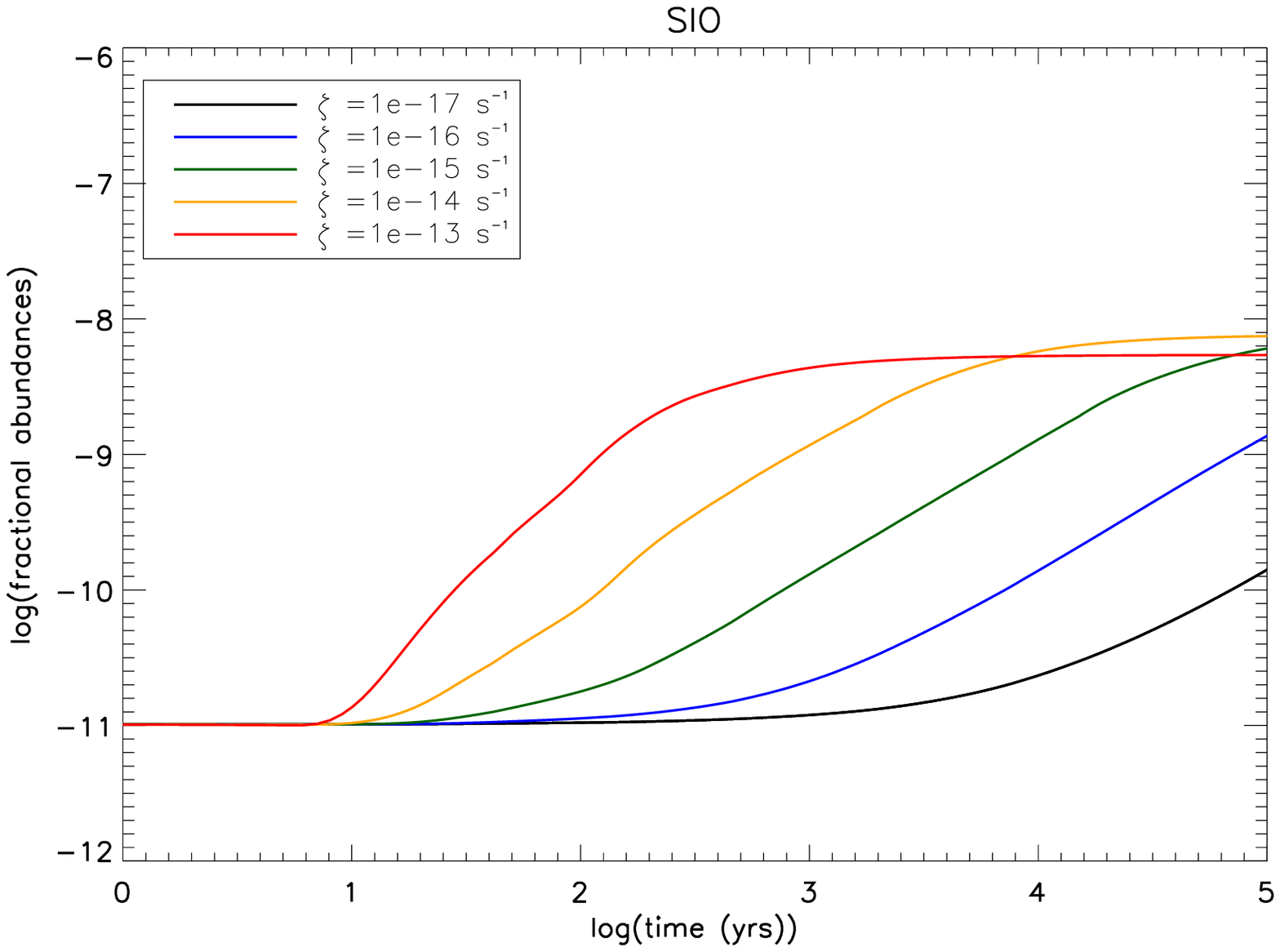}}
\centerline{\includegraphics[angle=0,width=.4\textwidth]{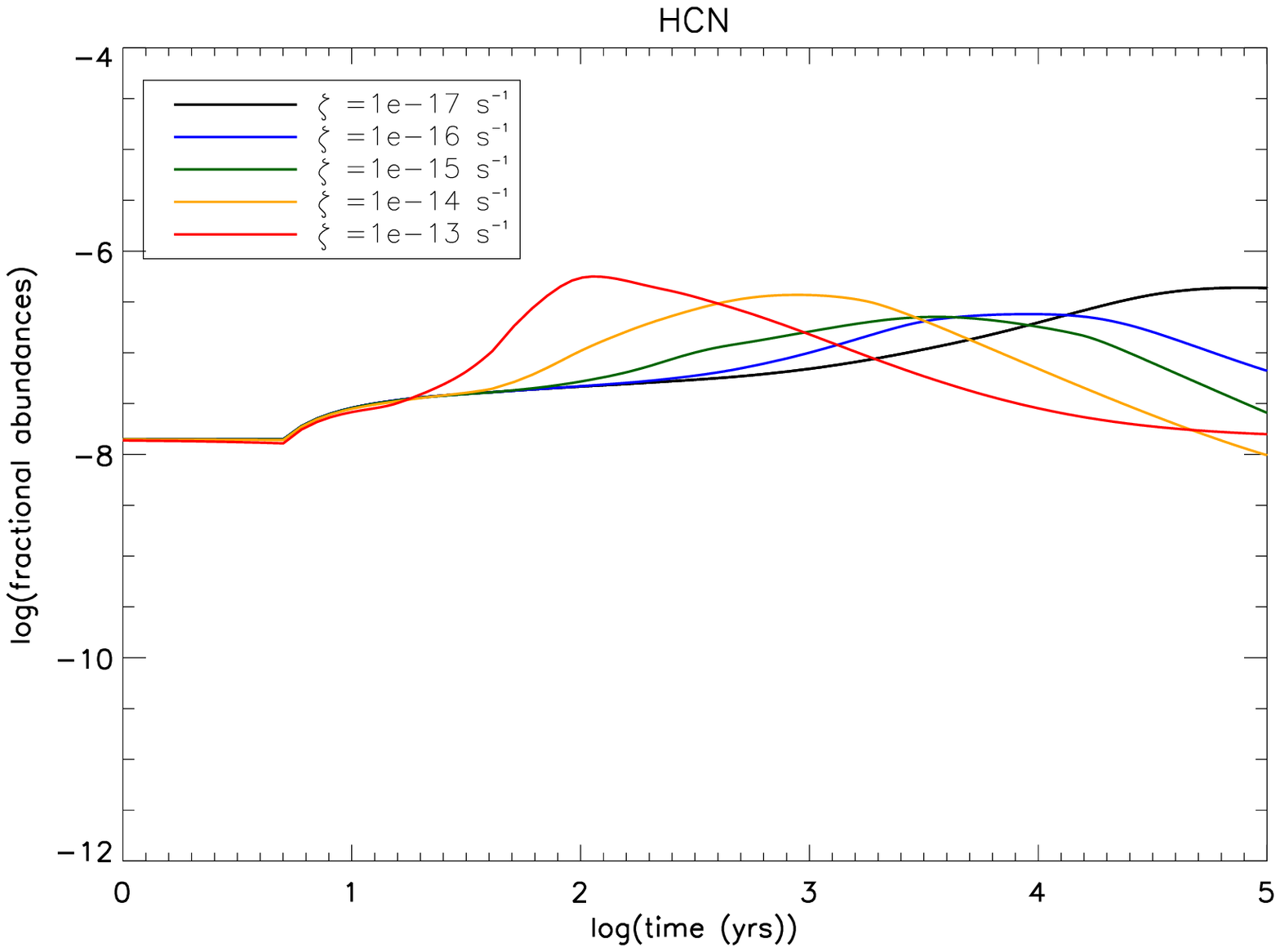}~~
\includegraphics[angle=0,width=.4\textwidth]{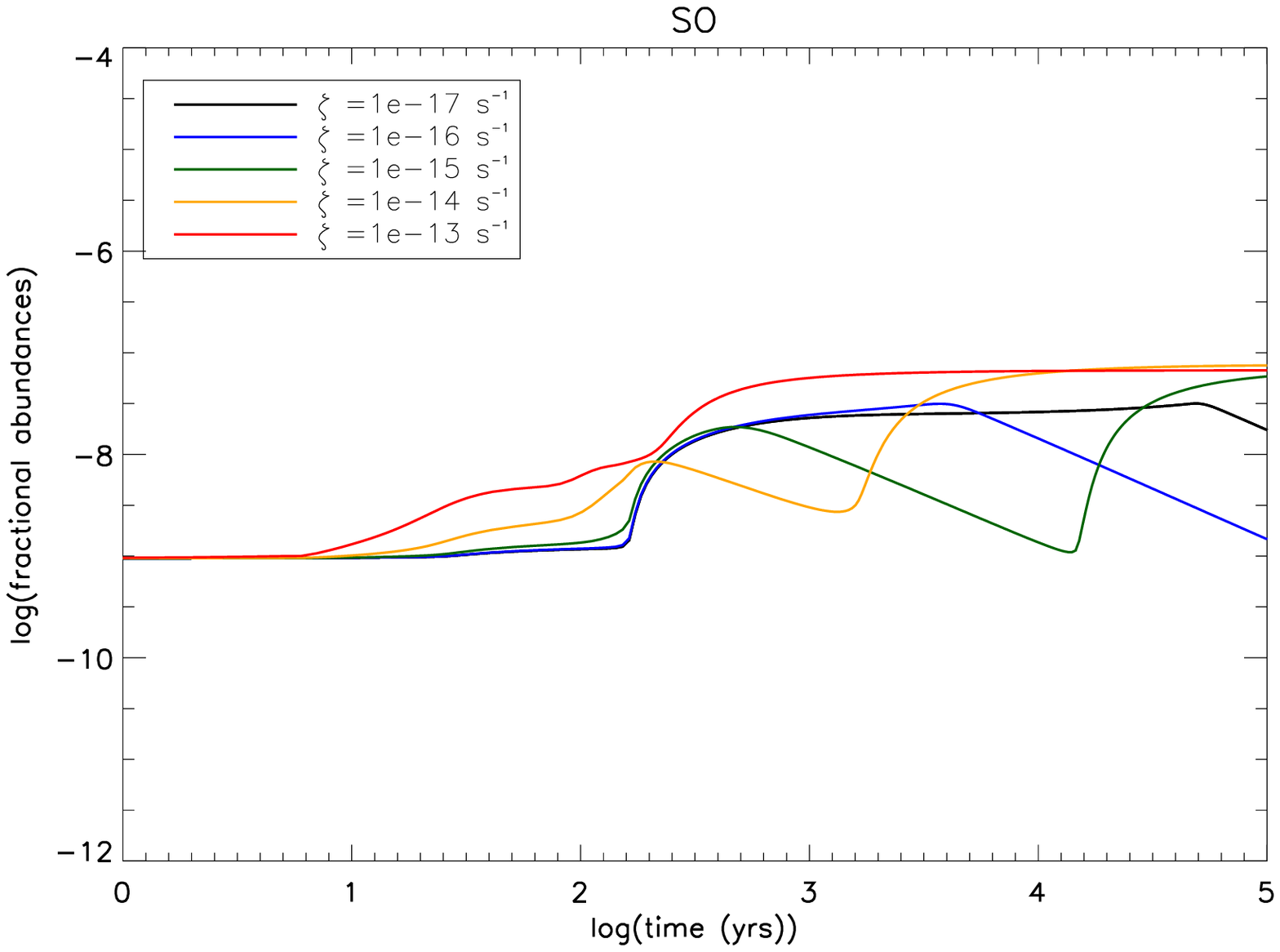}}
\centerline{\includegraphics[angle=0,width=.4\textwidth]{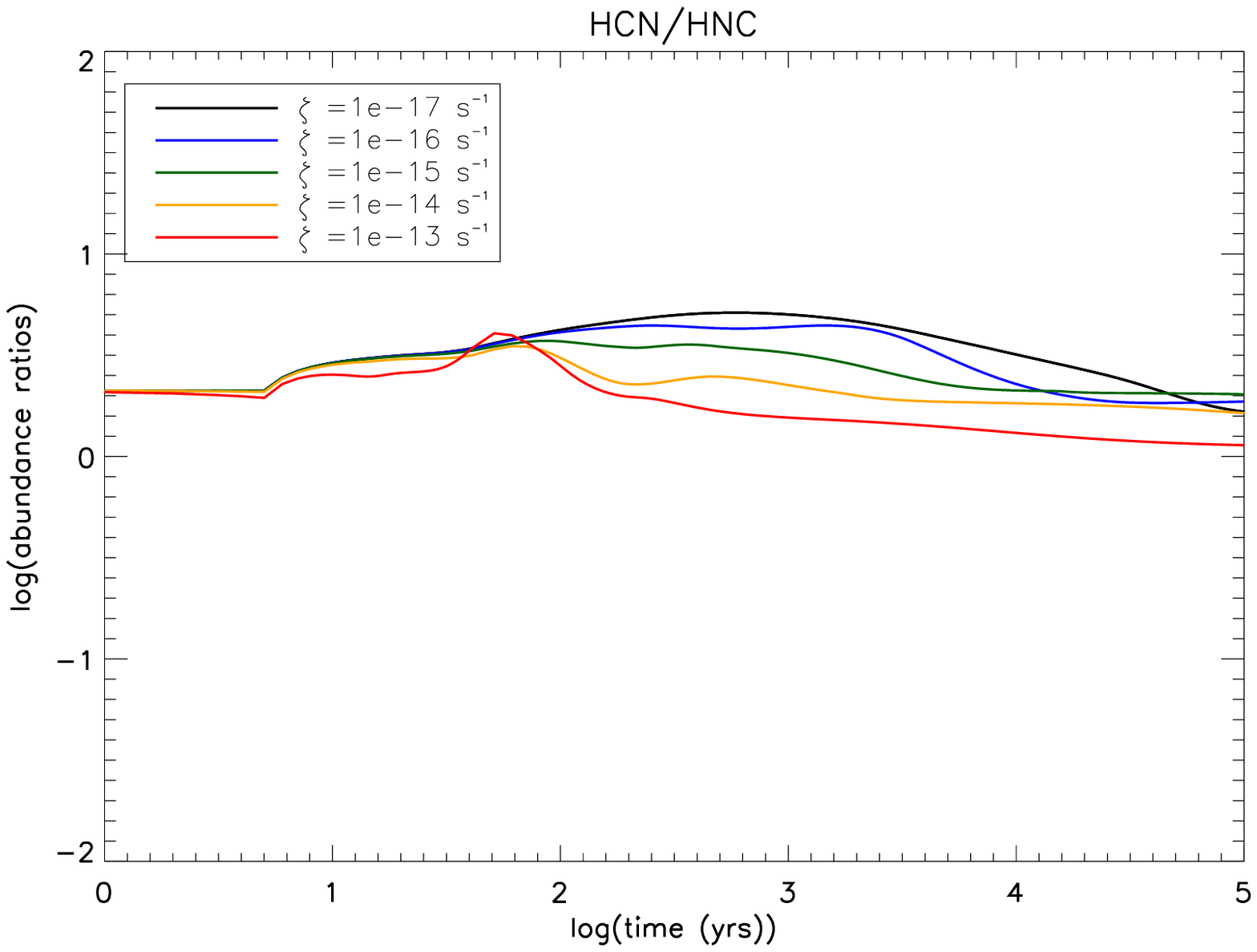}~~
\includegraphics[angle=0,width=.4\textwidth]{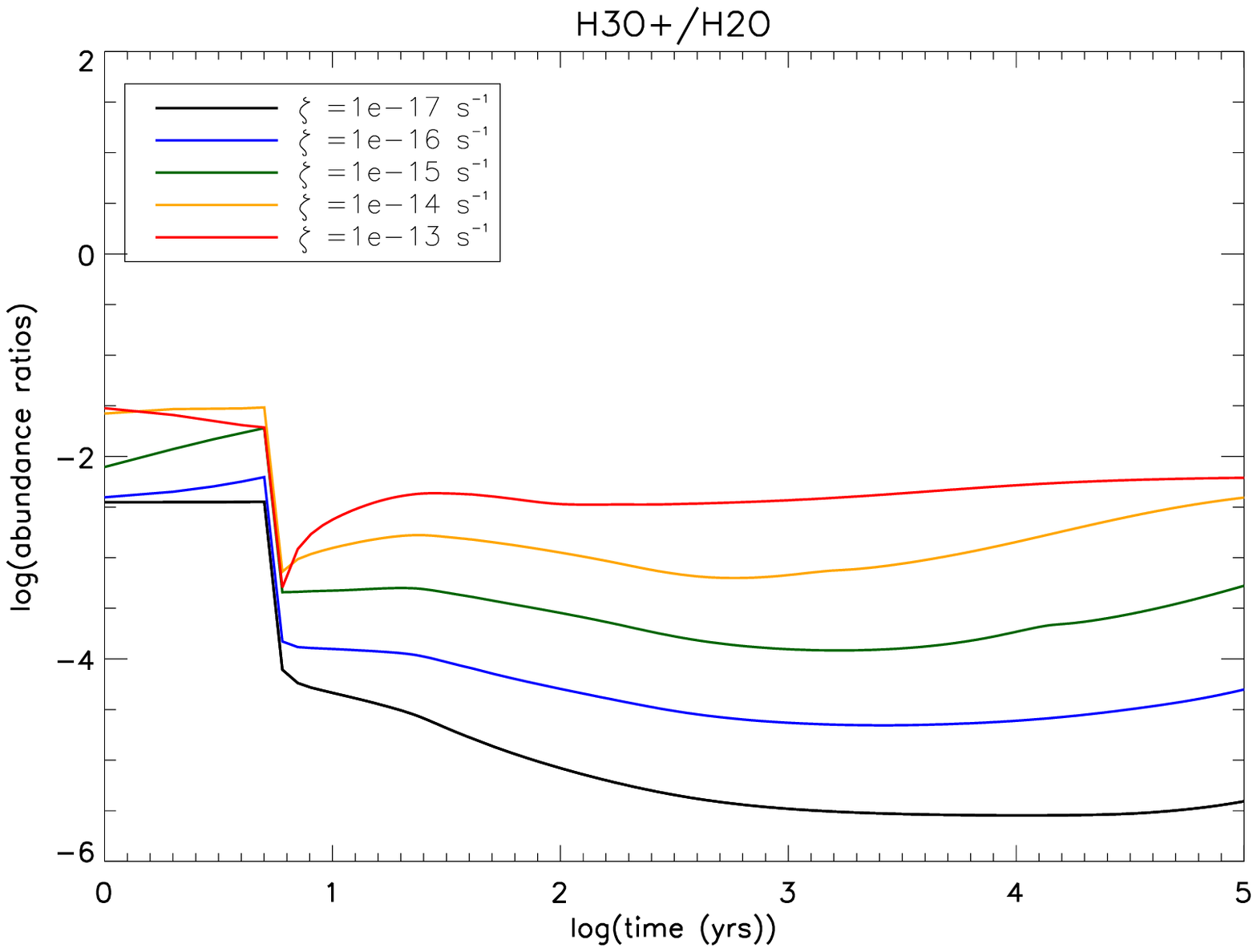}}
\caption{Same as Figure \ref{fig:n2e5_z-16_1}, but for other different species. The bottom panels are showing the abundance ratios instead of fractional abundances.}
\label{fig:n2e5_v20_em2_2}
\end{figure*}

\begin{figure*}
\centerline{\includegraphics[angle=0,width=.4\textwidth]{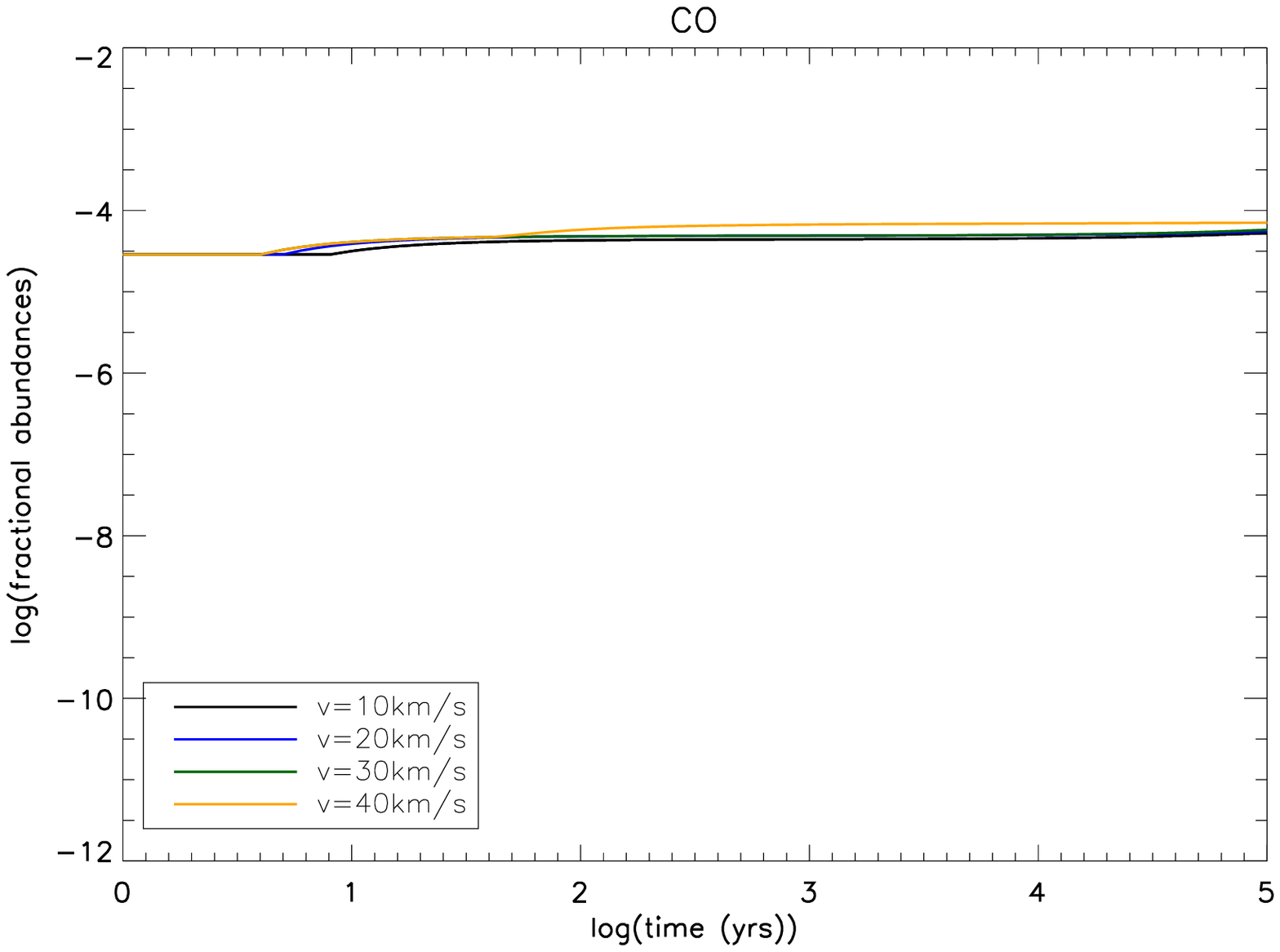}~~
\includegraphics[angle=0,width=.4\textwidth]{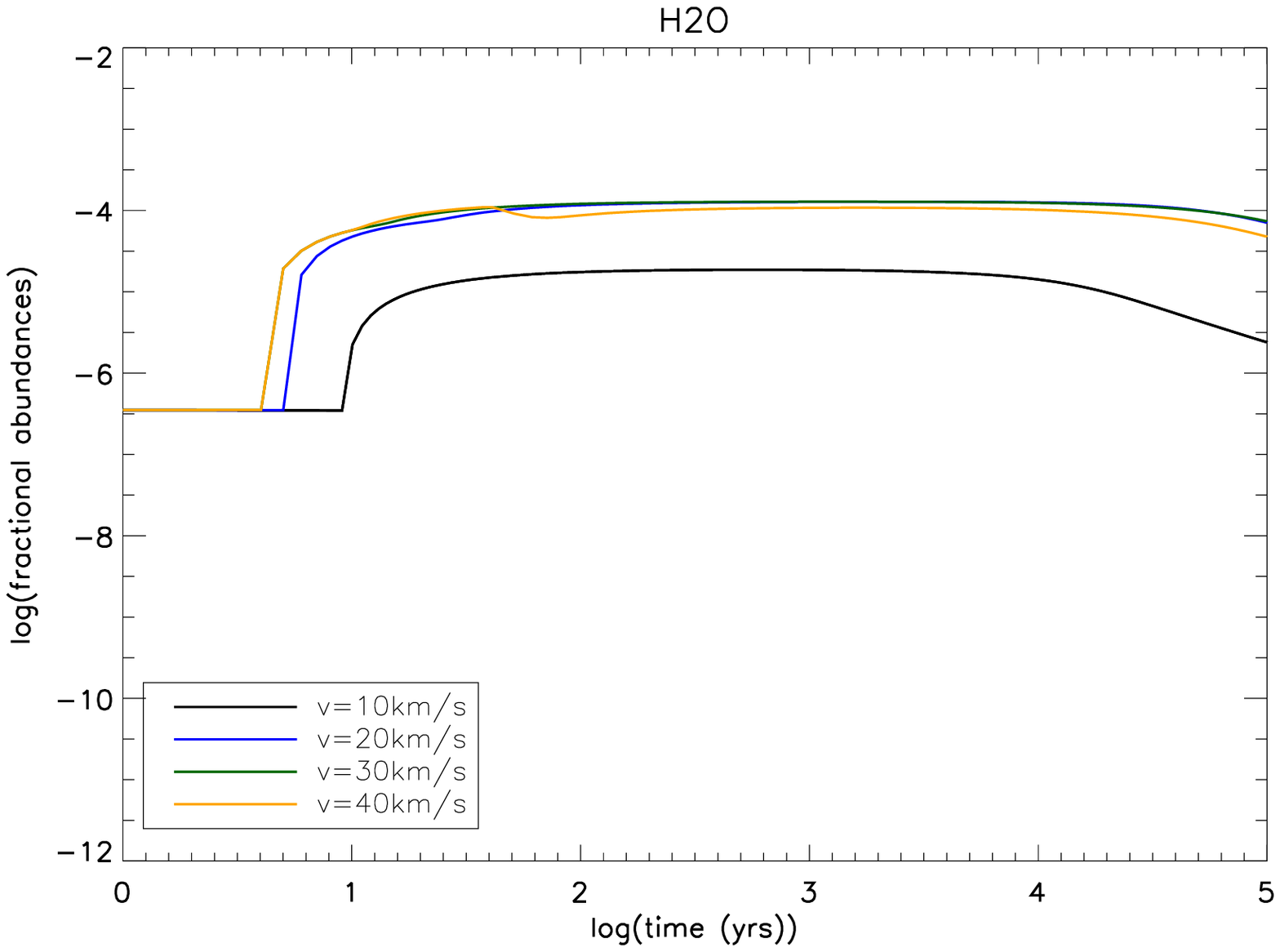}}
\centerline{\includegraphics[angle=0,width=.4\textwidth]{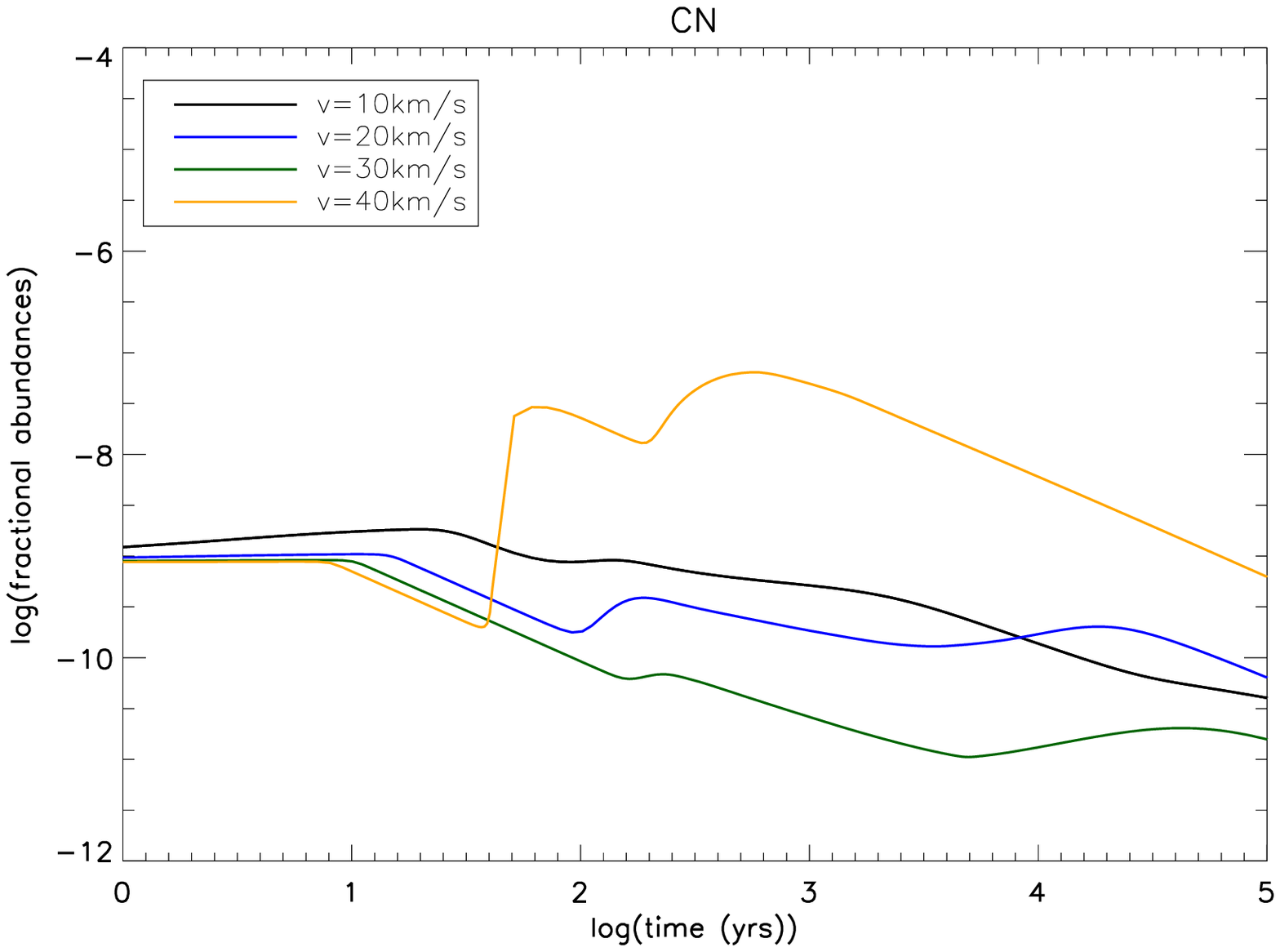}~~
\includegraphics[angle=0,width=.4\textwidth]{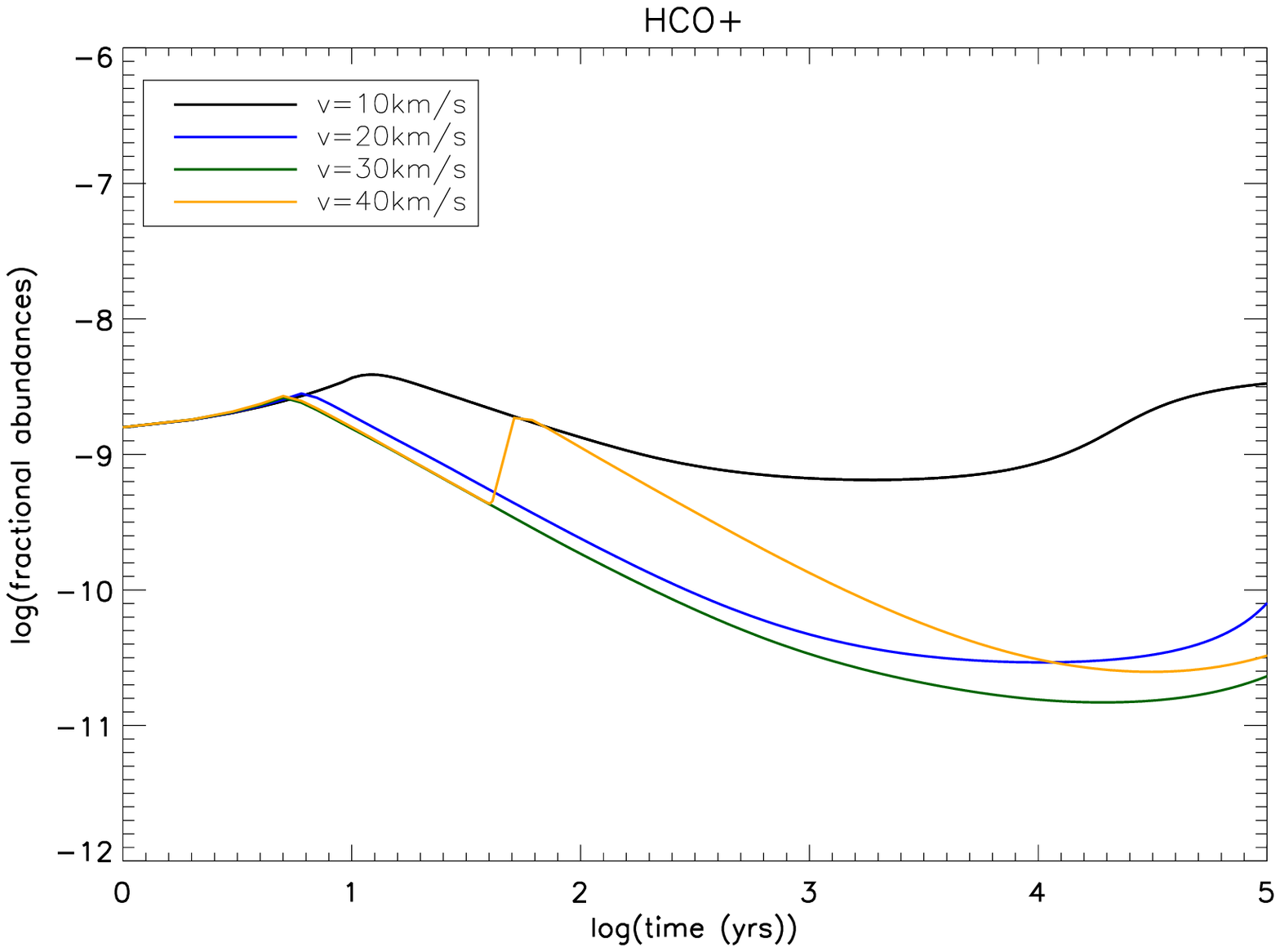}}
\centerline{\includegraphics[angle=0,width=.4\textwidth]{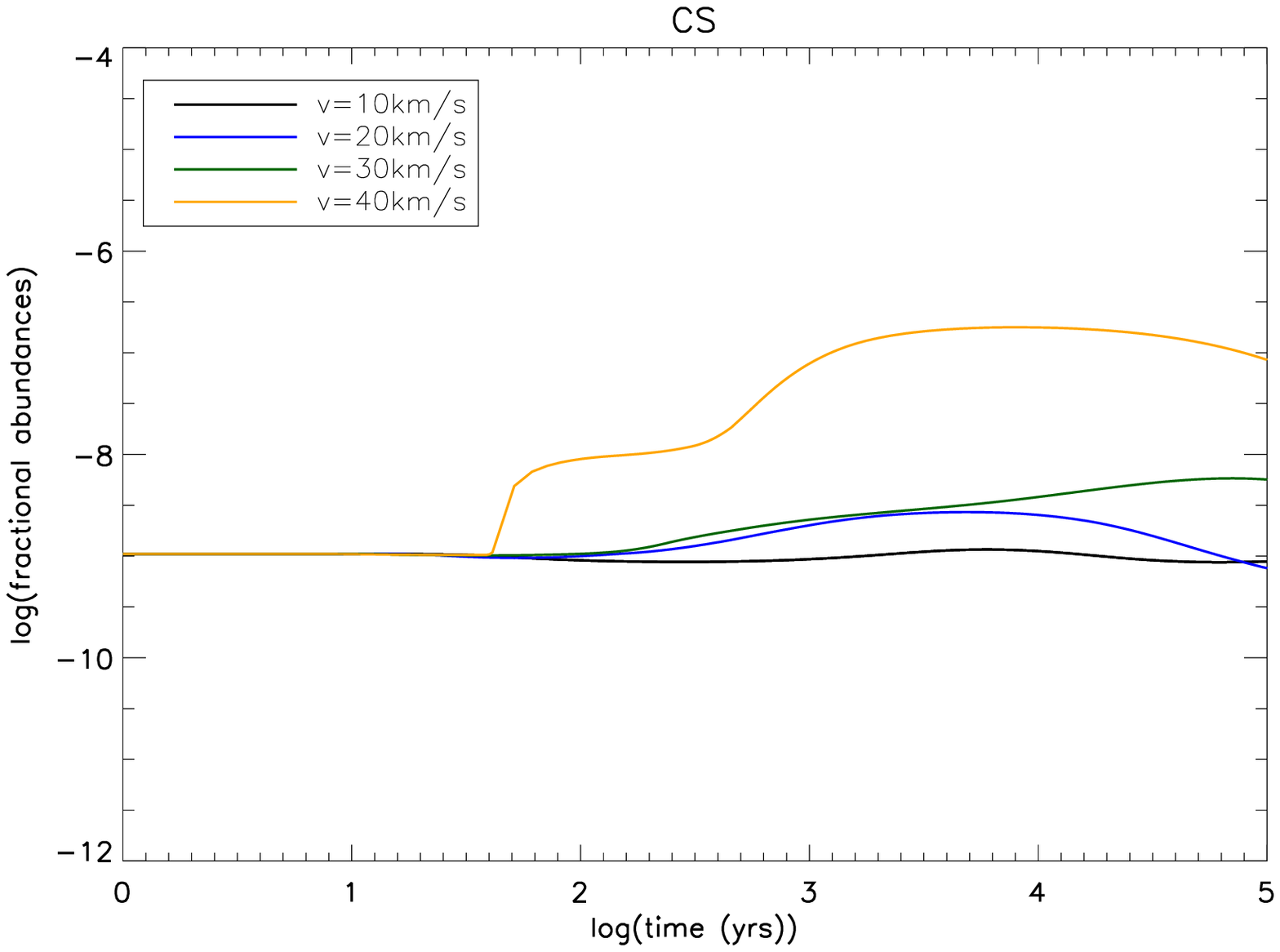}~~
\includegraphics[angle=0,width=.4\textwidth]{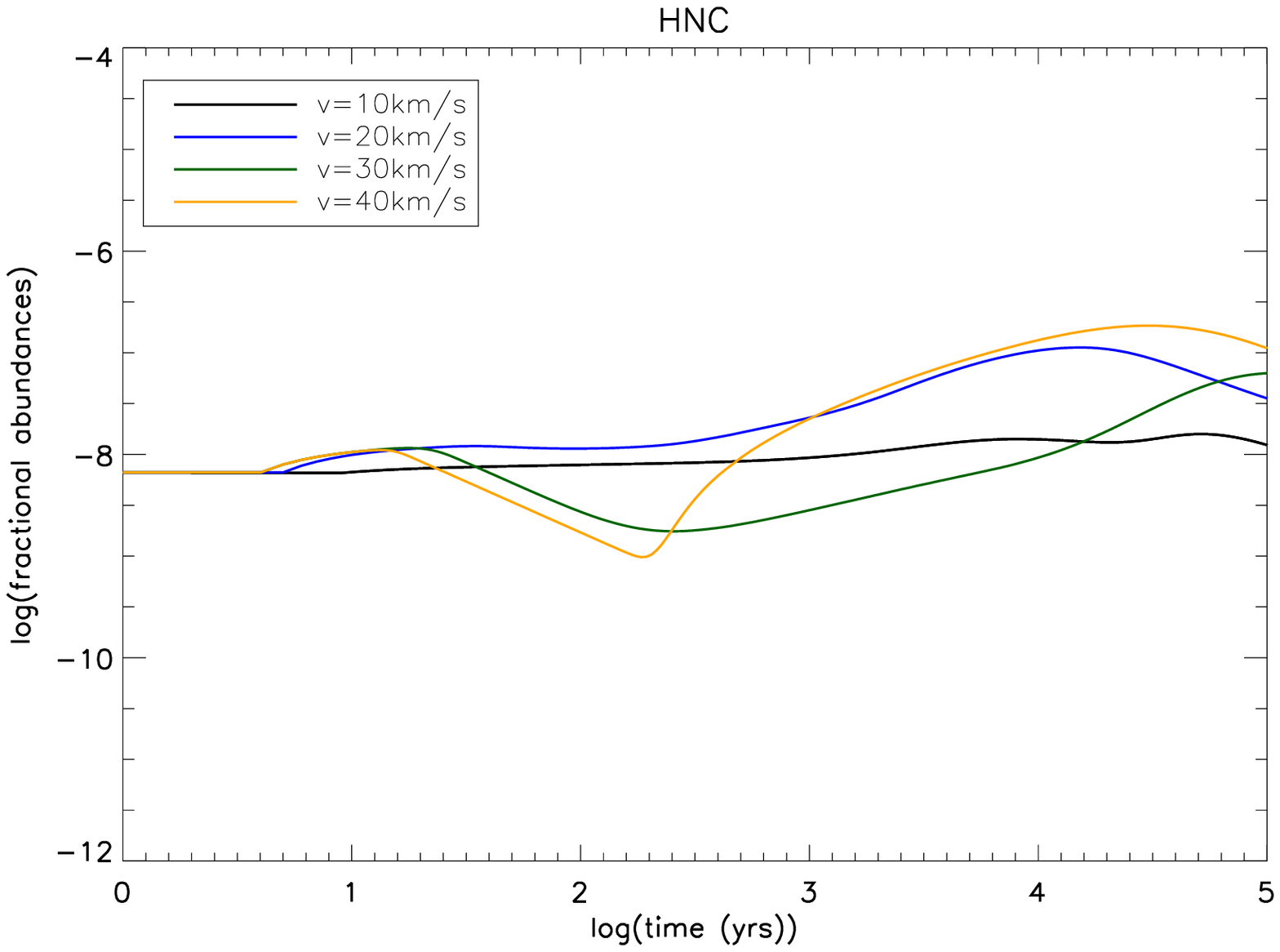}}
\centerline{\includegraphics[angle=0,width=.4\textwidth]{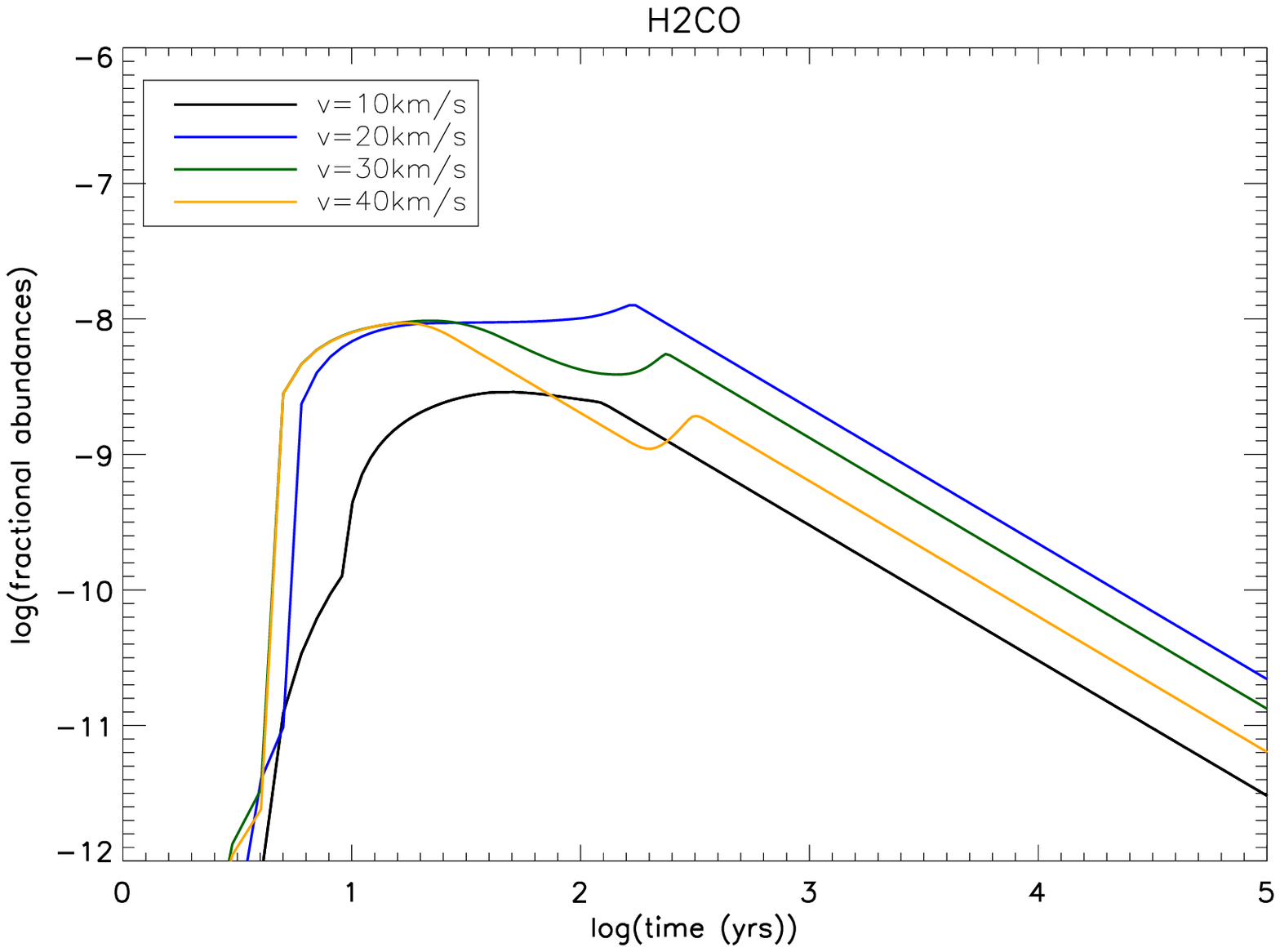}~~
\includegraphics[angle=0,width=.4\textwidth]{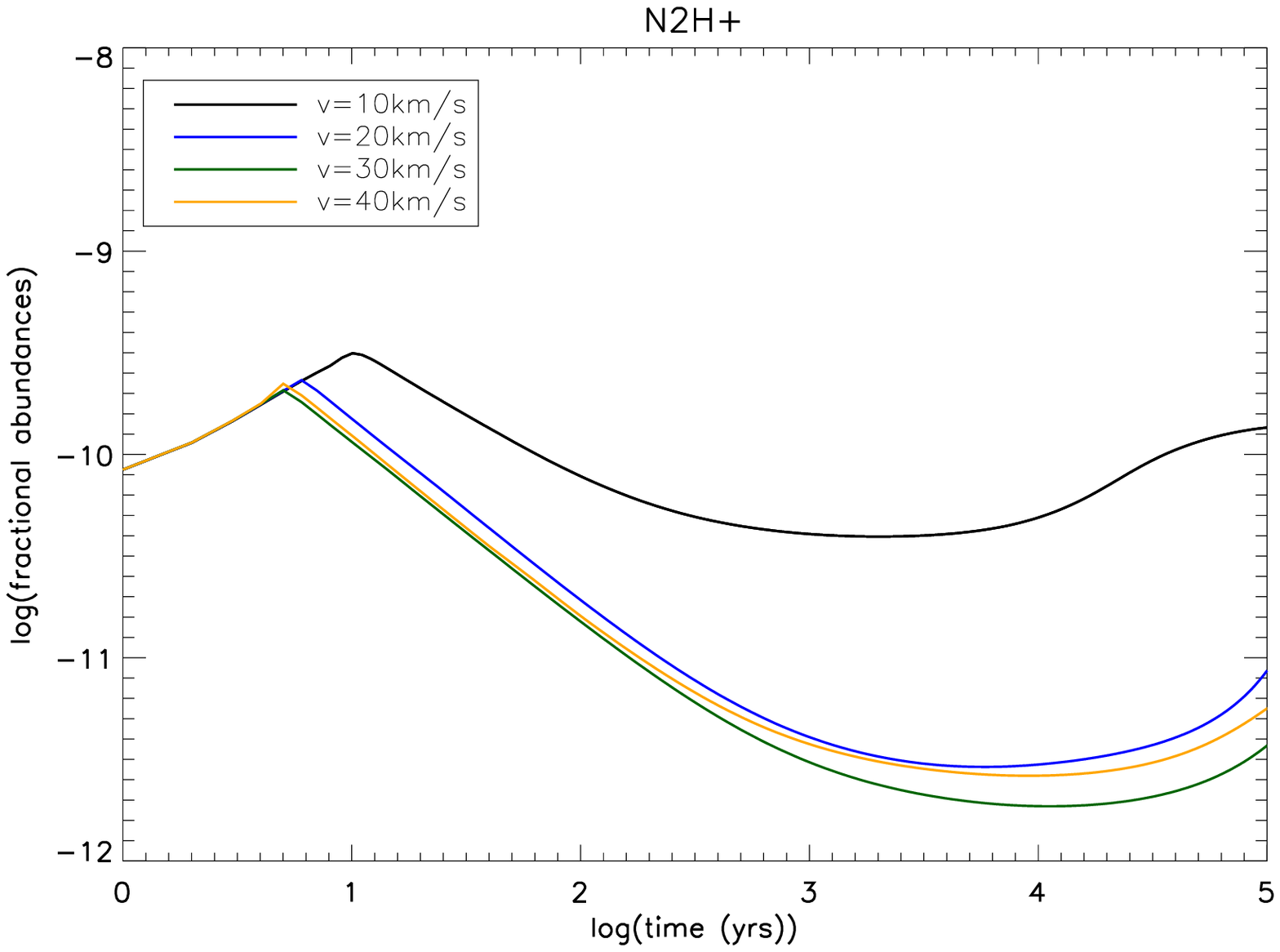}}
\caption{ Time-averaged fractional abundances of observed species are shown for $n=10^5\,$cm$^{-3}$, a cosmic-ray ionization rate of $\zeta = 1 \times 10^{-16}\,$s$^{-1}$, with varying shock velocities of 10, 20, 30, and 40 km s$^{-1}$.}
\label{fig:n2e5_z-16_1}
\end{figure*}

\begin{figure*}
\centerline{\includegraphics[angle=0,width=.4\textwidth]{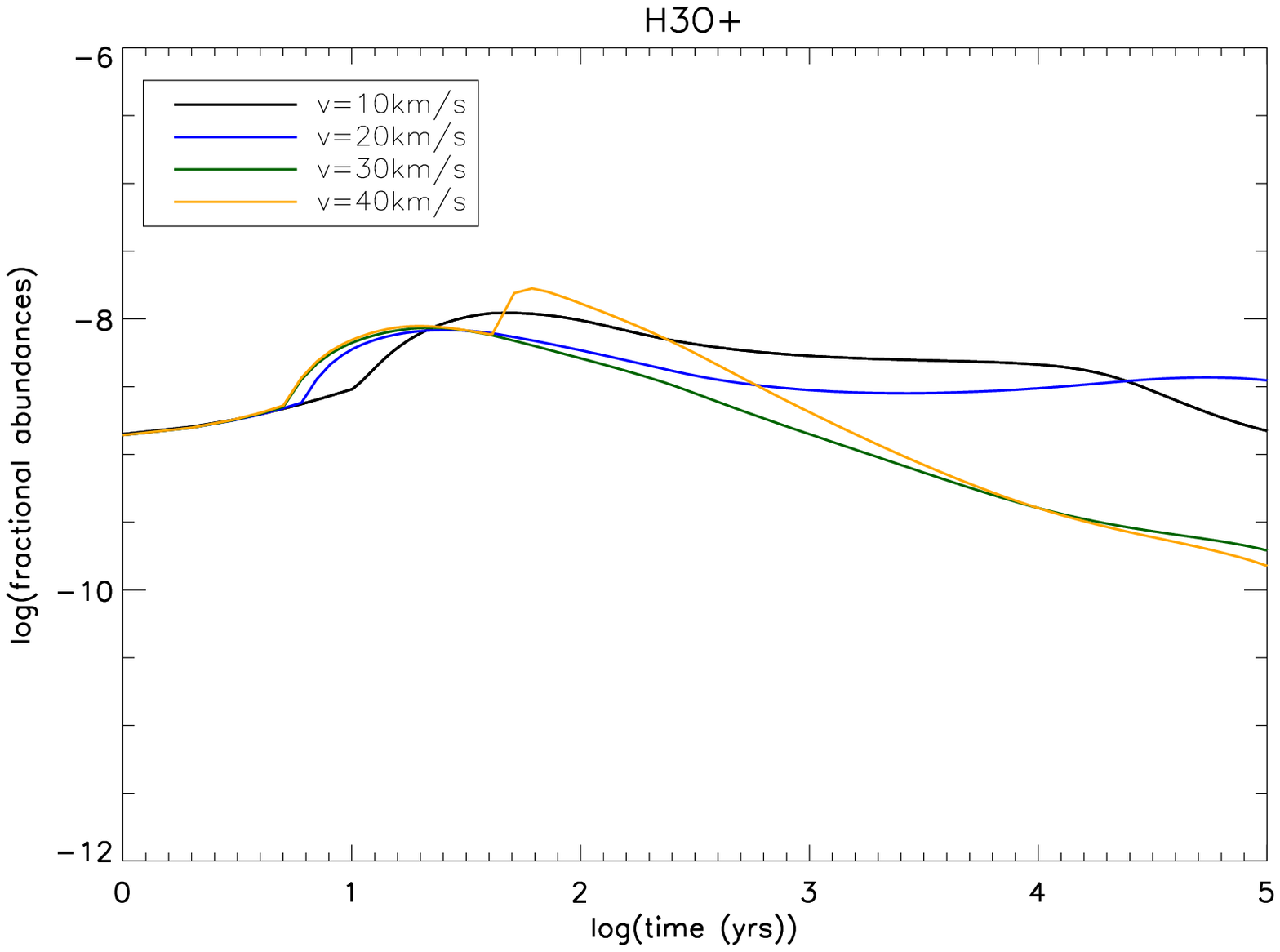}~~
\includegraphics[angle=0,width=.4\textwidth]{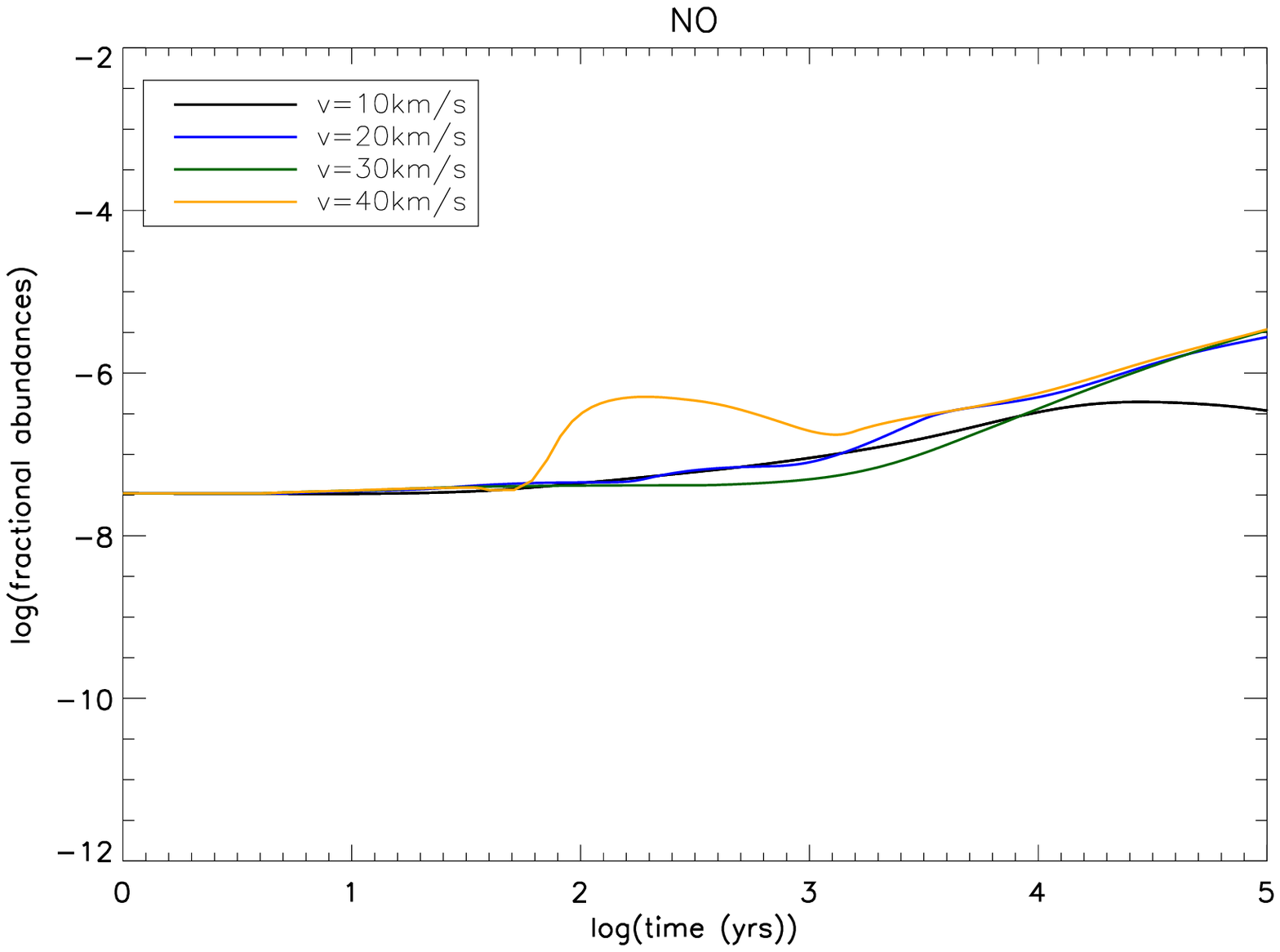}}
\centerline{\includegraphics[angle=0,width=.4\textwidth]{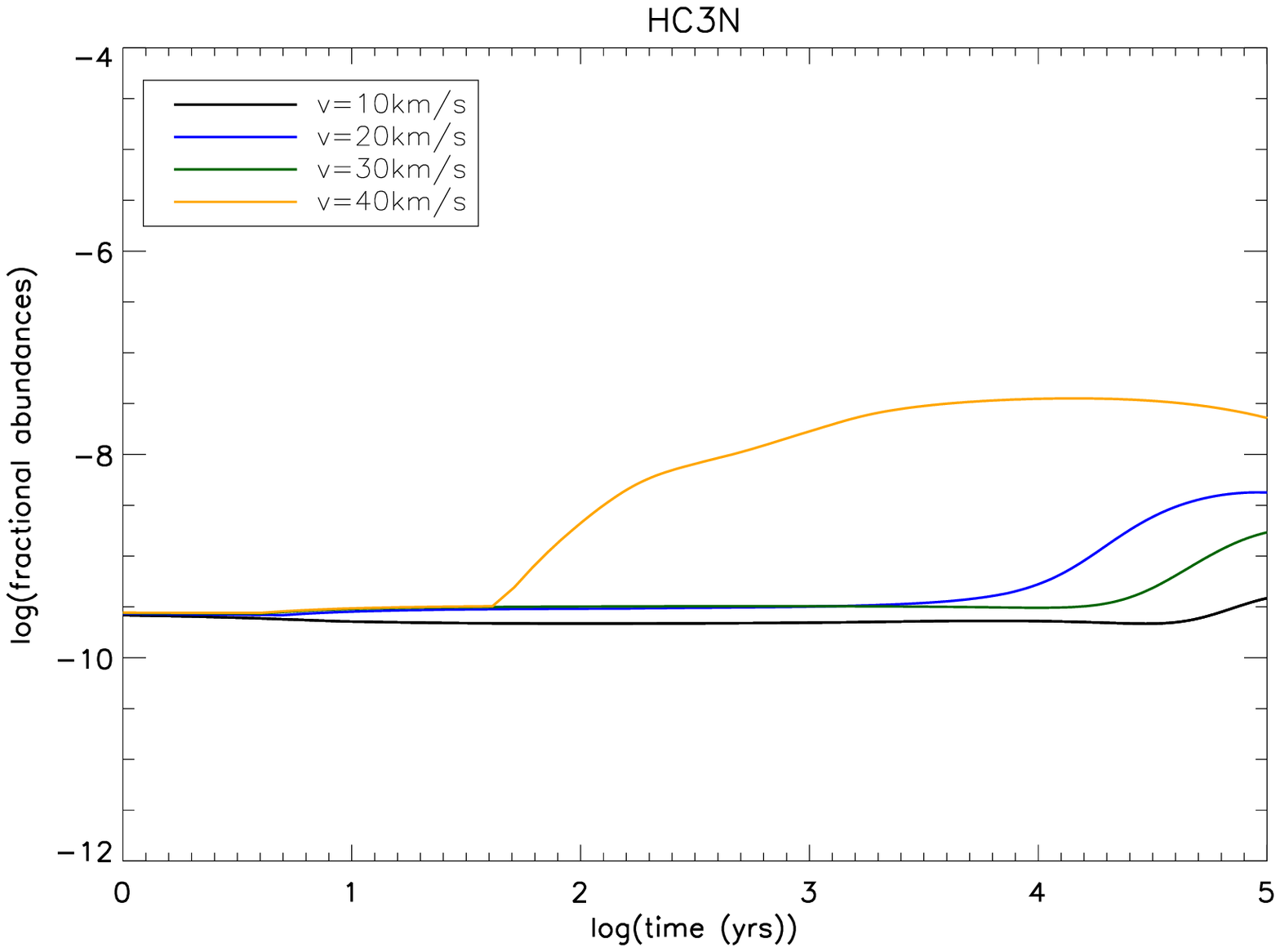}~~
\includegraphics[angle=0,width=.4\textwidth]{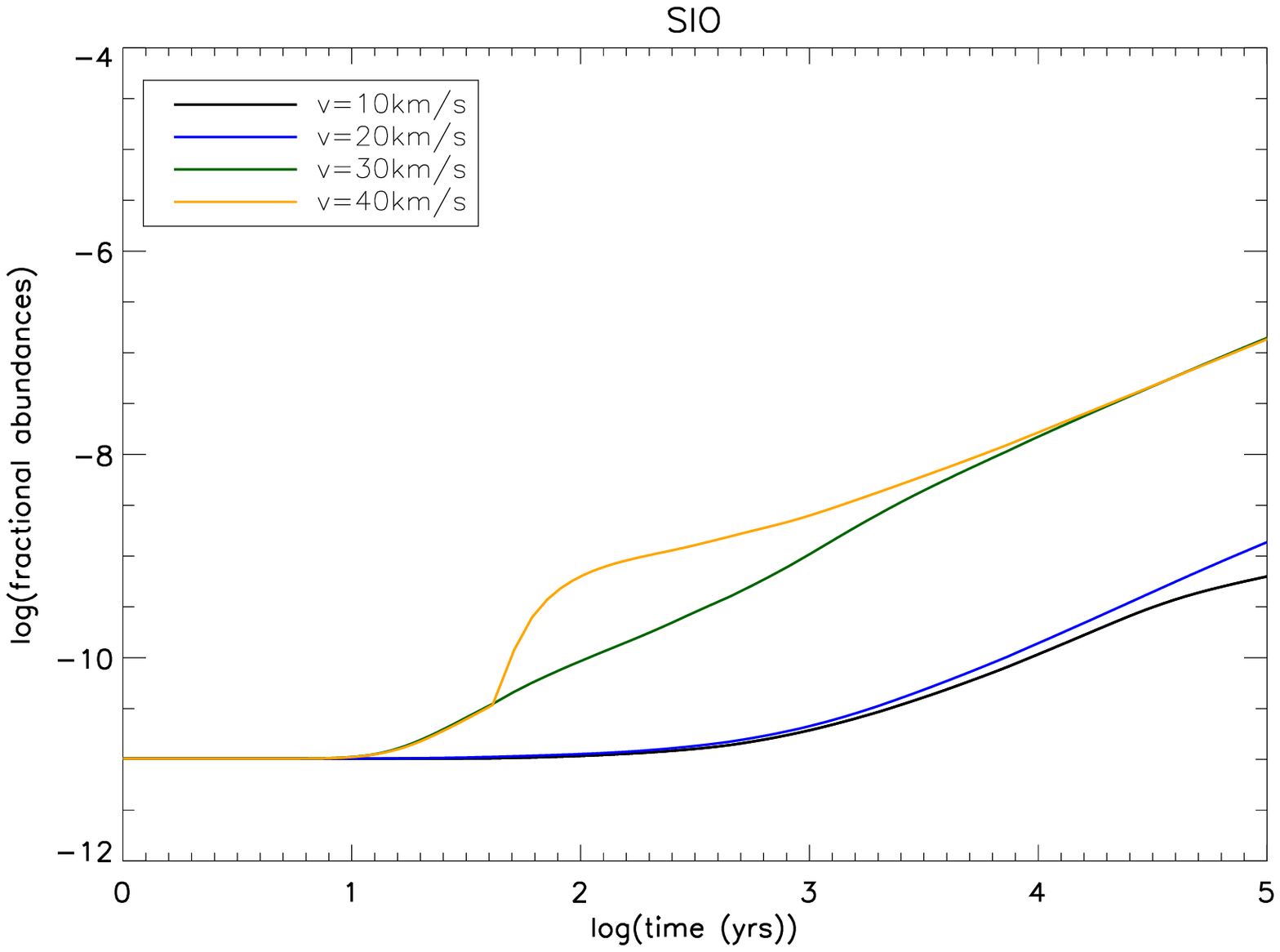}}
\centerline{\includegraphics[angle=0,width=.4\textwidth]{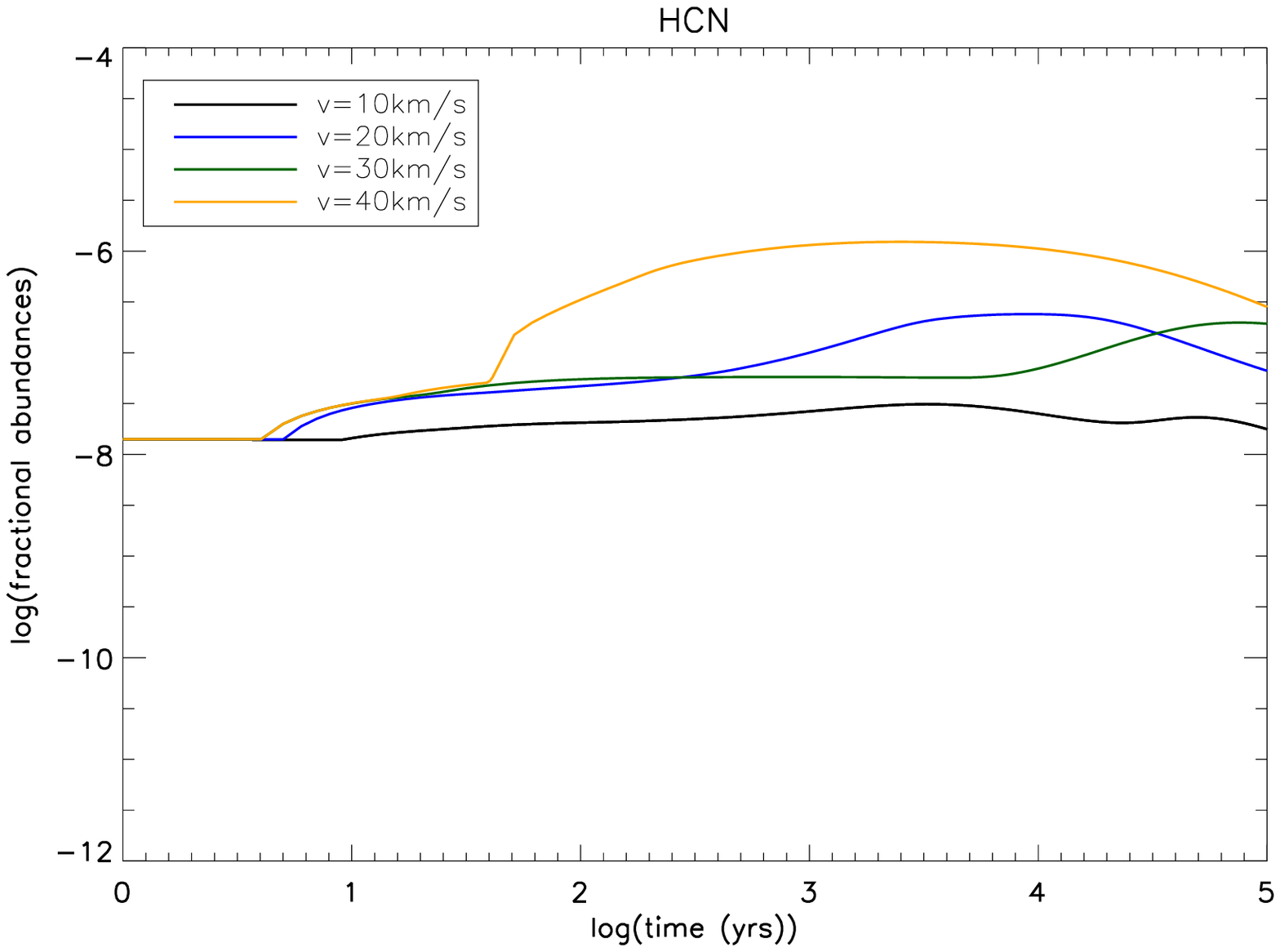}~~
\includegraphics[angle=0,width=.4\textwidth]{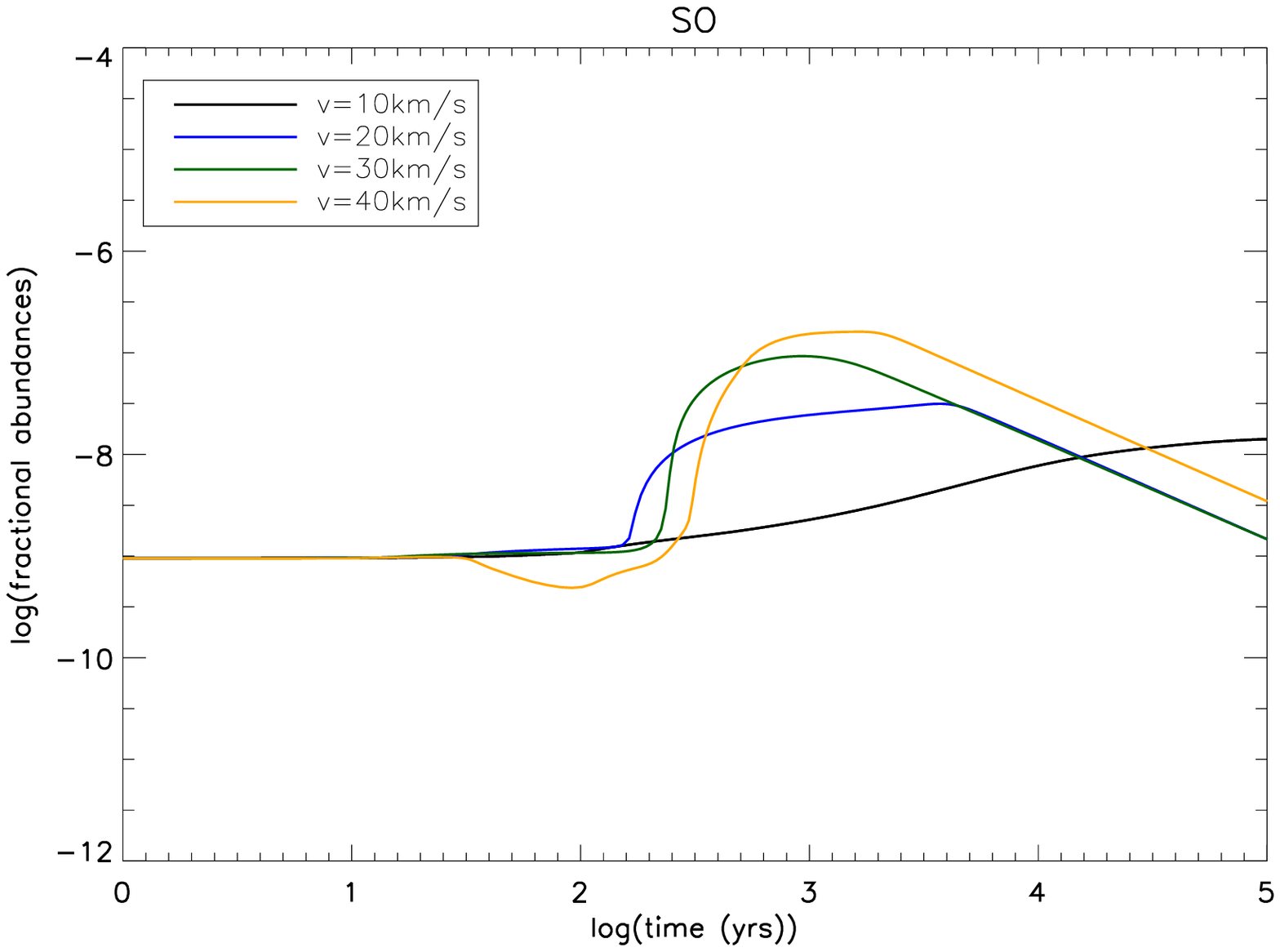}}
\centerline{\includegraphics[angle=0,width=.4\textwidth]{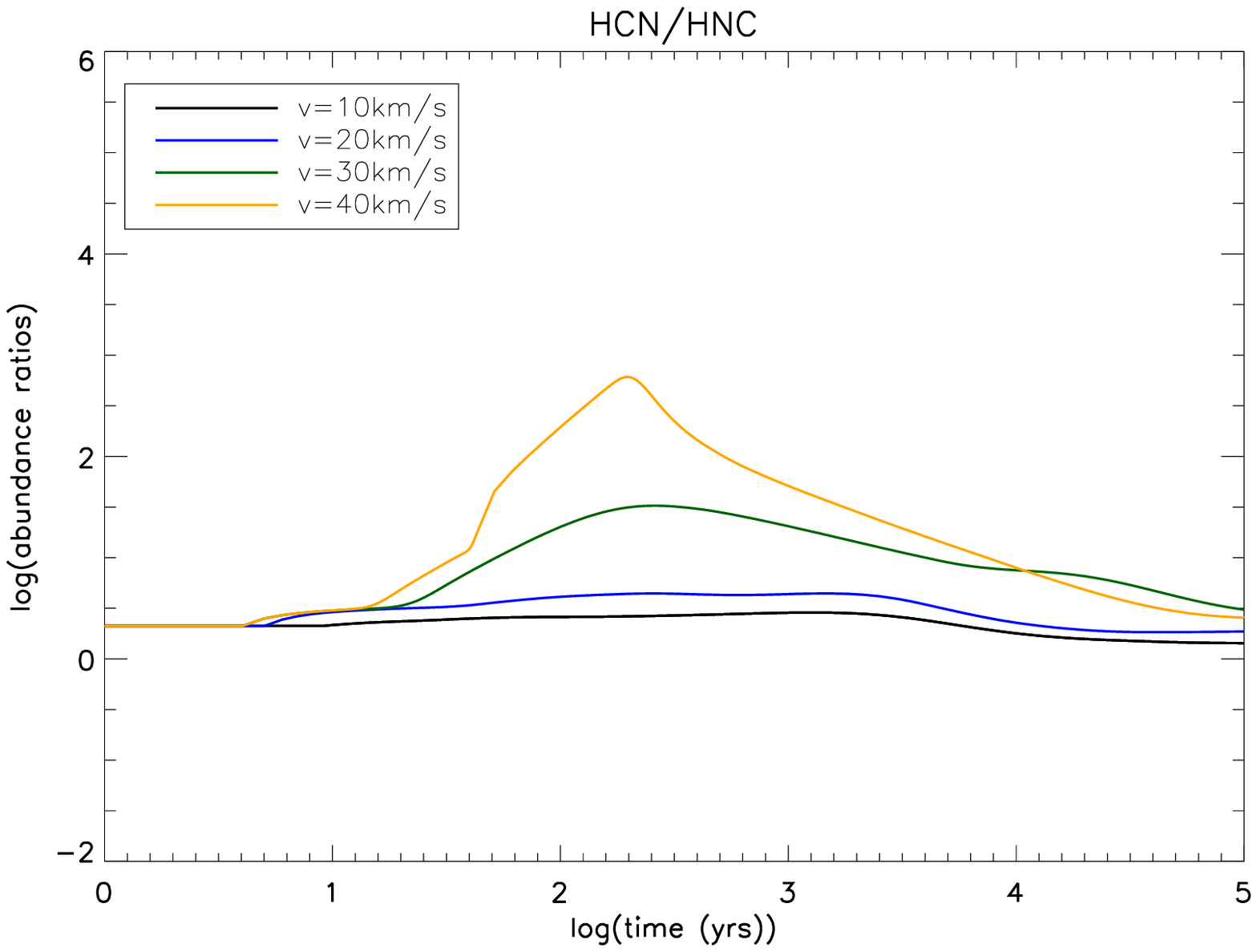}~~
\includegraphics[angle=0,width=.4\textwidth]{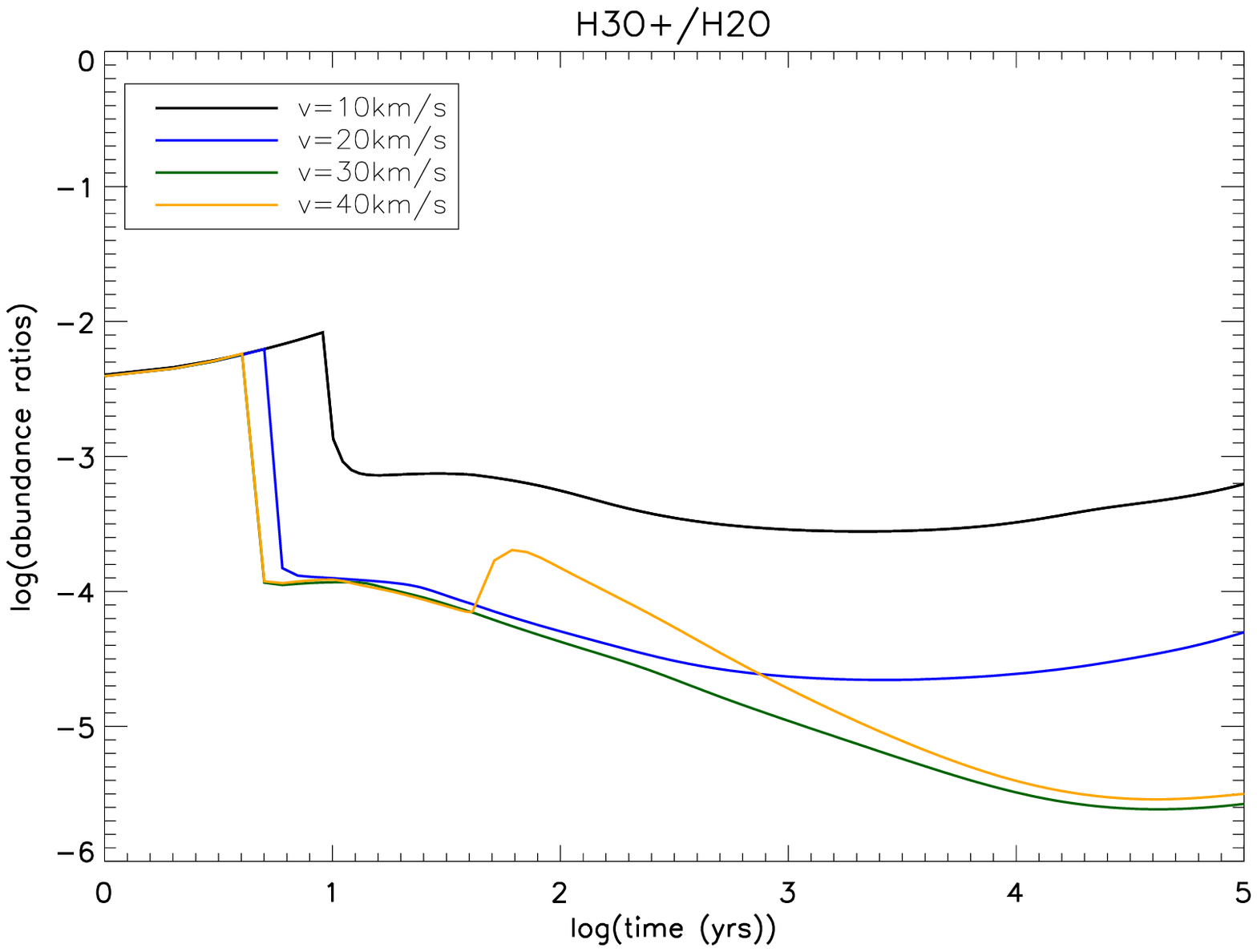}}
\caption{Same as Figure \ref{fig:n2e5_z-16_1}, but for different species. The bottom panels are showing the abundance ratios instead of fractional abundances.}
\label{fig:n2e5_z-16_2}
\end{figure*}

\begin{figure*}
\centerline{\includegraphics[angle=90,width=.4\textwidth]{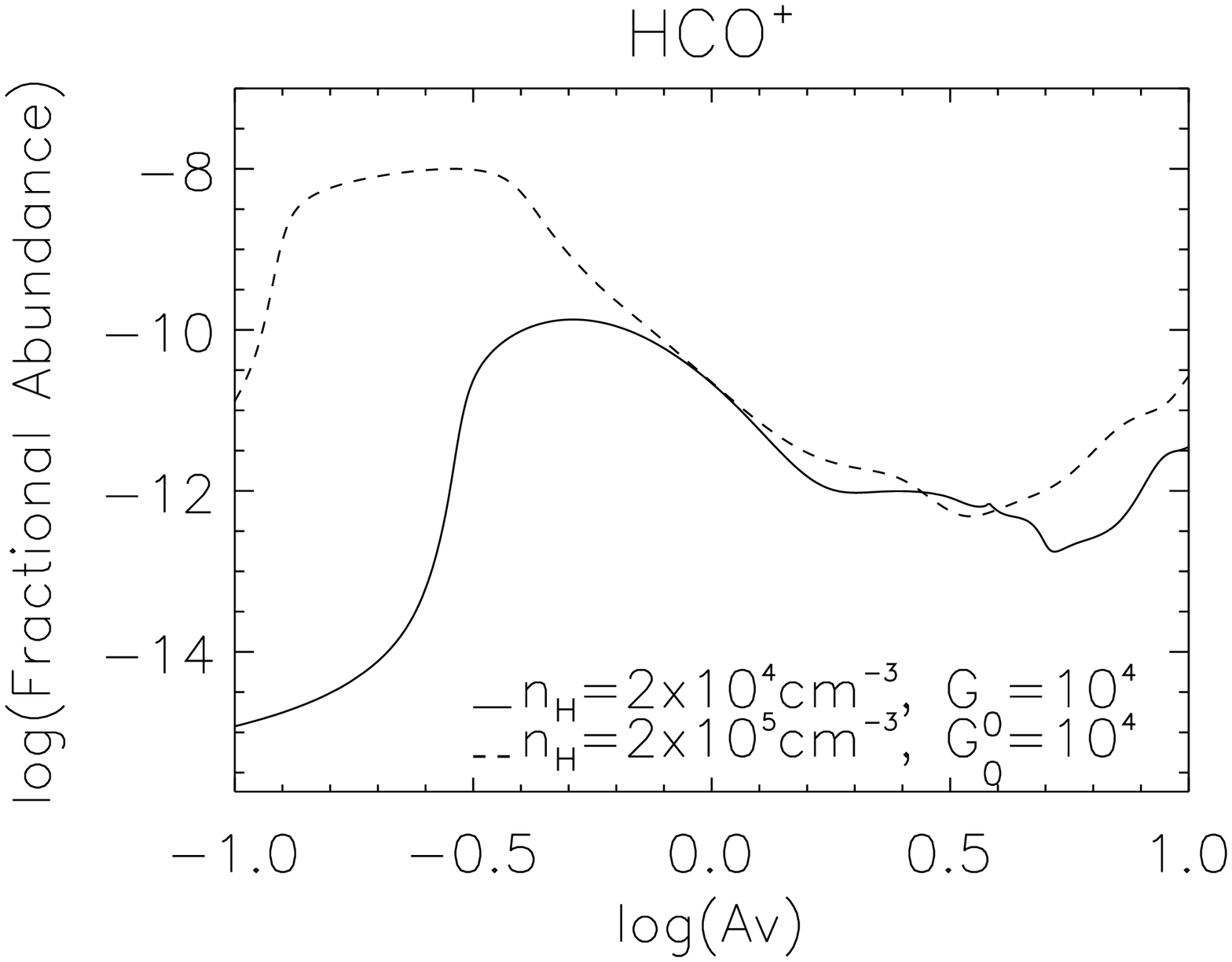}~~
\includegraphics[angle=90,width=.4\textwidth]{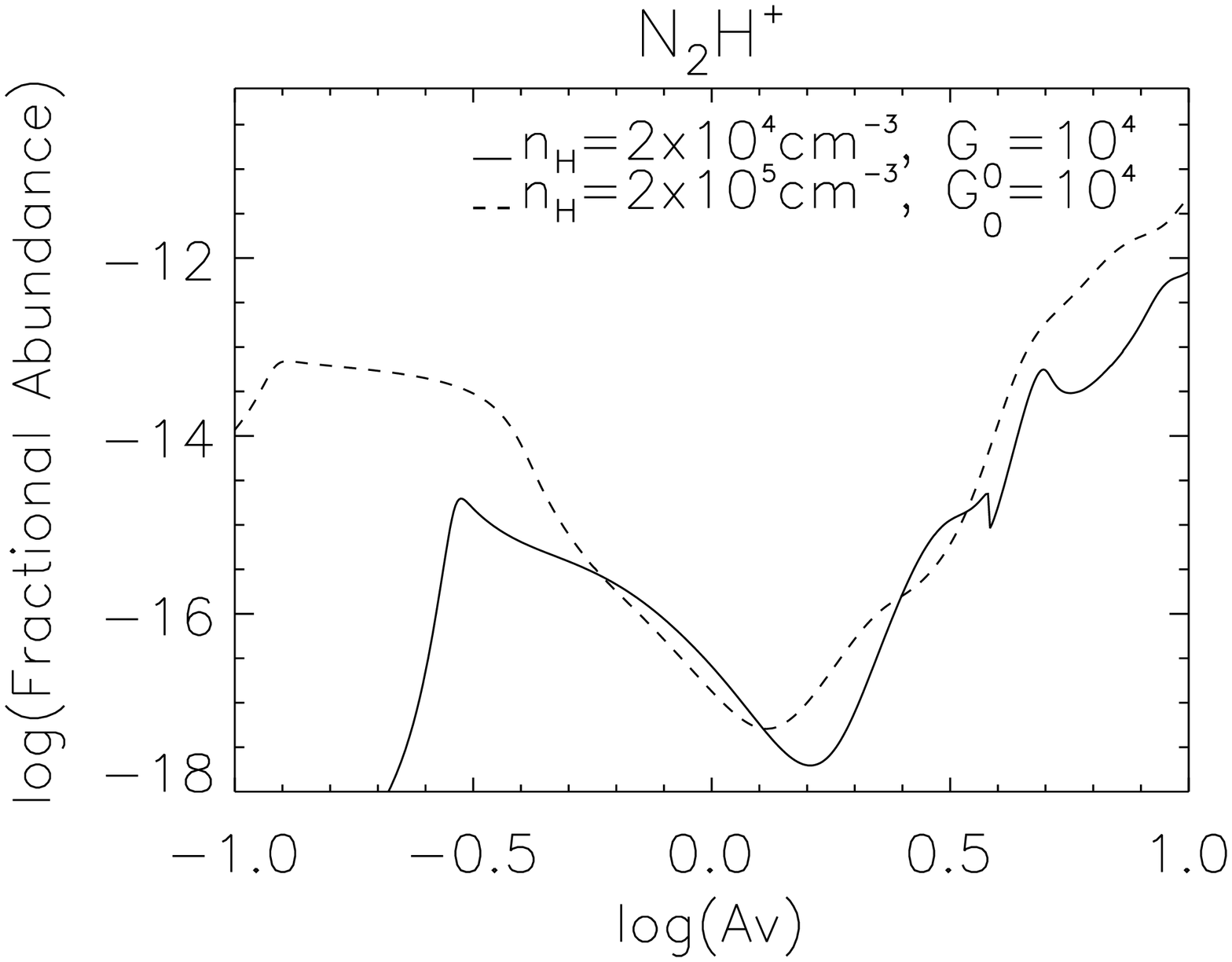}}
\centerline{\includegraphics[angle=90,width=.4\textwidth]{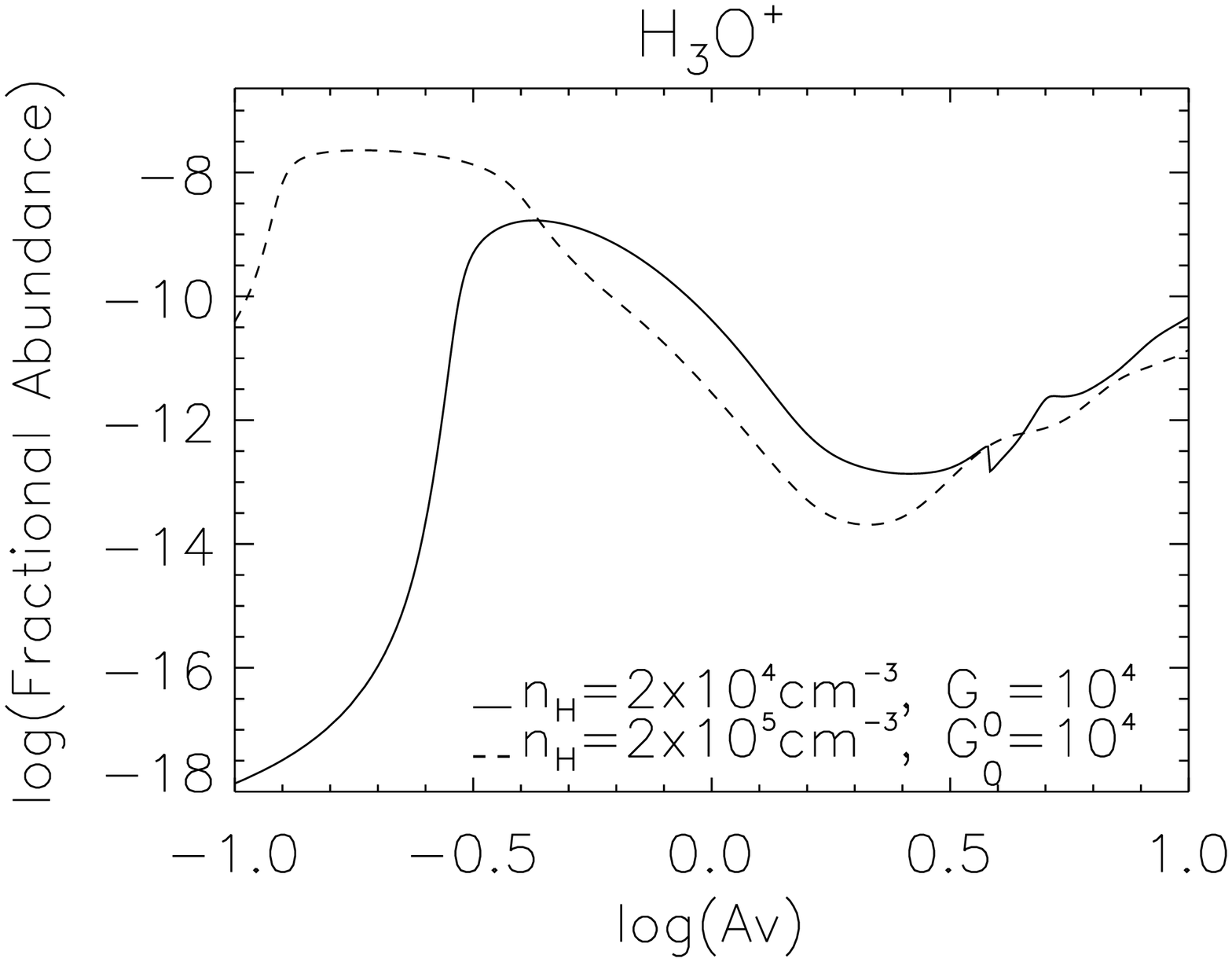}~~
\includegraphics[angle=90,width=.4\textwidth]{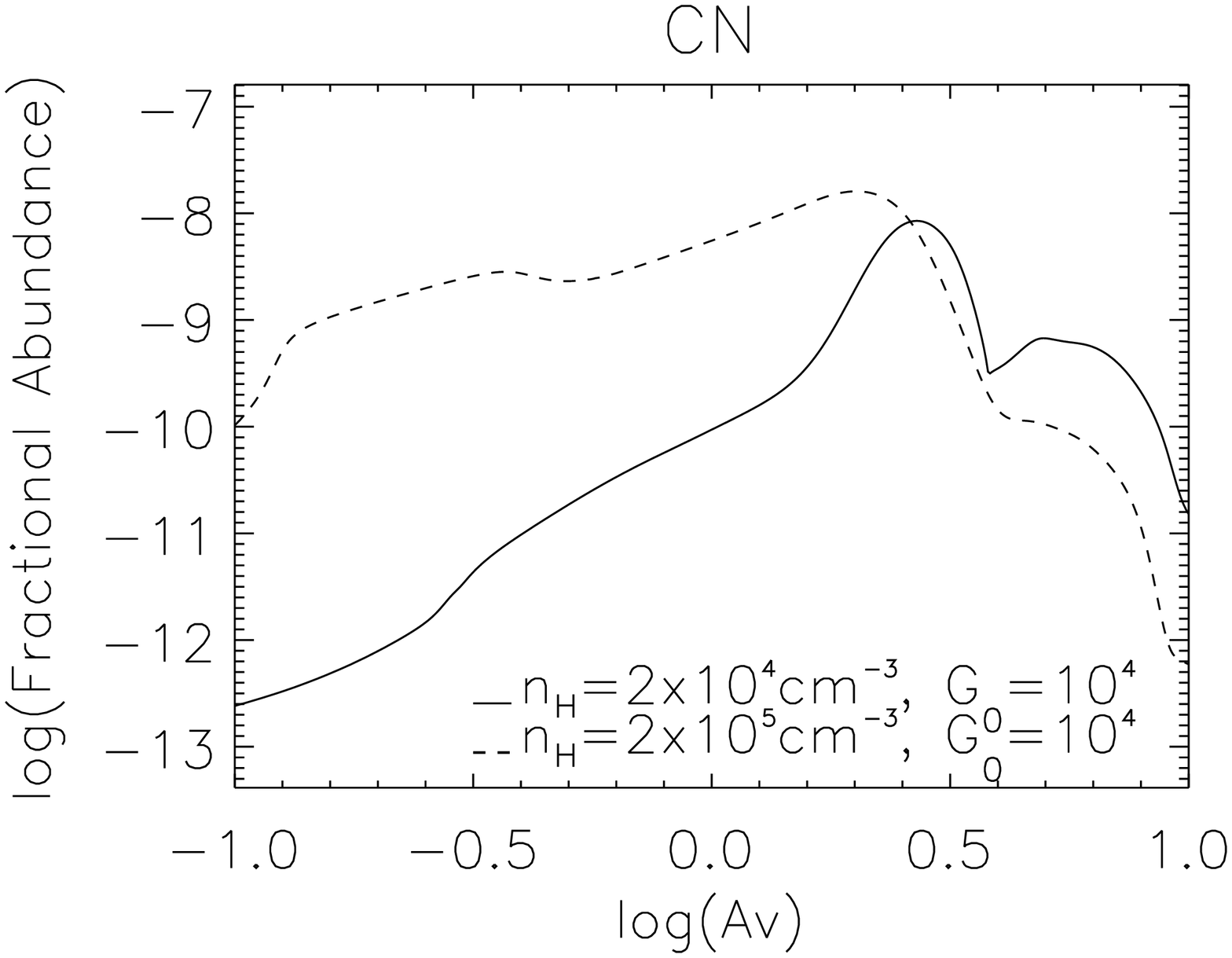}}
\caption{Fractional abundances of HCO$^{+}$, N$_2$H$^{+}$, H$_{3}$O$^{+}$, and CN derived by Meudon PDR code for $n_H = 2\times10^4$ and $2\times10^5$\,cm$^{-3}$, $G_0=10^4$, and $\zeta=1\times10^{-17}\,$s$^{-1}$.}
\label{fig:pdr}
\end{figure*}

\begin{figure*}
\centerline{\includegraphics[angle=0,width=.4\textwidth]{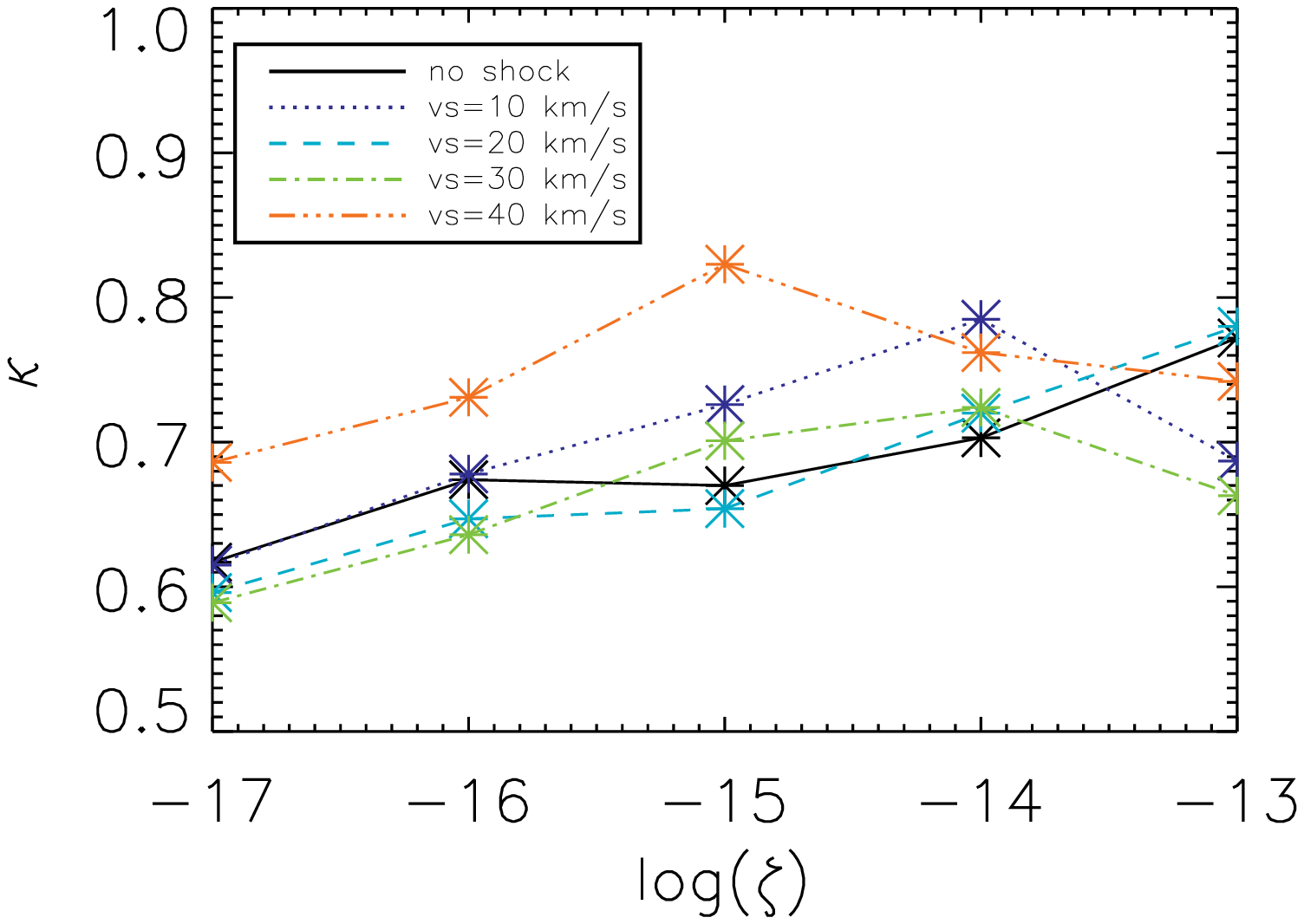}
\includegraphics[angle=0,width=.4\textwidth]{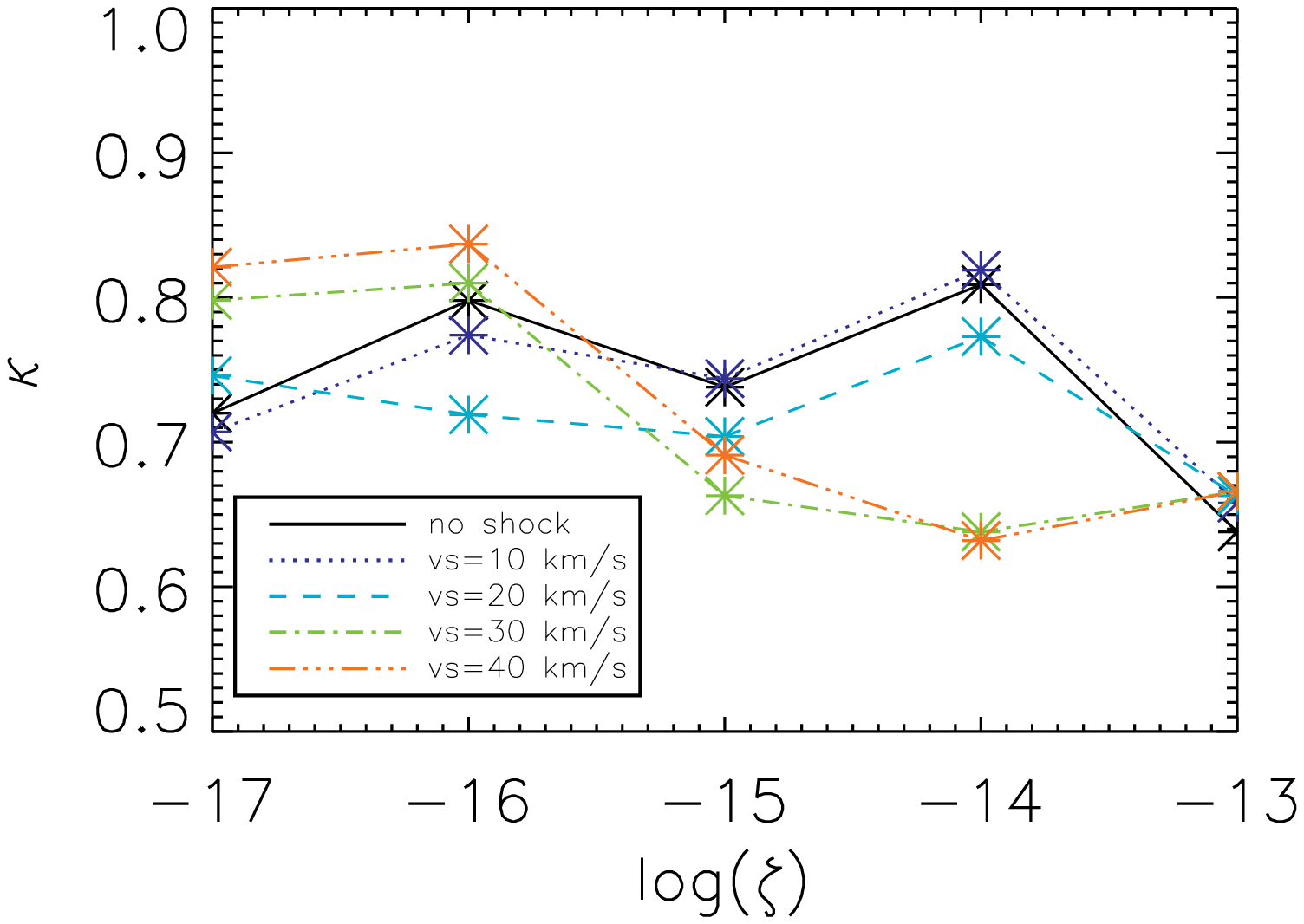}}
\caption{Values of confidence level as a function of cosmic-ray ionization rate are shown for $n_H=2\times 10^5\,$cm$^{-3}$({\it left panel}) and $n_H=2\times 10^4\,$cm$^{-3}$({\it right panel}).}
\label{fig:fit}
\end{figure*}

\begin{figure*}
\centerline{\includegraphics[angle=0,width=.3\textwidth]{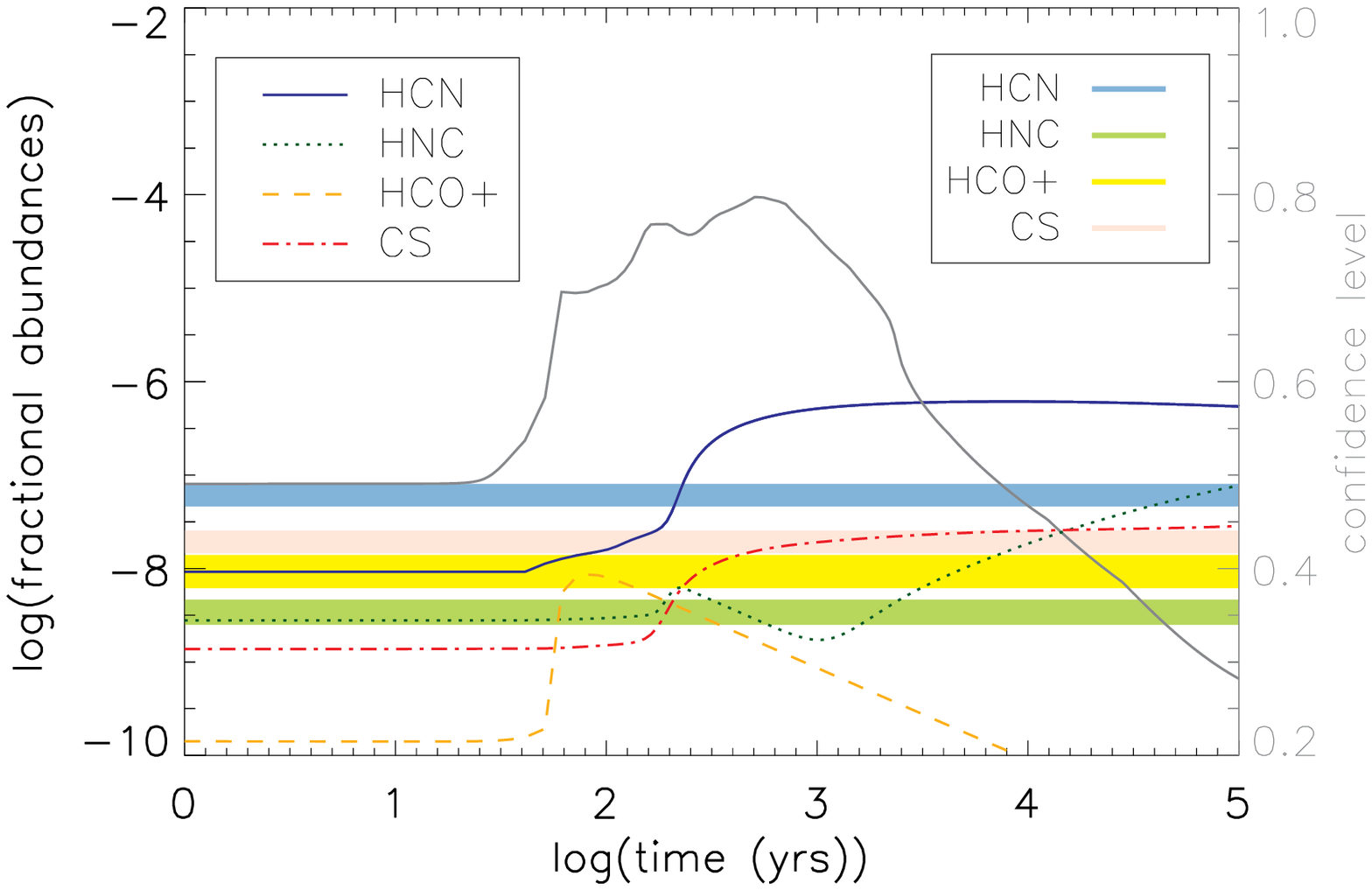}
\includegraphics[angle=0,width=.3\textwidth]{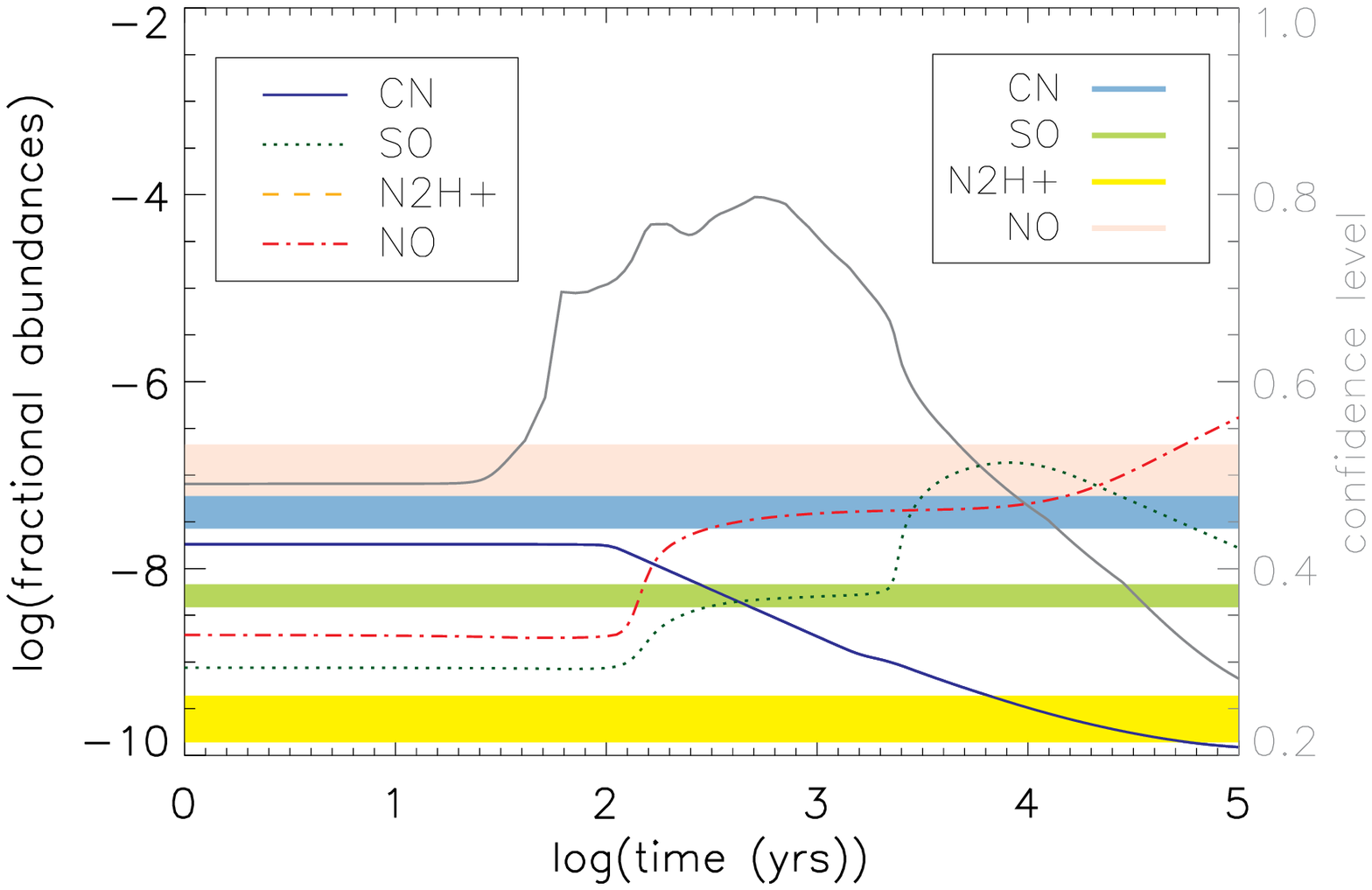}
\includegraphics[angle=0,width=.3\textwidth]{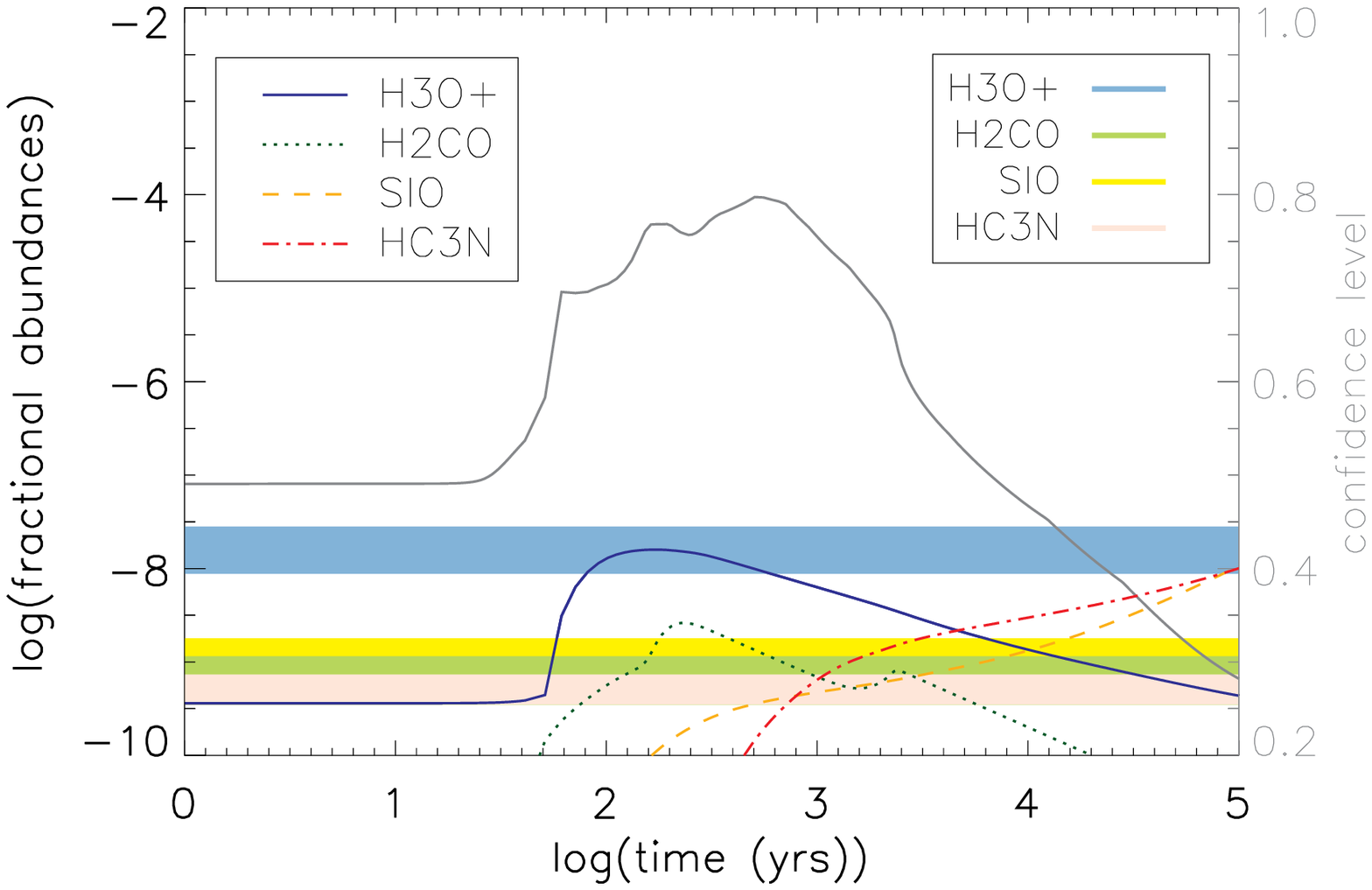}}
\caption{Time-averaged fractional abundances $<X(t)> = \int X(t') dt'/t$ are shown for $n_H=2\times10^4\,$cm$^{-3}$, $\zeta = 10^{-17}\,$s$^{-1}$, and $v_s = 30\,$km s$^{-1}$ with solid lines. Observed abundance ratios over CO are shown as filled colors. The grey line with the right ordinate scale shows the value of confidence level as a function of time.}
\label{fig:n2e4_model2}
\end{figure*}

\begin{figure*}
\centerline{\includegraphics[angle=0,width=.3\textwidth]{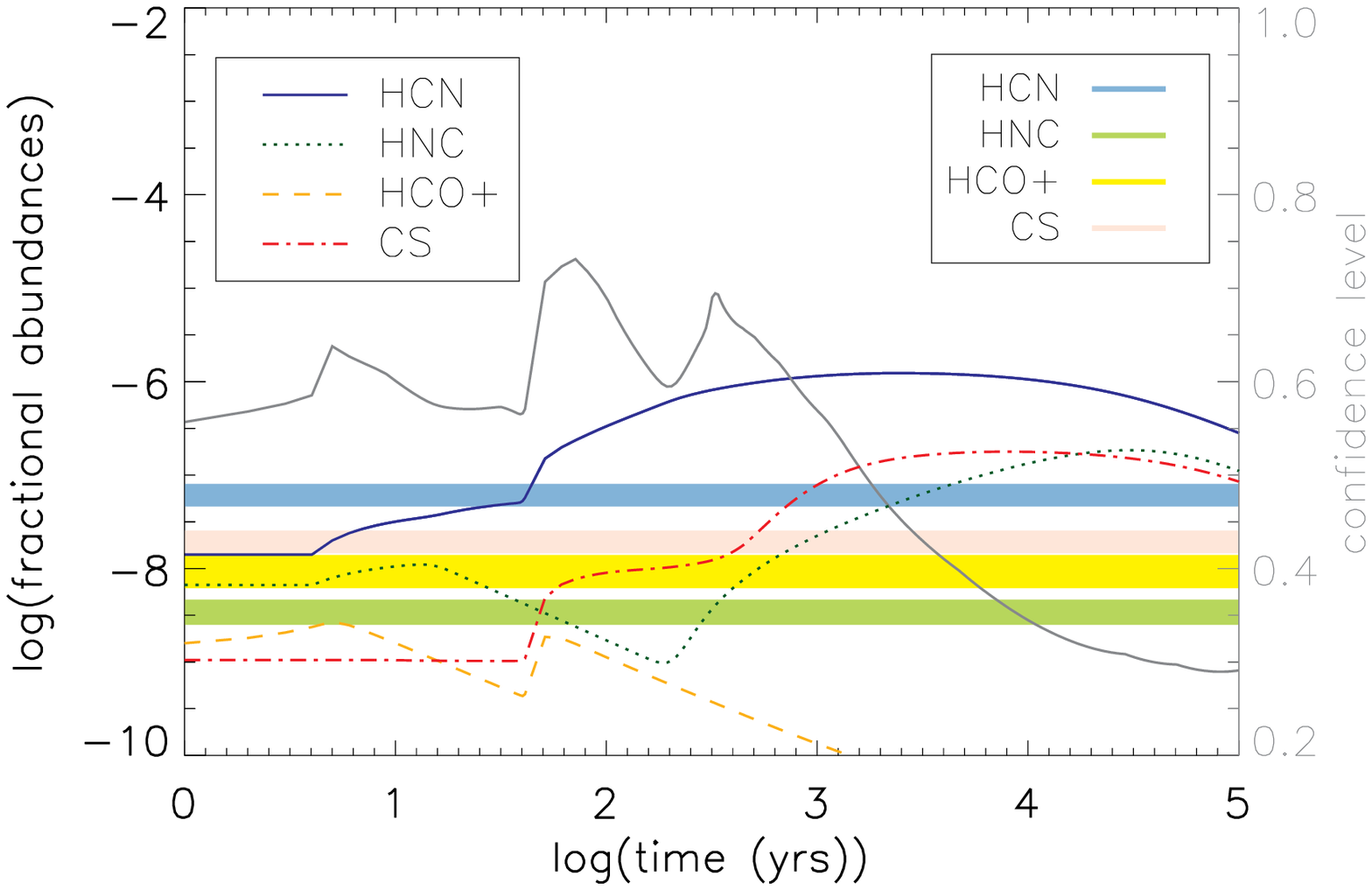}
\includegraphics[angle=0,width=.3\textwidth]{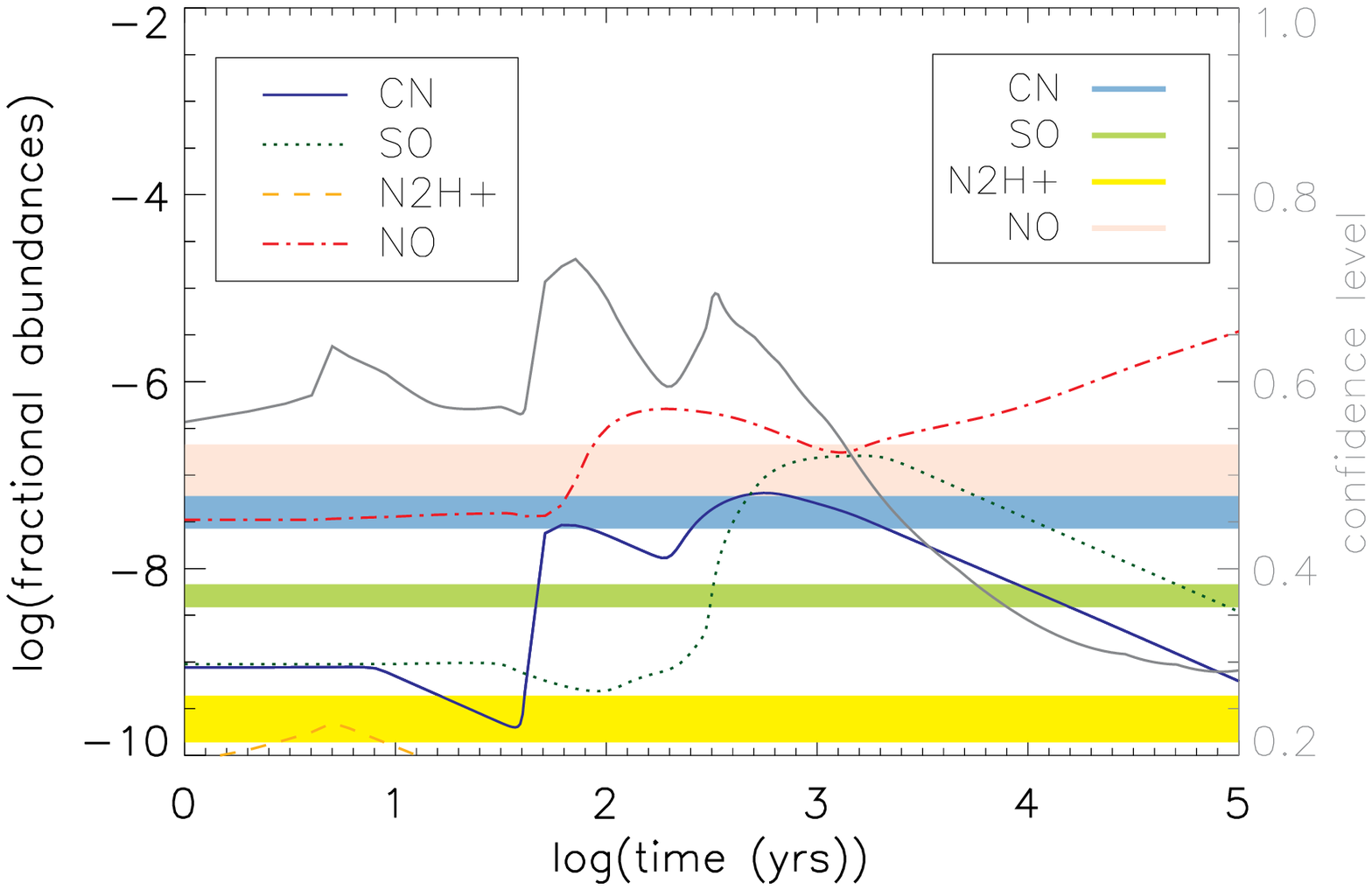}
\includegraphics[angle=0,width=.3\textwidth]{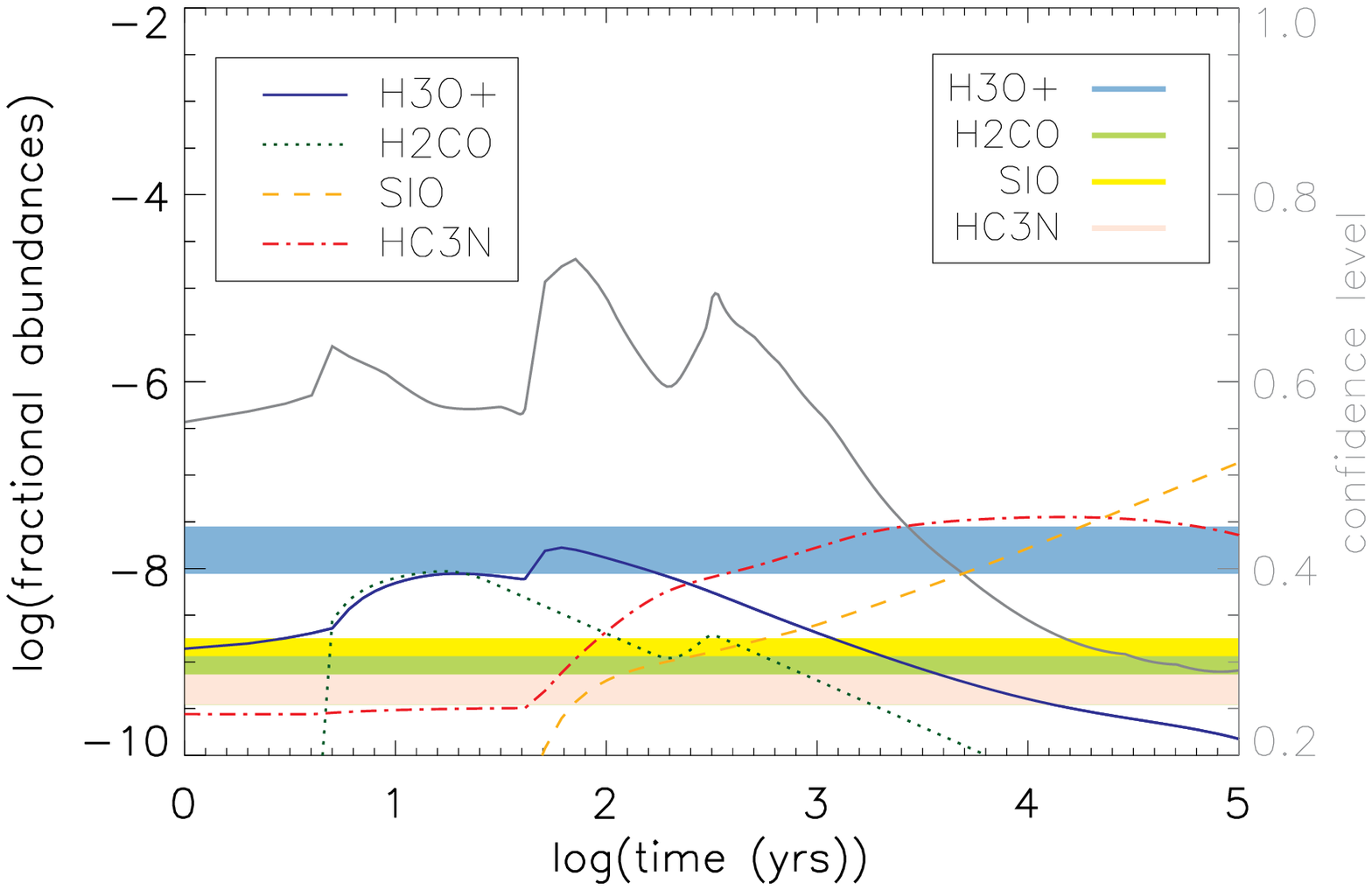}}
\caption{Same as  Figure \ref{fig:n2e4_model2} for $n_H=2\times10^5\,$cm$^{-3}$, $\zeta = 10^{-16}\,$s$^{-1}$, and $v_s = 40\,$km s$^{-1}$.}
\label{fig:n2e5_model7}
\end{figure*}

\begin{figure*}
\centerline{\includegraphics[angle=0,width=.3\textwidth]{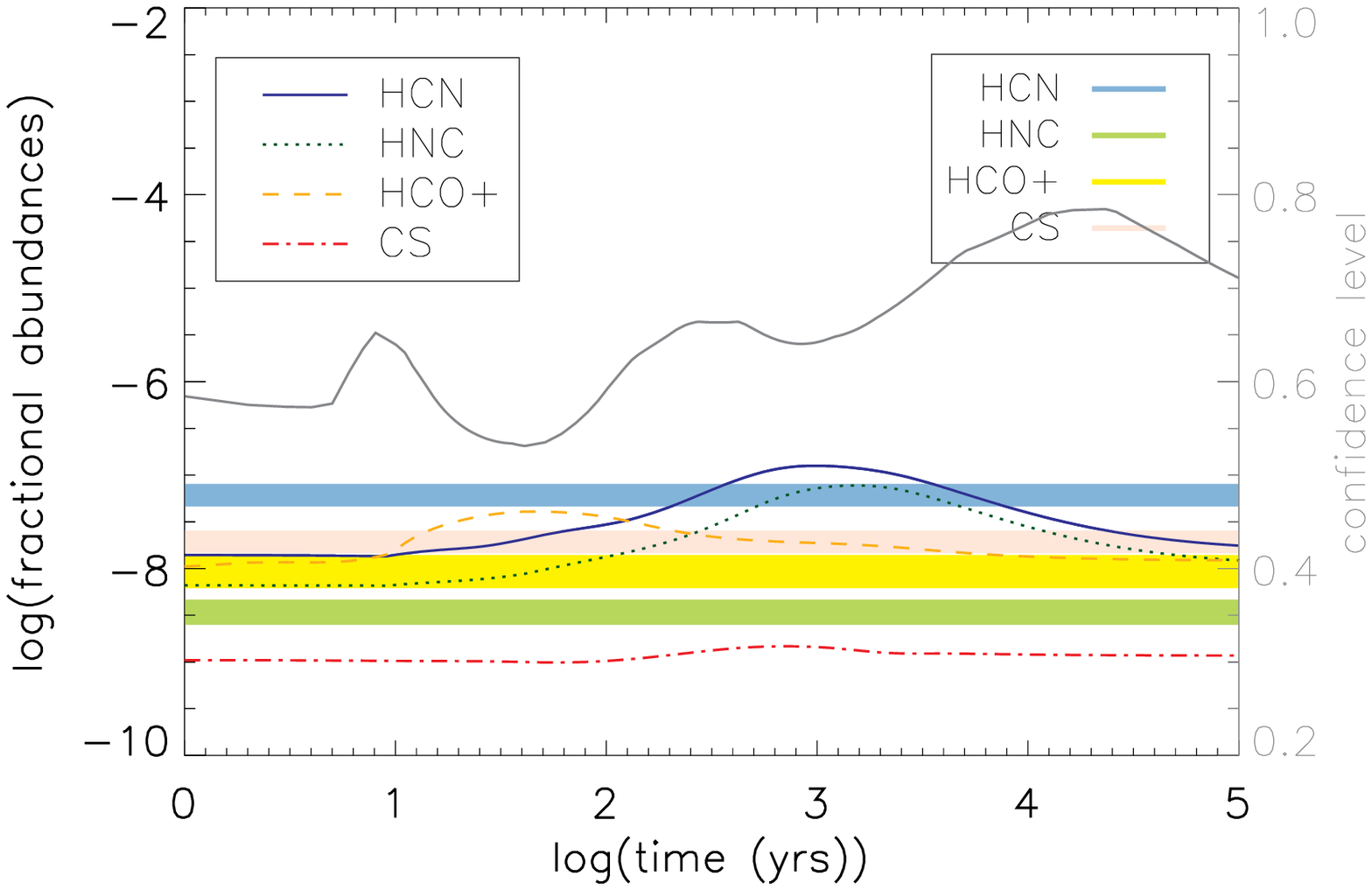}
\includegraphics[angle=0,width=.3\textwidth]{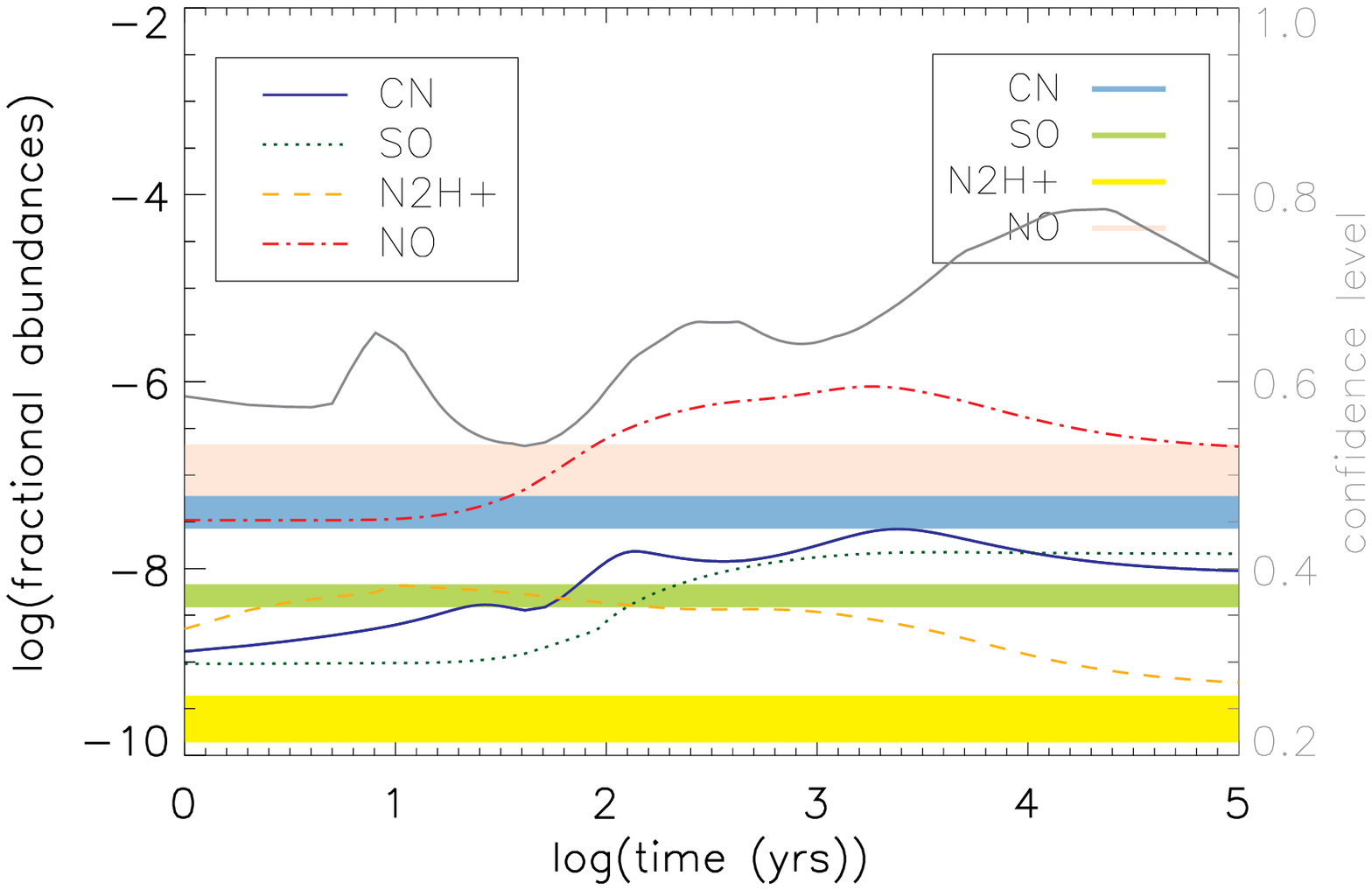}
\includegraphics[angle=0,width=.3\textwidth]{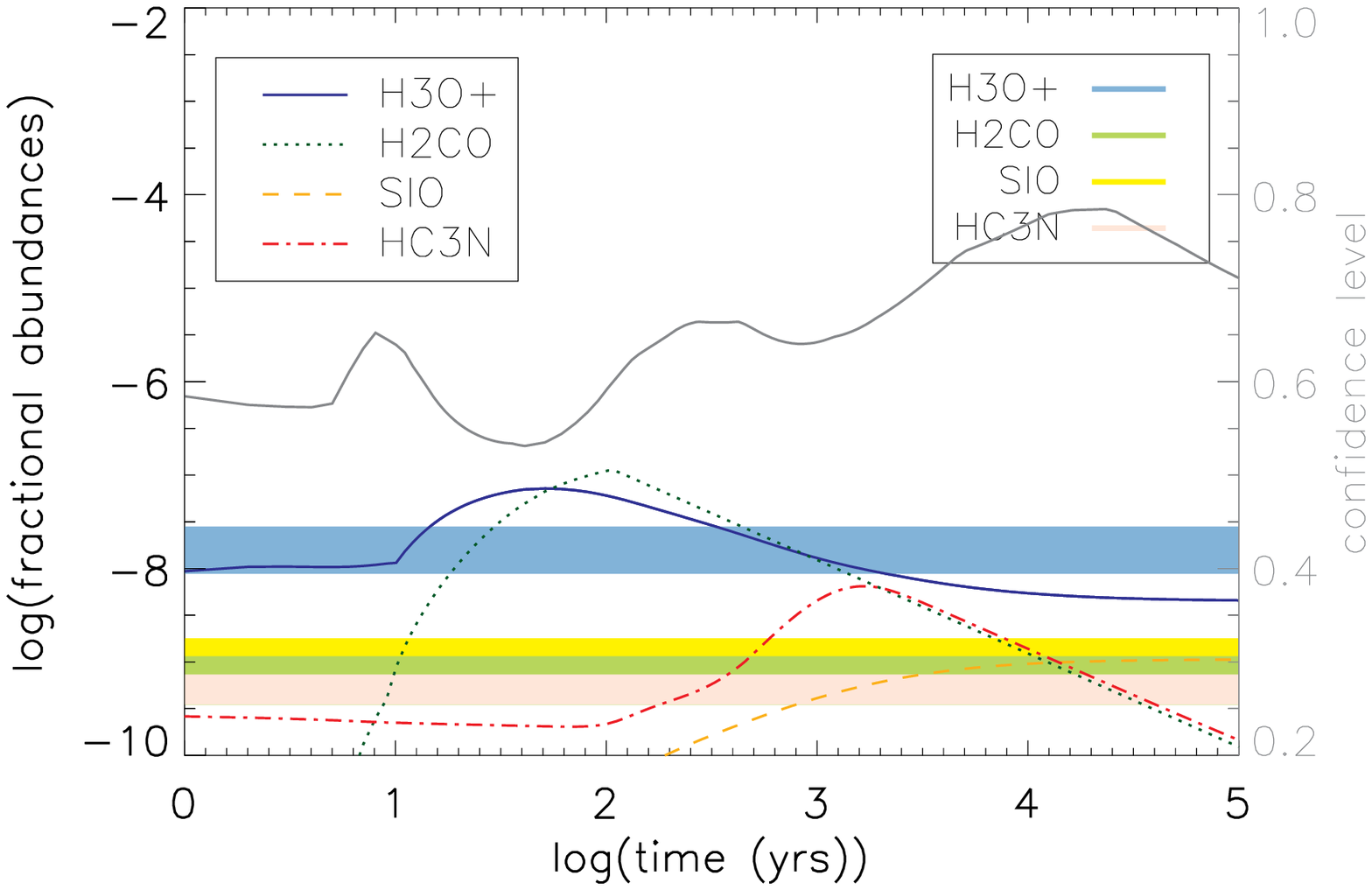}}
\caption{Same as Figure \ref{fig:n2e4_model2} for $n_H=2\times10^5\,$cm$^{-3}$, $\zeta = 10^{-14}\,$s$^{-1}$, and $v_s = 10\,$km s$^{-1}$.}
\label{fig:n2e5_model12}
\end{figure*}

Our models do not include X-rays, but their effect on the chemistry is very similar to that of cosmic-rays, and variation of the X-ray ionization rate can be mimicked by that of the cosmic-ray ionization rate. The difference in penetration depth may help differentiating between the effect of X-rays and cosmic-rays. As shown in Figure \ref{fig:zeta_x}, the X-ray ionization rate goes down below $\zeta_X =10^{-16}\,$s$^{-1}$ quickly after the shielding column density of $N_H=10^{21}\,$cm$^{-2}$ since we use a soft X-ray spectrum assuming most of the emission comes from stellar components as in \citet{2013JPCA..117.9919G}. Majority of the ionization is caused by these soft components. If the density is $ >10 ^{5}\,$cm$^{-3}$, this column density corresponds to 0.003pc, which is 1\% of the size of the molecular cloud in length. Although the observed chemical composition agrees with the models with ionization rate $\zeta > 10^{-15}\,$s$^{-1}$, X-rays are not likely the source of ionization if the average X-ray flux is similar to the value observed currently from Sgr A*. As mentioned in Section \ref{sec:phys_cond}, the X-ray activity may have been higher in the past. Although this higher (and harder spectrum) X-ray flux may have increased the X-ray ionization rate, the hard spectrum coming from the AGN may not significantly affect the ionization rate because of the smaller cross section. Also, the flares may be too short-lived to cause significant change in chemistry.

 \subsection{Possibility of no shock}
 As stated in Section \ref{sec:cmp_obs}, models without shocks yield good agreement with observations if the cosmic-ray ionization rate is high. This is due to the desorption of ice induced by cosmic rays described in \citet{2007MNRAS.382..733R}, and it can increase gas-phase abundances of shock-related molecules such as SiO. However, this mechanism of cosmic-ray induced photo-desorption has a large uncertainty in parameters. In addition, observations of ubiquitous complex organic molecules in the Galactic Center \citep{2008ApJ...672..352R} indicates that shock waves are common in the Galactic Center. Therefore, it is very likely that shocks are present in the CND.  
 
  \subsection{Other factors}
 The models discussed in this paper assume an enhancement of the cosmic-ray ionization rate and the contemporaneous passage of a shock wave. There are of course other possibilities. The shock wave may have passed earlier/later than the enhancement of the cosmic-ray ionization rate. There might have been multiple shock waves in the past $10^4\,$ yrs or so. The frequency of shock waves should be considered in future work with the help of hydrodynamic models of the CND. The pre-shock conditions have also some uncertainties. In our models, we run the pre-shock conditions with $\zeta=1\times 10^{-17}\,$s$^{-1}$ and $G_0=1$. It is likely that the enhancement of UV-field and cosmic-ray ionization rate has occurred before the shock passage, although it is almost impossible to trace the exact time evolution of these quantities.  
 
 The effects of turbulence may also need to be considered; it is possible that ionized or atomic gas might be fed into the molecular region continuously due to turbulence. The ionized gas may induce the chemical reactions similar to those in regions of high X-ray/cosmic-ray ionization rate. The effect of turbulence needs to be further investigated.

\section{Summary}
We have analysed the line survey data taken with the IRAM 30-meter telescope and the 12-meter APEX telescope toward a position in the southwestern lobe of the circumnuclear disk of the Galactic Center. Using an LVG radiative transfer code to predict the emission from 15 different molecular species we have derived column densities and H$_{2}$ volume densities implied by their excitation. A combined analysis of the excitation and the chemical modeling suggests that models with lower density ($n_{\rm H}\sim 2 \times 10^4\,$cm$^{-3}$), higher cosmic-ray ionization rate ($\zeta > 10^{-15}\,$s$^{-1}$), or high-velocity shocks ($v_s > 40\,$km s$^{-1}$) give good agreement with the observations. Considering the density derived from our observations and the estimated temperature in previous studies, the scenario of high cosmic-ray ionization rate is favored, but chemical compositions can also be partially affected by UV-photons or X-rays if the structure of the cloud is clumpy or filamentary.  Future study is needed both in terms of modeling and observation. Since the source is very heterogeneous, one-dimensional model that include XDRs and PDRs may be necessary. Observationally, high spatial resolution interferometeric maps that resolve the different structures are essential for the interpretation of this complex source. ALMA observations of the CND by Mills et al. (in preparation) should reveal such morphology of physical conditions in the CND.

\begin{appendix}
\section{Estimation of Physical Conditions}
\subsection{X-ray ionization rate}\label{sec:appen_xray}
As for the calculation of X-ray ionization rate, we followed the procedure described in \citet{2013JPCA..117.9919G}. From the X-ray luminosity measured in the 2-10 keV range by \citet{2003ApJ...591..891B} $L_X=2\times 10^{33}\,$ erg s$^{-1}$, X-ray flux in 0.1-2 keV range was estimated from the ATOMDB program \citep{2012ApJ...756..128F} with the plasma temperature of 1.9 keV. The ionization rate was taken to be 
\begin{equation}
\zeta_X = \int \sigma_{\rm tot}(E) \frac{E}{W(1{\rm keV})} F(E) dE,
\end{equation}
where $\sigma_{tot}$(E) is the ionizing cross section including the photo-ionization from heavy elements and Compton scattering at energy $E$, $W$ is the mean energy required for ionization \citep{1999ApJS..125..237D}, and $F$(E) is the flux at energy $E$. For $W$, a value at $E=1$ keV was used. In addition to the calculation by \citet{2013JPCA..117.9919G}, we calculated the dependence of X-ray ionization rate on the shielding hydrogen column densities $N$. The flux at energy $E$ becomes $F(E,N)=F_0 {\rm exp}(-\sigma_{\rm tot}(E) N)$. We also used 2 pc as the distance to the source. The calculated X-ray ionization rate is shown in Figure \ref{fig:zeta_x}. 

\subsection{UV-photon flux}\label{sec:app_uv}
\citet{1980ApJ...241..132L} estimated the number of ionizing photon
 flux to be $2\times10^{50}\,$s$^{-1}$ with the effective temperature of $T_{\rm eff}=31,000-35,000\,$K. From the stellar parameters listed in \citet{2005pcim.book.....T}, we
 consider spectral types of the stars in the central cluster to be similar to B0 types. The total number of ionizing photons
  is translated into $\sim140$ stars, and the total luminosity becomes $1.1\times10^{7}\,L_{\odot}$. Using Equation (9.1) in
   \citet{2005pcim.book.....T}, the radiation field strength $G_{0}$ becomes
\begin{equation}
G_0=625\frac{L_{*}\chi}{4\pi d^2}=1.3\times10^5(\chi/1)(d/1{\rm pc})^{-2},
\end{equation}
where $L_*$ is the luminosity of the star, $\chi$ is the fraction of luminosity above 6\,eV, and $d$ is the distance from the star.
\end{appendix}
\clearpage
\begin{table}[htp]
\centering
\begin{tabular}{cccc}
\hline
Molecule & Transition & Freq.  &  $\int^{-90}_{-120} T_{MB} dv$ \\
& &(GHz) &(K km s$^{-1}$)\\
\hline
CO &4-3  &461.04    &501.83\\
&6-5&691.47&895.68\\
&7-6&806.65&877.8\\
HCN & 1-0 &88.632  &56.860\\
 & 4-3 &354.51 &89.725\\
 &5-4 &443.116  &88.629\\
HNC & 1-0 &90.664  &8.0164\\
& 3-2 &271.981  &10.561\\
& 4-3 &362.630  &8.5620\\
HCO$^{+}$ &1-0 &89.1885 &23.681\\
&4-3 &356.7343 &45.648\\
SiO &2-1 &86.847  &2.0543 \\
&8-7 &347.331  &1.8295 \\
&9-8 &390.729 &1.7224 \\
CS &2-1 &97.981  &12.130\\
&6-5 &293.912  &22.196\\
&7-6 &342.883  &22.423\\
&8-7 &391.847  &22.484\\
p-H$_2$CO &2$_{0,2}$ - 1$_{0,1}$ &145.603  &2.1535\\
&4$_{0,4}  -3_{0,3}$ &290.623  &0.95264\\
N$_2$H$^+$ &3-2 &279.512  &1.6709\\ 
&4-3 &372.673   &1.7412\\
&5-4 &465.825  &1.2520\\
SO &$2_3   -   1_2$ &99.300  &1.0909\\
 &$7_8   -   6_7$ &304.077   &3.7813\\
 &$8_9   -   7_8$ &346.527  &3.7220\\
 &$9_{10}   -   8_9 $&389.120   &3.9306\\
CN &$1_{1/2}-0_{1/2}$ &113.169  &14.283\\
&$1_{3/2}-0_{1/2}$ &113.495   &14.004\\
&$3_{7/2}-2_{5/2}$ &340.249 &29.820\\
HC$_3$N &10-9 &90.979   &0.54742\\
 &18-17 &163.753  &1.0548\\
NO &4-3&350.69&2.4023 \\
p-H$_3$O$^+$ &$3_2+-2_2-$&364.797 &2.5981\\
&$3_1+ -2_1- $&388.459 &2.1160\\
$^{13}$CO &3-2 &330.588 &25.334\\
H$^{13}$CN &1-0 &86.3399 &5.0428\\
&2-1&172.6778&10.346\\
&4-3&345.3398&9.0259\\
H$^{13}$CO$^{+}$ &1-0 &86.7543 &1.1076\\
&2-1 &173.5067&1.4072\\
&4-3 &346.9983 &1.3180\\
\hline\\
\end{tabular}
\caption{A list of spectral lines used for chemical analysis. Their transitions, frequencies, and velocity-integrated intensities in the range of -120 to -90 km s$^{-1}$ are shown.}
\label{tab:lines}
\end{table}

\begin{table}[ht]
\centering
\begin{tabular}{ccc}
\hline\\
Species &Column Densities (cm$^{-2}$) &Hydrogen Densities (cm$^{-3}$)\\
\hline\\
CS &$(1.07\pm  0.13)\times 10^{15}$  &$(7.26   \pm   1.18)\times 10^{5}$\\
CN &$(5.48\pm  3.75)\times 10^{15}$  &---\\
H$_2$CO &$(5.01\pm  2.76)\times 10^{13}$  &---\\
SO &$(4.46\pm  0.61)\times 10^{14}$  &$(1.10  \pm   0. 21)\times 10^{6}$\\
N$_2$H$^+$ &$(1.47\pm  0.58)\times 10^{13}$ & ---\\
H$_3$O$^+$ &$(9.44\pm  0.38)\times 10^{14}$  &$(7.18  \pm    3.32)\times 10^{4}$\\
SiO &$(7.95\pm  0.49)\times 10^{13}$  &$(5.04  \pm   0.32)\times 10^{5}$\\
HCN &$(3.38 \pm 0.40)\times 10^{15}$  &$(8.70 \pm 1.76)\times 10^5$\\
HCO$^+$ &$(5.26\pm  1.28)\times 10^{14}$  &$(2.34 \pm  0.84)\times 10^{5}$\\
HNC &$(1.90\pm  0.29)\times 10^{14}$  &$(5.46 \pm   0.88)\times 10^{5}$\\
HC$_3$N &$(2.87\pm  0.58)\times 10^{13}$  &---\\
NO &$(3.02\pm  0.61)\times 10^{15}$  &---\\
CO &$(2.23\pm  0.35)\times 10^{18}$  &$(2.28  \pm 0.52)\times 10^{4}$\\
H$^{13}$CN &$(9.70\pm 2.30)\times 10^{13}$ &$(4.44 \pm   2.00)\times 10^{5}$\\
H$^{13}$CO$^+$ &$(4.02 \pm 2.55)\times 10^{13}$  &---\\
\hline\\
\end{tabular}
\caption{A list of best-fit column densities and number densities derived from the LVG analysis. Densities for some species are not shown due to the larger uncertainty considered as a consequence of the partial source coverage by the beam (see Section \ref{sec:obs}). Because CO and HCN are likely to be optically thick, $^{13}$C isotopologues are also used in the LVG analysis assuming the $^{12}C/^{13}C$ ratio of 25.}
\label{tab:obs_spc}
\end{table}

\begin{table}[ht]
\centering
\begin{tabular}{ccccc}
\hline
$n_{\rm H, pre}$(cm$^{-3}$) &$n_{\rm H, post}$(cm$^{-3}$) &$v_s$ (km/s) &$T_{max}$ (K)& $t_{sat}$ (yr) \\
\hline\\

$2\times10^4$&$6.1\times10^4$&10&300&95.4\\
$2\times10^4$&$1.0\times10^5$&20&900&57.0\\
$2\times10^4$&$1.3\times10^5$&30&1800&44.0\\
$2\times10^4$&$1.6\times10^5$&40&2200&45.5\\

$2\times10^5$&$6.3\times10^5$&10&300&10.5\\
$2\times10^5$&$1.0\times10^6$&20&800&5.7\\
$2\times10^5$&$1.4\times10^6$&30&2000&4.4\\
$2\times10^5$&$1.7\times10^6$&40&4000&4.6\\

$2\times10^6$&$6.3\times10^6$&10&300&1.0\\
$2\times10^6$&$1.0\times10^7$&20&800&0.6\\
$2\times10^6$&$1.4\times10^7$&30&2000&0.4\\
$2\times10^6$&$1.7\times10^7$&40&4000&0.5\\
\hline\\
\end{tabular}
\caption{Shock velocities used in this model and corresponding saturation times and maximum temperatures taken from \citet{2008A&A...482..549J}. In their paper, $t_{\rm sat}$ is defined as the time when sputtering of dust grains "saturates" the silicon abundance (i.e., when the silicon abundance changes between two consecutive time steps by less than 10 \%). Symbols $n_{\rm pre}$ and $n_{\rm post}$ stand for pre-shock and post-shock densities, respectively.} 
\label{tab:param}
\end{table}
\begin{table}[ht]
\centering
\begin{tabular}{ccccc}
\hline\\
Element &E1  &E2 \\
\hline\\
He  &0.14 &0.14\\
C  &7.3e-5 &7.3e-5\\
N  &2.1e-5 &2.1e-5\\
O  &1.8e-4&1.8e-4\\
S  &8.0e-8 &4.5e-7\\
Si  &8.0e-9&1.0e-6\\
Mg  &7.0e-9&2.4e-6\\
\hline\\
\end{tabular}
\caption{Elemental abundances used in the models with respect to the total hydrogen abundances. E1 is taken from the low-metal abundance case adopted in \citet{2008ApJ...680..371W} and taken from \citet{1982ApJS...48..321G} and \citet{1974ApJ...193L..35M}. For selected elements in E1, Si and Mg, we considered 3 \% of sputtering from the abundances in \citet{1989GeCoA..53..197A}, and other elements are taken from EA2 and EA3 of \citet{2008ApJ...680..371W}.}
\label{tab:elem}
\end{table}
\begin{table}[ht]
\centering
\begin{tabular}{ccccc}
\hline\\
 & \multicolumn{3}{c}{Pre-shock densities (cm$^{-3}$)}\\
Species  &$2\times 10^4$ &$2\times 10^5$ &$2\times 10^6$\\
\hline
 HCN     &  3.6(-9)     &  2.8(-8)     &  6.1(-10)\\
  HCN (ice)    &  8.5(-9)      &  2.1(-8)    &  3.5(-8)\\
 HNC     &  7.9(-10)    &  1.2(-8)     &  3.8(-10)\\
  HNC (ice)    &  5.4(-11)    &  3.0(-9)     &  9.8(-9)\\
 CN      &  3.5(-8)       &  4.4(-9)      &  9.6(-12)\\
 HCO$^+$    &  3.3(-11)     &  1.4(-9)     &  7.8(-10)\\
 CO      &  9.0(-6)    &  3.6(-5)     &  5.9(-6)\\
 CO (ice)     &  3.0(-7)    &  9.3(-6)     &  4.2(-5)\\
 CS      &  8.4(-10)      &  1.9(-9)        &  1.2(-10)\\
 CS (ice)     &  2.1(-14)    &  2.2(-14)      &  2.2(-14)\\
 SO      &  8.1(-14)      &  1.4(-12)       &  6.7(-11)\\
 SO (ice)     &  4.5(-14)    &  6.7(-13)    &  6.9(-11)\\
 H$_2$S (ice)  &  7.0(-8)    &  7.6(-8)    &  7.8(-8)\\
 N$_2$H$^+$    &  2.9(-11)    &  6.9(-11)    &  7.0(-11)\\
 N$_2$ (ice)    &  1.9(-7)     &  2.0(-6)     &  6.3(-6)\\
 NH$_3$ (ice)   &  2.6(-6)    &  4.8(-6)    &  6.5(-6)\\
 NO      &  2.0(-9)      &  3.1(-9)       &  8.4(-8)\\
 NO (ice)    &  8.4(-11)    &  8.1(-10)     &  1.3(-7)\\
 H$_3$O$^+$    &  6.9(-11)    &  1.7(-9)    &  3.9(-10)\\
 H$_2$O     &  9.2(-9)    &  2.9(-7)    &  8.9(-8)\\
 H$_2$O (ice)   &  2.9(-5)    &  7.3(-5)     &  1.2(-4)\\
 H$_2$CO    &  9.8(-10)    &  3.5(-10)     &  2.4(-10)\\
 H$_2$CO (ice)  &  1.0(-10)   &  5.4(-9)   &  3.8(-8)\\
 SiO     &  6.2(-12)    &  1.4(-11)    &  1.7(-12)\\
 SiH$_4$ (ice)  &  7.6(-9)  &  7.9(-9)     &  7.9(-9) \\
 HC$_3$N    &  5.3(-14)   &  2.8(-11)    &  4.8(-11)\\
  HC$_3$N (ice)  &  2.0(-15)   &  2.1(-12)    &  2.1(-10)\\
  CH$_{4}$ (ice) &6.5(-6) &2.0(-5) &2.2(-5) \\
\hline\\
\end{tabular}
\caption{Pre-shock fractional abundances of observed species and other related species in different pre-shock densities of $n=2\times 10^4$ cm$^{-3}$, $2\times 10^5$ cm$^{-3}$, and $2\times 10^6$ cm$^{-3}$. Species on grain surfaces are indicated with "(ice)" while species without any note mean that they are in the gas phase. A value of $a\times 10^{-b}$ is expressed as a(-b). }
\label{tab:preshock}
\end{table}

\clearpage


\begin{acknowledgements}
NH thanks Miwa Goto for sharing her results on X-ray ionization rate in her paper. IJ-S acknowledges funding from the People Programme (Marie Curie Actions) of the European Union's Seventh Framework Programme (FP7/2007-2013) under REA grant
agreement number PIIF-GA-2011-301538.
\end{acknowledgements}
\bibliographystyle{aa}
\bibliography{gc}

\begin{thebibliography}{64}
\expandafter\ifx\csname natexlab\endcsname\relax\def\natexlab#1{#1}\fi

\bibitem[{{Amo-Baladr{\'o}n} {et~al.}(2011){Amo-Baladr{\'o}n},
  {Mart{\'{\i}}n-Pintado}, \& {Mart{\'{\i}}n}}]{2011A&A...526A..54A}
{Amo-Baladr{\'o}n}, M.~A., {Mart{\'{\i}}n-Pintado}, J., \& {Mart{\'{\i}}n}, S.
  2011, \aap, 526, A54

\bibitem[{{Anders} \& {Grevesse}(1989)}]{1989GeCoA..53..197A}
{Anders}, E. \& {Grevesse}, N. 1989, \gca, 53, 197

\bibitem[{{Ao} {et~al.}(2013){Ao}, {Henkel}, {Menten}, {Requena-Torres},
  {Stanke}, {Mauersberger}, {Aalto}, {M{\"u}hle}, \&
  {Mangum}}]{2013A&A...550A.135A}
{Ao}, Y., {Henkel}, C., {Menten}, K.~M., {et~al.} 2013, \aap, 550, A135

\bibitem[{{Baganoff} {et~al.}(2003){Baganoff}, {Maeda}, {Morris}, {Bautz},
  {Brandt}, {Cui}, {Doty}, {Feigelson}, {Garmire}, {Pravdo}, {Ricker}, \&
  {Townsley}}]{2003ApJ...591..891B}
{Baganoff}, F.~K., {Maeda}, Y., {Morris}, M., {et~al.} 2003, \apj, 591, 891

\bibitem[{{Bradford} {et~al.}(2005){Bradford}, {Stacey}, {Nikola}, {Bolatto},
  {Jackson}, {Savage}, \& {Davidson}}]{2005ApJ...623..866B}
{Bradford}, C.~M., {Stacey}, G.~J., {Nikola}, T., {et~al.} 2005, \apj, 623, 866

\bibitem[{{Capelli} {et~al.}(2012){Capelli}, {Warwick}, {Porquet}, {Gillessen},
  \& {Predehl}}]{2012A&A...545A..35C}
{Capelli}, R., {Warwick}, R.~S., {Porquet}, D., {Gillessen}, S., \& {Predehl},
  P. 2012, \aap, 545, A35

\bibitem[{{Chernyakova} {et~al.}(2011){Chernyakova}, {Malyshev}, {Aharonian},
  {Crocker}, \& {Jones}}]{2011ApJ...726...60C}
{Chernyakova}, M., {Malyshev}, D., {Aharonian}, F.~A., {Crocker}, R.~M., \&
  {Jones}, D.~I. 2011, \apj, 726, 60

\bibitem[{{Christopher} {et~al.}(2005){Christopher}, {Scoville}, {Stolovy}, \&
  {Yun}}]{2005ApJ...622..346C}
{Christopher}, M.~H., {Scoville}, N.~Z., {Stolovy}, S.~R., \& {Yun}, M.~S.
  2005, \apj, 622, 346

\bibitem[{{Dalgarno} {et~al.}(1999){Dalgarno}, {Yan}, \&
  {Liu}}]{1999ApJS..125..237D}
{Dalgarno}, A., {Yan}, M., \& {Liu}, W. 1999, \apjs, 125, 237

\bibitem[{{Flower} \& {Pineau des For{\^e}ts}(2012)}]{2012MNRAS.421.2786F}
{Flower}, D.~R. \& {Pineau des For{\^e}ts}, G. 2012, \mnras, 421, 2786

\bibitem[{{Foster} {et~al.}(2012){Foster}, {Ji}, {Smith}, \&
  {Brickhouse}}]{2012ApJ...756..128F}
{Foster}, A.~R., {Ji}, L., {Smith}, R.~K., \& {Brickhouse}, N.~S. 2012, \apj,
  756, 128

\bibitem[{{Garrod} {et~al.}(2007){Garrod}, {Wakelam}, \&
  {Herbst}}]{2007A&A...467.1103G}
{Garrod}, R.~T., {Wakelam}, V., \& {Herbst}, E. 2007, \aap, 467, 1103

\bibitem[{{Goto} {et~al.}(2014){Goto}, {Geballe}, {Indriolo}, {Yusef-Zadeh},
  {Usuda}, {Henning}, \& {Oka}}]{2014ApJ...786...96G}
{Goto}, M., {Geballe}, T.~R., {Indriolo}, N., {et~al.} 2014, \apj, 786, 96

\bibitem[{{Goto} {et~al.}(2013){Goto}, {Indriolo}, {Geballe}, \&
  {Usuda}}]{2013JPCA..117.9919G}
{Goto}, M., {Indriolo}, N., {Geballe}, T.~R., \& {Usuda}, T. 2013, Journal of
  Physical Chemistry A, 117, 9919

\bibitem[{{Goto} {et~al.}(2008){Goto}, {Usuda}, {Nagata}, {Geballe}, {McCall},
  {Indriolo}, {Suto}, {Henning}, {Morong}, \& {Oka}}]{2008ApJ...688..306G}
{Goto}, M., {Usuda}, T., {Nagata}, T., {et~al.} 2008, \apj, 688, 306

\bibitem[{{Graedel} {et~al.}(1982){Graedel}, {Langer}, \&
  {Frerking}}]{1982ApJS...48..321G}
{Graedel}, T.~E., {Langer}, W.~D., \& {Frerking}, M.~A. 1982, \apjs, 48, 321

\bibitem[{{Graninger} {et~al.}(2014){Graninger}, {Herbst}, {{\"O}berg}, \&
  {Vasyunin}}]{2014ApJ...787...74G}
{Graninger}, D.~M., {Herbst}, E., {{\"O}berg}, K.~I., \& {Vasyunin}, A.~I.
  2014, \apj, 787, 74

\bibitem[{{Gusdorf} {et~al.}(2008){Gusdorf}, {Cabrit}, {Flower}, \& {Pineau Des
  For{\^e}ts}}]{2008A&A...482..809G}
{Gusdorf}, A., {Cabrit}, S., {Flower}, D.~R., \& {Pineau Des For{\^e}ts}, G.
  2008, \aap, 482, 809

\bibitem[{{G\"usten} {et~al.}(1987){G\"usten}, {Genzel}, {Wright}, {Jaffe},
  {Stutzki}, \& {Harris}}]{1987ApJ...318..124G}
{G\"usten}, R., {Genzel}, R., {Wright}, M.~C.~H., {et~al.} 1987, \apj, 318, 124

\bibitem[{{G{\"u}sten} {et~al.}(2006){G{\"u}sten}, {Nyman}, {Schilke},
  {Menten}, {Cesarsky}, \& {Booth}}]{2006A&A...454L..13G}
{G{\"u}sten}, R., {Nyman}, L.~{\AA}., {Schilke}, P., {et~al.} 2006, \aap, 454,
  L13

\bibitem[{{Indriolo} {et~al.}(2009){Indriolo}, {Fields}, \&
  {McCall}}]{2009ApJ...694..257I}
{Indriolo}, N., {Fields}, B.~D., \& {McCall}, B.~J. 2009, \apj, 694, 257

\bibitem[{{Jim{\'e}nez-Serra} {et~al.}(2008){Jim{\'e}nez-Serra}, {Caselli},
  {Mart{\'{\i}}n-Pintado}, \& {Hartquist}}]{2008A&A...482..549J}
{Jim{\'e}nez-Serra}, I., {Caselli}, P., {Mart{\'{\i}}n-Pintado}, J., \&
  {Hartquist}, T.~W. 2008, \aap, 482, 549

\bibitem[{{Kasemann} {et~al.}(2006){Kasemann}, {G{\"u}sten}, {Heyminck},
  {Klein}, {Klein}, {Philipp}, {Korn}, {Schneider}, {Henseler}, {Baryshev}, \&
  {Klapwijk}}]{2006SPIE.6275E..0NK}
{Kasemann}, C., {G{\"u}sten}, R., {Heyminck}, S., {et~al.} 2006, in Society of
  Photo-Optical Instrumentation Engineers (SPIE) Conference Series, Vol. 6275,
  Society of Photo-Optical Instrumentation Engineers (SPIE) Conference Series

\bibitem[{{Klein} {et~al.}(2014){Klein}, {Ciechanowicz}, {Leinz}, {Heyminck},
  {G\"usten}, {Kasemann}, {Wunsch}, {Maier}, \& {Sekimoto}}]{klein_flash}
{Klein}, T., {Ciechanowicz}, M., {Leinz}, C., {et~al.} 2014, IEEE Trans. on
  Terahertz Science and Technology, 4, 588

\bibitem[{{Koyama} {et~al.}(2007){Koyama}, {Hyodo}, {Inui}, {Nakajima},
  {Matsumoto}, {Tsuru}, {Takahashi}, {Maeda}, {Yamazaki}, {Murakami},
  {Yamauchi}, {Tsuboi}, {Senda}, {Kataoka}, {Takahashi}, {Holt}, \&
  {Brown}}]{2007PASJ...59S.245K}
{Koyama}, K., {Hyodo}, Y., {Inui}, T., {et~al.} 2007, \pasj, 59, 245

\bibitem[{{Lacy} {et~al.}(1980){Lacy}, {Townes}, {Geballe}, \&
  {Hollenbach}}]{1980ApJ...241..132L}
{Lacy}, J.~H., {Townes}, C.~H., {Geballe}, T.~R., \& {Hollenbach}, D.~J. 1980,
  \apj, 241, 132

\bibitem[{{Le Petit} {et~al.}(2006){Le Petit}, {Nehm{\'e}}, {Le Bourlot}, \&
  {Roueff}}]{2006ApJS..164..506L}
{Le Petit}, F., {Nehm{\'e}}, C., {Le Bourlot}, J., \& {Roueff}, E. 2006, \apjs,
  164, 506

\bibitem[{{Linden} {et~al.}(2012){Linden}, {Lovegrove}, \&
  {Profumo}}]{2012ApJ...753...41L}
{Linden}, T., {Lovegrove}, E., \& {Profumo}, S. 2012, \apj, 753, 41

\bibitem[{{Lo}(1986)}]{1986Sci...233.1394L}
{Lo}, K.~Y. 1986, Science, 233, 1394

\bibitem[{{Mart{\'{\i}}n} {et~al.}(2012){Mart{\'{\i}}n},
  {Mart{\'{\i}}n-Pintado}, {Montero-Casta{\~n}o}, {Ho}, \&
  {Blundell}}]{2012A&A...539A..29M}
{Mart{\'{\i}}n}, S., {Mart{\'{\i}}n-Pintado}, J., {Montero-Casta{\~n}o}, M.,
  {Ho}, P.~T.~P., \& {Blundell}, R. 2012, \aap, 539, A29

\bibitem[{{Mart{\'{\i}}n-Pintado} {et~al.}(1997){Mart{\'{\i}}n-Pintado}, {de
  Vicente}, {Fuente}, \& {Planesas}}]{1997ApJ...482L..45M}
{Mart{\'{\i}}n-Pintado}, J., {de Vicente}, P., {Fuente}, A., \& {Planesas}, P.
  1997, \apjl, 482, L45

\bibitem[{{Mills} {et~al.}(2013){Mills}, {G{\"u}sten}, {Requena-Torres}, \&
  {Morris}}]{2013ApJ...779...47M}
{Mills}, E.~A.~C., {G{\"u}sten}, R., {Requena-Torres}, M.~A., \& {Morris},
  M.~R. 2013, \apj, 779, 47

\bibitem[{{Montero-Casta{\~n}o} {et~al.}(2009){Montero-Casta{\~n}o},
  {Herrnstein}, \& {Ho}}]{2009ApJ...695.1477M}
{Montero-Casta{\~n}o}, M., {Herrnstein}, R.~M., \& {Ho}, P.~T.~P. 2009, \apj,
  695, 1477

\bibitem[{{Morton}(1974)}]{1974ApJ...193L..35M}
{Morton}, D.~C. 1974, \apjl, 193, L35

\bibitem[{{Nobukawa} {et~al.}(2010){Nobukawa}, {Koyama}, {Tsuru}, {Ryu}, \&
  {Tatischeff}}]{2010PASJ...62..423N}
{Nobukawa}, M., {Koyama}, K., {Tsuru}, T.~G., {Ryu}, S.~G., \& {Tatischeff}, V.
  2010, \pasj, 62, 423

\bibitem[{{Oka} {et~al.}(2011){Oka}, {Nagai}, {Kamegai}, \&
  {Tanaka}}]{2011ApJ...732..120O}
{Oka}, T., {Nagai}, M., {Kamegai}, K., \& {Tanaka}, K. 2011, \apj, 732, 120

\bibitem[{{Padovani} {et~al.}(2009){Padovani}, {Galli}, \&
  {Glassgold}}]{2009A&A...501..619P}
{Padovani}, M., {Galli}, D., \& {Glassgold}, A.~E. 2009, \aap, 501, 619

\bibitem[{{Paumard} {et~al.}(2006){Paumard}, {Genzel}, {Martins}, {Nayakshin},
  {Beloborodov}, {Levin}, {Trippe}, {Eisenhauer}, {Ott}, {Gillessen}, {Abuter},
  {Cuadra}, {Alexander}, \& {Sternberg}}]{2006ApJ...643.1011P}
{Paumard}, T., {Genzel}, R., {Martins}, F., {et~al.} 2006, \apj, 643, 1011

\bibitem[{{Ponti} {et~al.}(2013){Ponti}, {Morris}, {Terrier}, \&
  {Goldwurm}}]{2013ASSP...34..331P}
{Ponti}, G., {Morris}, M.~R., {Terrier}, R., \& {Goldwurm}, A. 2013, in
  Advances in Solid State Physics, Vol.~34, Cosmic Rays in Star-Forming
  Environments, ed. D.~F. {Torres} \& O.~{Reimer}, 331

\bibitem[{{Ponti} {et~al.}(2010){Ponti}, {Terrier}, {Goldwurm}, {Belanger}, \&
  {Trap}}]{2010ApJ...714..732P}
{Ponti}, G., {Terrier}, R., {Goldwurm}, A., {Belanger}, G., \& {Trap}, G. 2010,
  \apj, 714, 732

\bibitem[{{Prasad} \& {Tarafdar}(1983)}]{1983ApJ...267..603P}
{Prasad}, S.~S. \& {Tarafdar}, S.~P. 1983, \apj, 267, 603

\bibitem[{{Reid} {et~al.}(2009){Reid}, {Menten}, {Zheng}, {Brunthaler},
  {Moscadelli}, {Xu}, {Zhang}, {Sato}, {Honma}, {Hirota}, {Hachisuka}, {Choi},
  {Moellenbrock}, \& {Bartkiewicz}}]{2009ApJ...700..137R}
{Reid}, M.~J., {Menten}, K.~M., {Zheng}, X.~W., {et~al.} 2009, \apj, 700, 137

\bibitem[{{Requena-Torres} {et~al.}(2012){Requena-Torres}, {G{\"u}sten},
  {Wei{\ss}}, {Harris}, {Mart{\'{\i}}n-Pintado}, {Stutzki}, {Klein},
  {Heyminck}, \& {Risacher}}]{2012A&A...542L..21R}
{Requena-Torres}, M.~A., {G{\"u}sten}, R., {Wei{\ss}}, A., {et~al.} 2012, \aap,
  542, L21

\bibitem[{{Requena-Torres} {et~al.}(2008){Requena-Torres},
  {Mart{\'{\i}}n-Pintado}, {Mart{\'{\i}}n}, \& {Morris}}]{2008ApJ...672..352R}
{Requena-Torres}, M.~A., {Mart{\'{\i}}n-Pintado}, J., {Mart{\'{\i}}n}, S., \&
  {Morris}, M.~R. 2008, \apj, 672, 352

\bibitem[{{Requena-Torres} {et~al.}(2006){Requena-Torres},
  {Mart{\'{\i}}n-Pintado}, {Rodr{\'{\i}}guez-Franco}, {Mart{\'{\i}}n},
  {Rodr{\'{\i}}guez-Fern{\'a}ndez}, \& {de Vicente}}]{2006A&A...455..971R}
{Requena-Torres}, M.~A., {Mart{\'{\i}}n-Pintado}, J.,
  {Rodr{\'{\i}}guez-Franco}, A., {et~al.} 2006, \aap, 455, 971

\bibitem[{{Revnivtsev} {et~al.}(2004){Revnivtsev}, {Churazov}, {Sazonov},
  {Sunyaev}, {Lutovinov}, {Gilfanov}, {Vikhlinin}, {Shtykovsky}, \&
  {Pavlinsky}}]{2004A&A...425L..49R}
{Revnivtsev}, M.~G., {Churazov}, E.~M., {Sazonov}, S.~Y., {et~al.} 2004, \aap,
  425, L49

\bibitem[{{Roberts} {et~al.}(2007){Roberts}, {Rawlings}, {Viti}, \&
  {Williams}}]{2007MNRAS.382..733R}
{Roberts}, J.~F., {Rawlings}, J.~M.~C., {Viti}, S., \& {Williams}, D.~A. 2007,
  \mnras, 382, 733

\bibitem[{{Sakai} {et~al.}(2008){Sakai}, {Sakai}, {Hirota}, \&
  {Yamamoto}}]{2008ApJ...672..371S}
{Sakai}, N., {Sakai}, T., {Hirota}, T., \& {Yamamoto}, S. 2008, \apj, 672, 371

\bibitem[{{Schilke} {et~al.}(1997){Schilke}, {Walmsley}, {Pineau des Forets},
  \& {Flower}}]{1997A&A...321..293S}
{Schilke}, P., {Walmsley}, C.~M., {Pineau des Forets}, G., \& {Flower}, D.~R.
  1997, \aap, 321, 293

\bibitem[{{Sch{\"o}del} {et~al.}(2009){Sch{\"o}del}, {Merritt}, \&
  {Eckart}}]{2009A&A...502...91S}
{Sch{\"o}del}, R., {Merritt}, D., \& {Eckart}, A. 2009, \aap, 502, 91

\bibitem[{{Serabyn} {et~al.}(1986){Serabyn}, {Guesten}, {Walmsley}, {Wink}, \&
  {Zylka}}]{1986A&A...169...85S}
{Serabyn}, E., {Guesten}, R., {Walmsley}, J.~E., {Wink}, J.~E., \& {Zylka}, R.
  1986, \aap, 169, 85

\bibitem[{{Smith} \& {Wardle}(2014)}]{2014MNRAS.437.3159S}
{Smith}, I.~L. \& {Wardle}, M. 2014, \mnras, 437, 3159

\bibitem[{{Takekawa} {et~al.}(2014){Takekawa}, {Oka}, {Tanaka}, {Matsumura},
  {Miura}, \& {Sakai}}]{2014ApJS..214....2T}
{Takekawa}, S., {Oka}, T., {Tanaka}, K., {et~al.} 2014, \apjs, 214, 2

\bibitem[{{Terrier} {et~al.}(2010){Terrier}, {Ponti}, {B{\'e}langer},
  {Decourchelle}, {Tatischeff}, {Goldwurm}, {Trap}, {Morris}, \&
  {Warwick}}]{2010ApJ...719..143T}
{Terrier}, R., {Ponti}, G., {B{\'e}langer}, G., {et~al.} 2010, \apj, 719, 143

\bibitem[{{Tielens}(2005)}]{2005pcim.book.....T}
{Tielens}, A.~G.~G.~M. 2005, {The Physics and Chemistry of the Interstellar
  Medium} (Cambridge University Press)

\bibitem[{{van der Tak} {et~al.}(2006){van der Tak}, {Belloche}, {Schilke},
  {G{\"u}sten}, {Philipp}, {Comito}, {Bergman}, \&
  {Nyman}}]{2006A&A...454L..99V}
{van der Tak}, F.~F.~S., {Belloche}, A., {Schilke}, P., {et~al.} 2006, \aap,
  454, L99

\bibitem[{{van der Tak} {et~al.}(2007){van der Tak}, {Black}, {Sch{\"o}ier},
  {Jansen}, \& {van Dishoeck}}]{2007A&A...468..627V}
{van der Tak}, F.~F.~S., {Black}, J.~H., {Sch{\"o}ier}, F.~L., {Jansen}, D.~J.,
  \& {van Dishoeck}, E.~F. 2007, \aap, 468, 627

\bibitem[{{Viti} {et~al.}(2004){Viti}, {Collings}, {Dever}, {McCoustra}, \&
  {Williams}}]{2004MNRAS.354.1141V}
{Viti}, S., {Collings}, M.~P., {Dever}, J.~W., {McCoustra}, M.~R.~S., \&
  {Williams}, D.~A. 2004, \mnras, 354, 1141

\bibitem[{{Viti} {et~al.}(2011){Viti}, {Jimenez-Serra}, {Yates}, {Codella},
  {Vasta}, {Caselli}, {Lefloch}, \& {Ceccarelli}}]{2011ApJ...740L...3V}
{Viti}, S., {Jimenez-Serra}, I., {Yates}, J.~A., {et~al.} 2011, \apjl, 740, L3

\bibitem[{{Wakelam} \& {Herbst}(2008)}]{2008ApJ...680..371W}
{Wakelam}, V. \& {Herbst}, E. 2008, \apj, 680, 371

\bibitem[{{Woodall} {et~al.}(2007){Woodall}, {Ag{\'u}ndez}, {Markwick-Kemper},
  \& {Millar}}]{2007A&A...466.1197W}
{Woodall}, J., {Ag{\'u}ndez}, M., {Markwick-Kemper}, A.~J., \& {Millar}, T.~J.
  2007, \aap, 466, 1197

\bibitem[{{Yusef-Zadeh} {et~al.}(2013{\natexlab{a}}){Yusef-Zadeh}, {Cotton},
  {Viti}, {Wardle}, \& {Royster}}]{2013ApJ...764L..19Y}
{Yusef-Zadeh}, F., {Cotton}, W., {Viti}, S., {Wardle}, M., \& {Royster}, M.
  2013{\natexlab{a}}, \apjl, 764, L19

\bibitem[{{Yusef-Zadeh} {et~al.}(2013{\natexlab{b}}){Yusef-Zadeh}, {Hewitt},
  {Wardle}, {Tatischeff}, {Roberts}, {Cotton}, {Uchiyama}, {Nobukawa}, {Tsuru},
  {Heinke}, \& {Royster}}]{2013ApJ...762...33Y}
{Yusef-Zadeh}, F., {Hewitt}, J.~W., {Wardle}, M., {et~al.} 2013{\natexlab{b}},
  \apj, 762, 33

\bibitem[{{Yusef-Zadeh} {et~al.}(2007){Yusef-Zadeh}, {Muno}, {Wardle}, \&
  {Lis}}]{2007ApJ...656..847Y}
{Yusef-Zadeh}, F., {Muno}, M., {Wardle}, M., \& {Lis}, D.~C. 2007, \apj, 656,
  847

\end{thebibliography}

\end{document}